\documentclass[aps,prb,twocolumn,amsmath,amssymb,superscriptaddress]{revtex4-1}

\usepackage{amssymb}
\usepackage{amsmath}
\usepackage{graphicx}
\usepackage{textcomp}
\usepackage{color}
\usepackage{amsfonts}
\usepackage{epsfig}
\usepackage{hyperref}

\newcommand{\arctanh}[1]{\operatorname{arctan}}
\newcommand{\dtilde}[1]{\tilde{\tilde{#1}}}

\providecommand{\supi}[1]{\raisebox{0.5ex}{\footnotesize $#1$}}

\begin{document}

\title{Quantum transport simulation scheme including strong correlations and its application to organic radicals adsorbed on gold}
\author{Andrea Droghetti}
\email{andrea.droghetti@ehu.eus}
\affiliation{Nano-Bio Spectroscopy Group and European Theoretical Spectroscopy Facility (ETSF), 
Universidad del Pais Vasco CFM CSIC-UPV/EHU-MPC and DIPC,Av.Tolosa 72 ,20018 San Sebastian, Spain}
\author{Ivan Rungger}
\email{ivan.rungger@npl.co.uk}
\affiliation{National Physical Laboratory, Hampton Road, TW11 0LW, United Kingdom}
\begin{abstract}
We present a computational method to quantitatively describe the linear-response conductance of nanoscale devices in the Kondo regime. 
This method relies on a projection scheme to extract an Anderson impurity model from the results of density functional theory and non-equilibrium Green's functions calculations.
The Anderson impurity model is then solved by 
continuous time quantum Monte Carlo. The developed formalism allows us to separate the different contributions to the transport, 
including coherent or non-coherent transport channels, and also the quantum interference between impurity and background transmission.
We apply the method to a scanning tunneling microscope setup for the 1,3,5-triphenyl-6-oxoverdazyl (TOV) stable radical molecule adsorbed on gold. 
The TOV molecule has one unpaired electron, which when brought in contact with metal electrodes behaves like a prototypical single Anderson impurity.
We evaluate the Kondo temperature, the finite temperature spectral function and transport properties, finding good agreement with published experimental results.
  
\end{abstract}

\maketitle

\section{Introduction}
In recent years, much research effort has been dedicated to study the electronic transport through magnetic molecules and single atoms in order to combine 
molecular electronics with spintronics\cite{Alex,Bogani,Nature_nano}.
Experiments and theoretical works have demonstrated that molecular and atomic spin states can be inferred and sometimes switched 
through an electrical current\cite{Heersche,Das,Timm,Grose,Osorio,Meded,Prins,Candini,Urdampilleta,Komeda,Burzuri,Vincent,Baadji,Wagner,Urdampilleta2}.
Among the many interesting phenomena arising in devices comprising magnetic molecules, there is the Kondo effect\cite{Kondo,Hewson,Kouwenhoven,Scott}. 
Below the so called Kondo temperature, $\theta_\mathrm{K}$, characteristic of each system, the coupling between the electrons from the electrodes and the spin of the molecule 
promotes the formation of a many-body state with a fully or partially quenched magnetic moment. 
This results in a new resonant transport channel at the electrodes Fermi level.
To date the Kondo effect has been studied in a number of molecular devices
 \cite{Osorio,Meded,Burzuri,Wagner,Park2, Liang, Yu,Wahl,Yu2, Parks,Zhao,Iancu1,Iancu2,Gao,Roch,
Perera,Choi,Tsukahara,Miyamachi,DiLullo,Fu,BWHeinrich,Krull,Komeda2,Iancu3,Wu},
 and exotic manifestations, such as the orbital Kondo effect, have also been reported\cite{Jarillo}.
 
{\it Ab-initio} computational studies play a prominent role in molecular electronics and spintronics. 
 In particular density functional theory (DFT) combined with the non-equilibrium Green's functions (NEGF) formalism\cite{Datta}, 
 known as DFT+NEGF for short,
has become the dominant method to address electronic transport\cite{Taylor, Xue, Brandbyge, Palacios, Pecchia, Smeagol,SelfEnergies,gollum}. 
As initial step in a typical DFT+NEGF study of a molecular device, one optimizes the atomic configuration of the device active region, 
usually called ``scattering region'' or ``extended molecule''. 
Then the scattering region is joint to two semi-infinite left- and right-hand side leads (electrodes), 
whose effect on the states in the scattering region is taken into account by the so-called leads' self-energies.  
This approach allows to treat the extended molecule as an open system,
with the leads' self-energies that provide quantitative estimates of the molecule-electrode hybridization and of its dynamic character\cite{Andrea}.
Finally, the device transmission function and conductance are calculated within the Landauer framework\cite{Datta}, 
by considering the Kohn-Sham (KS) eigenvalues as the single-particle excitations. 
Although this assumption is formally not correct, since even in exact DFT only the
highest occupied molecular orbital (HOMO) can be rigorously associated to the negative of the ionization potential\cite{Levy,Almbladh,Perdew_Parr,Perdew_Levy,SICtransport}, 
it practically works well for metallic point-contacts\cite{singleAtom}, nanowires and nanotubes \cite{forcespaperPRB, XihuaDNA}, quasi two-dimensional systems\cite{Awadhesh_bise_step},
and tunnel junctions\cite{femgo,prbNuala2011}. 
In contrast, the calculation of transport properties via the KS eigenvalues encounters some drastic limitations in case of molecules. 
The fundamental gap between the HOMO and the lowest unoccupied molecular orbital (LUMO) 
is often severely underestimated by the KS gap computed with standard 
(semi-)local exchange-correlation density functionals. Furthermore, non-local correlation effects, 
such as the dynamical response of the electronic system to the addition of an electron or hole\cite{Lu, Neaton, Garcia, Perrin, Heimel},
are not captured. These shortcomings hinder the ability of DFT+NEGF to predict the correct energy level alignment between a molecule and the electrodes, which
 often results in overestimated values for the conductance. Consequently, different improvements have been proposed, such as 
corrections for self-interaction error\cite{Cormac,DAS_asic}, scissor operator schemes\cite{Amaury1,Quek,Quek2,Suarez,Strange1,Strange2, Cehovin} 
and constrained-DFT\cite{Amaury1,Sau,Souza_cdft, Amaury2,andrea_cdft}. Furthermore, in the recent years,
there have been several attempts to move beyond the DFT+NEGF method by using the $GW$ approximation 
of the many-body perturbation theory\cite{Darancet1,Thygesen1,Thygesen2,Darancet2}. 

The description of the Kondo effect, even at the qualitative level, still represents a challenge. 
On the one hand, electron correlations leading to the Kondo effect are beyond the $GW$ perturbative scheme. On the other hand, the DFT KS spectrum 
 fails to display any Kondo-related feature, and so does the conductance computed via the Landauer approach. 
 We note that in principle DFT is able to capture the Kondo effect in one-orbital lattice models\cite{Stefanucci, Troster, Bergfield}
if the exchange-correlation potential has the correct derivative discontinuity at integer number of electrons, and if the conductance 
 is computed from the density through the 
 Friedel sum rule\cite{Mera1,Mera2} and not from the KS states.

In order to describe the Kondo effect in real molecular systems and to overcome the limitations of DFT+NEGF and $GW$, recent studies have proposed to combine DFT 
with model calculations, thus extending to molecular electronics theoretical schemes originally proposed for the study of strongly correlated solid state materials. 
These schemes include the dynamical mean-field theory (DMFT)\cite{Kotliar}. 
The combination of DFT and models is typically achieved by partitioning the system of interest in two coupled subsystems, a weakly correlated one,
 whose electronic structure is well accounted for by DFT, and a strongly correlated one. 
Mathematically, this means that part of the system of interest is projected onto the correlated sub-system, and that the rest is integrated out as an effective bath. 
In case of molecular devices, this approach ultimately leads 
to the reformulation of the electronic structure and transport problem in terms of an effective Anderson impurity model (AIM),
which then has to be solved either exactly or within some approximations. 
The potential of this approach has been firstly demonstrated by comparing to photoemission experiments\cite{Gardonio} 
the computed spectral properties of single magnetic atoms\cite{Surer,Panda,Mazurenko, Dang} and molecules\cite{Korytar,kugel} 
on metallic surfaces. Then, the linear-response (i.e. zero-bias) transport properties 
have been addressed for example by Smogunov, Tosatti and co-workers\cite{Lucignano,Baruselli1,Baruselli2,Requist}, 
and Jacob and co-workers\cite{jacob1,jacob2,jacob3,jacob4,jacob5,jacob6}. 
We have recently contributed to further develop the DFT+NEGF scheme including DMFT to study spin transport in solid state devices such as multilayered heterostructures\cite{Liviu}.

\begin{figure}[t!]
\centering\includegraphics[width=0.2\textwidth,clip=true]{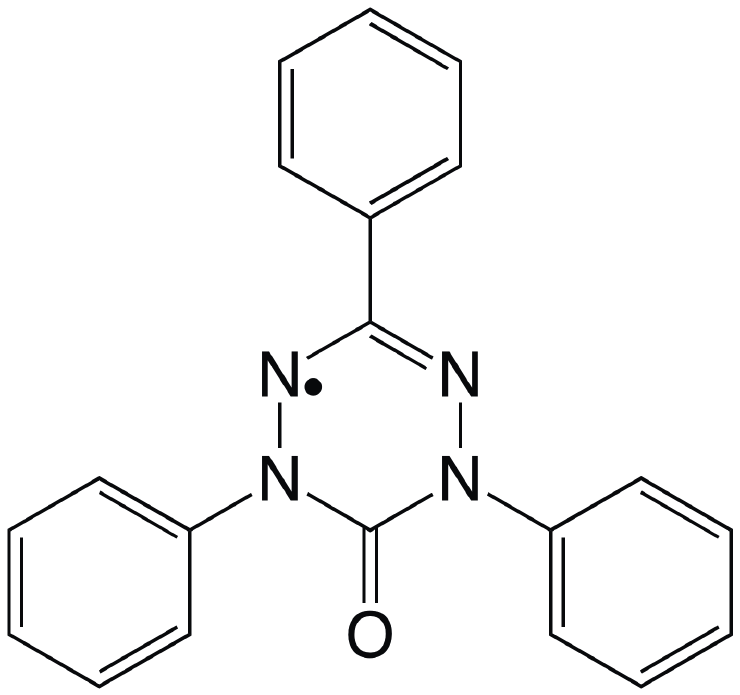}
\caption{Schematic of the 1,3,5-triphenyl-6-oxoverdazyl (TOV) molecule.}
\label{fig.TOV}
\end{figure}

In this article, we present our scheme to project out from DFT+NEGF an AIM, which is then solved numerically using continuous time quantum Monte Carlo (CTQMC)\cite{Gull}.
The developed method allows us to evaluate the temperature-dependent linear-response transport properties of magnetic molecules on metal surfaces. 
An important class of such molecules is formed by the stable organic radicals, which are 
are paramagnetic compounds presenting an unpaired electron in a singly occupied molecular orbital (SOMO)\cite{Nuria,Marta,Herrman,Simao,JLiu,Zhang,Mullegger,Frisenda,Rudnev}. 
In particular, here we consider the 1,3,5-triphenyl-6-oxoverdazyl (TOV, Fig.~\ref{fig.TOV}).
This is the first organic radical for 
which the Kondo effect was experimentally observed in a scanning tunneling microscope (STM) setup with a gold substrate, and  
the Kondo temperature was reported to be about 37 K\cite{JLiu}.
Since the system is well characterized by STM, the comparison of the calculated density of states and Kondo temperature with the experiment serves as a stringent test for the theory.
Additionally, we demonstrate how the different contributions to the conductance that originate from elastic, 
non-coherent and quantum interference effects can be disentangled. An open issue concerns the calculation of the electron-electron interaction energy
for the Anderson impurity. Here we suggest that a partially screened value should be used in order to reproduce the experimental Kondo temperature.

The article is separated in two main parts. In the first part (Sec. \ref{sec:method}) we describe the theoretical methods, while in the second (Secs. \ref{sec:comput} 
and \ref{sec:results}) we present the results for the TOV molecule on Au.
In the first part we initially summarize the projection scheme and outline how the different contributions to the transmission are computed (Secs. \ref{sec:negf_setup} to \ref{sec:teq}), and then
we describe the employed CTQMC algorithm (Sec. \ref{sec:ctqmc}) and the analytic continuation of the CTQMC data (Sec. \ref{sec:analyticCont}). 
In the second part, we start by listing the computational details of the DFT and NEGF calculations (Sec. \ref{sec:comput}). 
We present the DFT results for the gas phase TOV, for the TOV on gold (Sec. \ref{sec:geo}) and the DFT+NEGF results for the transport properties (Sec. \ref{sec:teq}). 
We then introduce the AIM and discuss its solution within the mean-field approximation (Sec. \ref{sec_MF}) and by CTQMC (Sec. \ref{sec_Kondo}). 
Finally, we present the transport properties computed via DFT+NEGF+CTQMC (Sec. \ref{sec:tctqmc}) and we conclude (Sec. \ref{sec:conclusions}).

\section{Method and implementation}
\label{sec:method}

\begin{figure}[t!]
\centering
\includegraphics[width=8.7cm,clip=true]{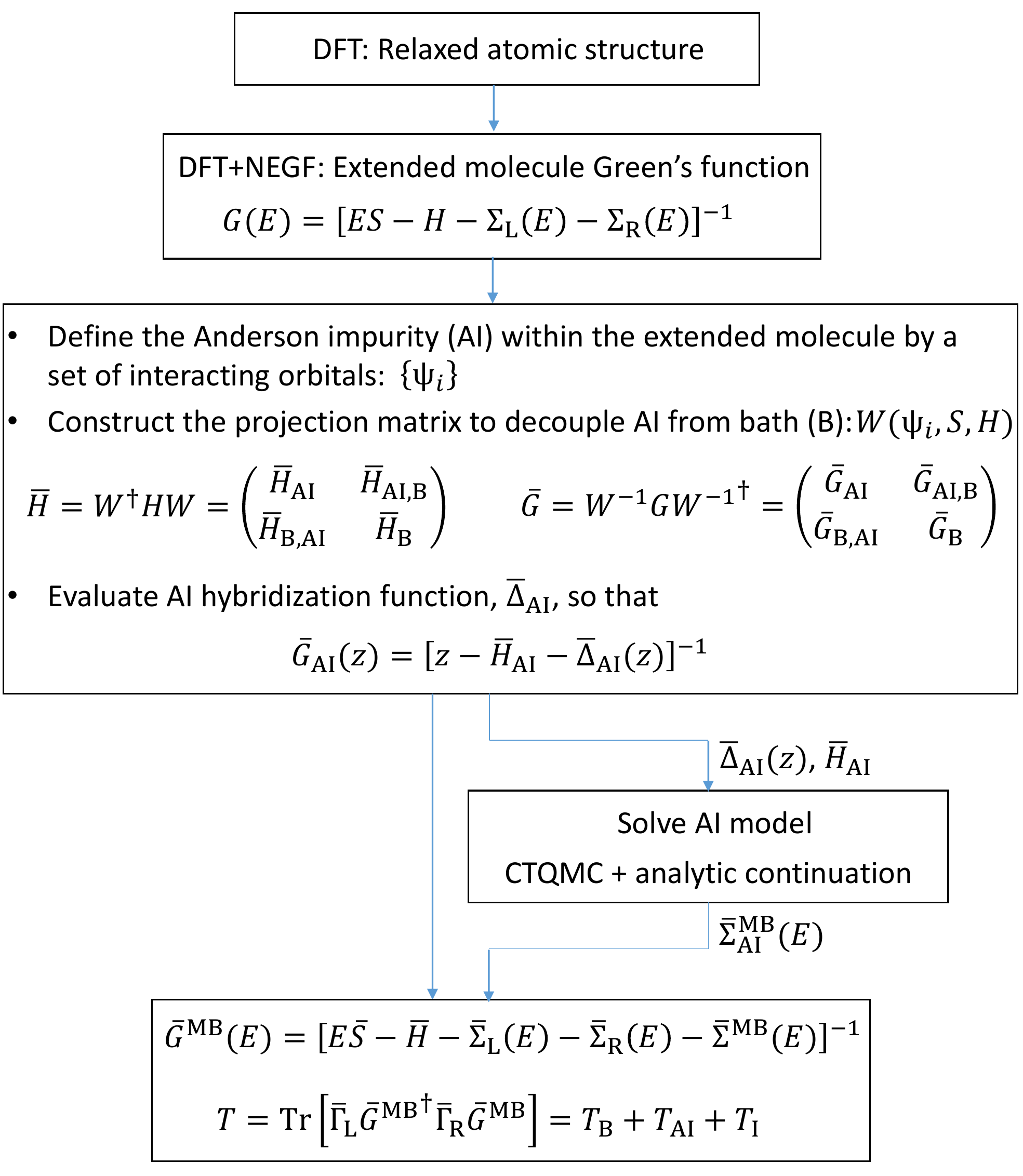}
\caption{Schematic representation of the method for the calculation of the transport properties in presence of interacting Anderson impurities.}
\label{fig:schematicmethod}
\end{figure}

The first step of the method is the separation of the Anderson impurity sub-system from the full system calculated in the DFT+NEGF setup. 
This is done by an appropriate projection scheme, outlined in subsections \ref{sec:negf_setup}, \ref{sec:projection}, and \ref{sec:hybridization},
and schematically summarized in Appendix \ref{sec:appendixsum}. Then, the AIM with an effective interaction term is introduced and solved. 
This means that the many-body Green's function and self-energy are computed (subsection \ref{sec:interaction}) 
so that the transport properties can be obtained (subsection \ref{sec:teq}). 
The method can, in principle, address both zero- and finite-bias transport problems 
provided the availability of a computationally efficient out-of-equilibrium solver for the AIM.
Nevertheless, in this work we only consider the zero-bias case, and therefore the solution of the AIM is achieved by using CTQMC 
for quantum systems in equilibrium at finite temperatures, as outlined in subsection \ref{sec:ctqmc}.
The use of CTQMC requires a scheme to carry out the analytic continuation of the many-body Green's function from the discrete imaginary Matsubara frequencies 
onto real energies This is presented in subsection \ref{sec:analyticCont}. A schematic summary of the whole method is shown in Fig. \ref{fig:schematicmethod}.

\subsection{NEGF transport setup}
\label{sec:negf_setup}

The method has been implemented in the NEGF code \textit{Smeagol}\cite{Smeagol}, 
which uses a linear combination of atomic orbitals (LCAO) basis set $\{\phi_{\mu}\}$, and obtains the KS Hamiltonian from the DFT package Siesta\cite{Siesta}. 
Note however that the method is general and can be readily used for any code based on the LCAO approach. 
Each basis orbital $\vert \phi_{\mu}\rangle$ in \textit{Smeagol} is characterized by its integer index $\mu$,
which is a collective index that includes the atom, $I$, the orbital $n$, and the angular momentum $(l, m)$ indices. 
The orbital index $n$ can run over different radial functions corresponding to the same angular momentum, 
according to the multiple-zetas scheme\cite{Szabo}. 
Any operator $\hat O$ can be expressed in this basis by using its matrix form $O$, with the matrix elements given by
\begin{eqnarray}
 (O)_{\mu\nu} =\langle \phi_\mu\vert \hat O\vert\phi_\nu\rangle.
\end{eqnarray}
As a matter of notation we remark that in general upper case symbols represent matrices, and to distinguish operators from matrices we explicitly add a hat on top of the symbol for operators.

\begin{figure}
\center
\includegraphics[width=8.5cm,clip=true]{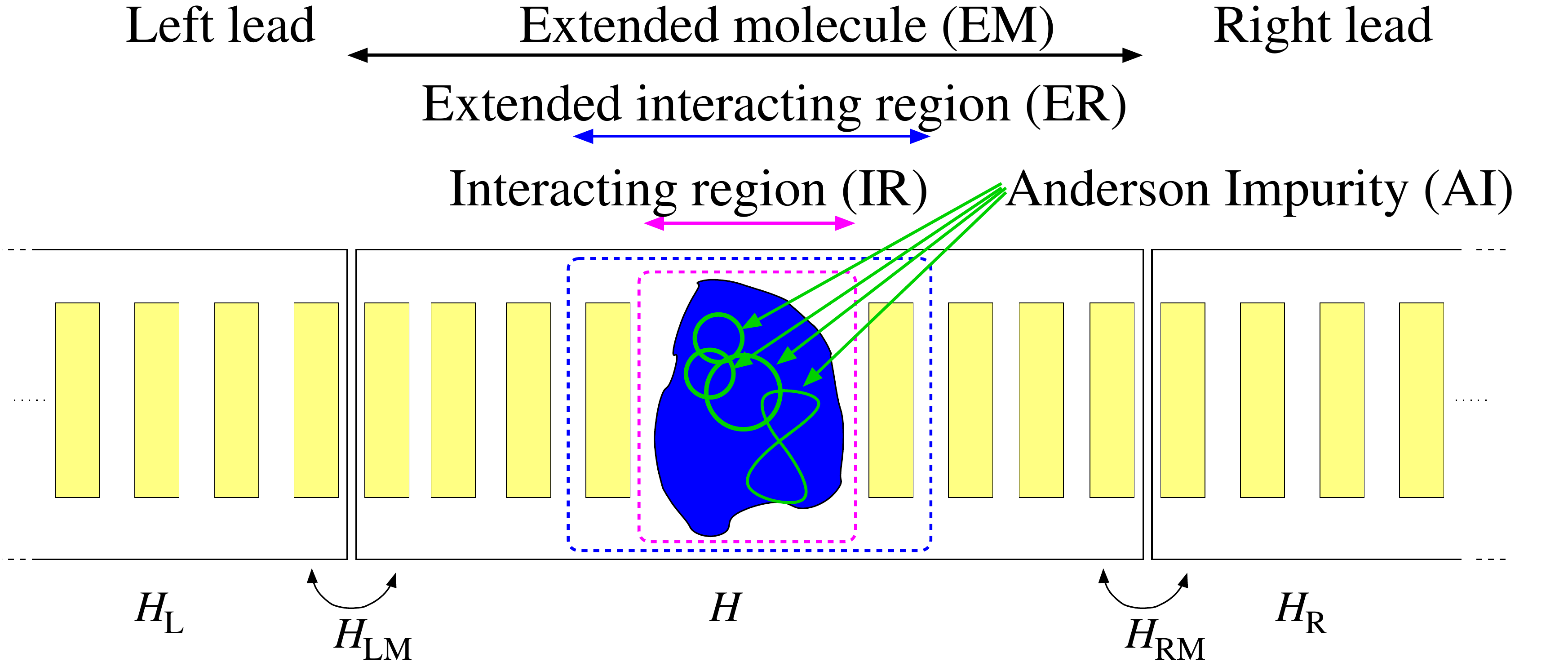}
\caption{(Color online) Schematic representation of the two-terminal device comprising the left lead, the {\it extended molecule} (EM), and the right lead. 
We denote the Hamiltonian matrix of the semi-infinite left (right) lead by $H_\mathrm{L}$ ($H_\mathrm{R}$), the one of the EM as $H$, 
and the coupling Hamiltonian matrix as $H_\mathrm{LM}$ ($H_\mathrm{RM}$).
The EM is further subdivided in the {\it interacting region} (IR) and the {\it extended interacting region} (ER). 
The IR includes a set of interacting molecular orbitals $\left\{\psi_i\right\}$, with $i\in[1,N_\mathrm{AI}]$ ($N_\mathrm{AI}$ 
is the number of interacting molecular orbitals; in the schematic figure $N_\mathrm{AI}=4$), 
forming the {\it Anderson impurity} (AI). The ER consists of all orbitals in the EM that have finite overlap with the IR orbitals, 
so that the IR orbitals are always also part of the ER. In total the system is therefore subdivided into 4 subspaces of decreasing size: EM, ER, IR, and AI.
}
\label{fig:p_schematic}
\end{figure}

The setup used in the calculations is schematically illustrated in Fig. \ref{fig:p_schematic}, 
where the basis set described above is used to expand the charge density\cite{Siesta,Smeagol}.
Like in standard electron transport simulations, the system is first split into a semi-infinite left lead, a so-called scattering region 
(or {\it extended molecule} (EM)), and the semi-infinite right lead\cite{Smeagol}. We denote the number of basis orbitals within the EM by $N$.
We can then introduce the Hamiltonian matrix of the EM as $(H)_{\mu\nu} =\langle \phi_\mu\vert \hat H\vert\phi_\nu\rangle$, 
where the basis orbitals span all orbitals within the EM, and where $\hat H$ is the Hamiltonian operator. Since in general the basis set is non-orthogonal, 
we also need to introduce the overlap matrix of the EM, $S$, given by
\begin{eqnarray}
 (S)_{\mu\nu} =\langle \phi_\mu\vert \phi_\nu\rangle.
\end{eqnarray}
We note here that we shift all energies of leads and EM in such a way to set the Fermi energy, $E_\mathrm{F}$, equal to 0. 
Such a global shift of the spectrum does not affect the properties of the system.
 
The EM is then further subdivided in a total of 3 subsystems. A set of pre-determined wave-functions $\psi_i$ ($i\in[1,N_\mathrm{AI}]$), 
defines the {\it Anderson impurity} (AI), with $N_\mathrm{AI}$ being to the number of interacting states. 
The {\it interacting region} (IR) includes all basis orbitals inside the EM that contribute to the AI, 
which corresponds to the collection of those basis orbitals of the EM where any of the $\{\psi_i\}$ is non-zero. 
The number of basis orbitals in the IR, $N_\mathrm{IR}$, is usually much larger than $N_\mathrm{AI}$. 
We require that the overlap and Hamiltonian matrix elements between orbitals within the IR and the orbitals of the leads outside the EM is zero, 
so that the EM has to be chosen large enough to ensure this.
As last subspace we introduce the {\it extended interacting region} (ER), which includes all basis orbitals inside the EM that have a finite overlap 
or Hamiltonian matrix element with the basis orbitals within the IR. 
We denote the number of orbitals within the ER as $N_\mathrm{ER}$, with $N_\mathrm{ER}\ge N_\mathrm{IR}$. 
Ordered by decreasing size, the 4 subsystems are then: EM, ER, IR, and AI. The AI is a subsystem of the IR, the IR is a subsystem of the ER, 
and the ER is a subsystem of the EM (see Fig. \ref{fig:p_schematic}). 

%*****************************************************************
%*****************************************************************

The first step is therefore to define the set of molecular orbitals, $\left\{\psi_i\right\}$, inside the EM that constitute the AI. 
The method outlined here is applicable for any arbitrary set of wave-functions, although in general the choice of the $\left\{\psi_i\right\}$ 
is based on physical intuition. In the simplest case one can simply take a combination of $d$ or $f$ orbitals for correlated magnetic atoms in the system, 
or else the wave-functions of the HOMO or LUMO of a molecule attached to metal electrodes. 
A practical way to construct such a set of interacting molecular orbitals is to extract from the full EM Hamiltonian, $H$, and overlap, $S$, 
matrices sub-blocks with basis orbitals on the molecule, which we denote as $H_\mathrm{M}$ and $S_\mathrm{M}$, respectively. 
We can then calculate the eigenvalues $\epsilon_i$ and eigenvectors $\psi_i$ of these sub-systems by solving $H_\mathrm{M} \psi_i=\epsilon_i S_\mathrm{M} \psi_i$. 
Note that each of the $\psi_i$ is a vector of dimension $N_\mathrm{IR}$. 

Once the set of wave-functions that defines the AI as $\left\{\psi_{1,\mathrm{IR}},\psi_{2,\mathrm{IR}},\dots,\psi_{N_\mathrm{AI},\mathrm{IR}}\right\}$ 
is chosen (here we have added the subscript "IR" to indicate explicitly that the wave-functions extend over the IR), 
the projection matrix onto the AI inside the IR, $U^\mathrm{IR}$, is then defined as
\begin{equation}
U_\mathrm{IR}=\left( \begin{array}{ccccc}
\psi_{1,\mathrm{IR}} & \psi_{2,\mathrm{IR}} & \psi_{3,\mathrm{IR}} & \dots & \psi_{N_\mathrm{AI},\mathrm{IR}}\\
\end{array} \right),
\end{equation}
which is of dimension $N_\mathrm{IR}\times N_\mathrm{AI}$.
We then construct the ER by adding to the IR all basis orbitals within the EM that have finite overlap with the IR basis orbitals, 
and construct the set of wave-functions $\left\{\psi_{i,\mathrm{ER}}\right\}$. Each $\psi_{i,\mathrm{ER}}$ is a vector of length $N_\mathrm{ER}$, 
and is equal to $\psi_{i}^\mathrm{IR}$ for its elements within the IR, and 0 for the others. 
These vectors therefore define the AI in the ER, and lead to the ER projection matrix onto the AI, $U_\mathrm{ER}$, given by
\begin{equation}
U_\mathrm{ER}=\left( \begin{array}{ccccc}
\psi_{1,\mathrm{ER}} & \psi_{2,\mathrm{ER}} & \psi_{3,\mathrm{ER}} & \dots & \psi_{N_\mathrm{AI},\mathrm{ER}}\\
\end{array} \right),
\label{eq:uer}
\end{equation}
which is of dimension $N_\mathrm{ER}\times N_\mathrm{AI}$. Note that the simulation setup in \textit{Smeagol} 
requires that one leads' unit cell is included at both the left and right ends of the EM. 
For the projection onto the AI we then further require that the orbitals of those cells cannot be part of the IR, while they are allowed to be part of the ER.

If the basis orbitals indices are approximately ordered from left to right in the EM, then the overlap matrix of the EM\cite{Siesta,Smeagol} 
can be written in the following general block-matrix structure
\begin{equation}
S=\left( \begin{array}{ccc}
S_{\mathrm{\alpha\alpha}}         & S_{\mathrm{\alpha, ER}} & S_{\alpha\beta}          \\
S_{\mathrm{\alpha, ER}}^\dagger  & S_{\mathrm{ER}}         & S_{\mathrm{\beta, ER}}^\dagger   \\
S_{\alpha\beta}^\dagger                 & S_{\mathrm{\beta, ER}}  & S_\mathrm{\beta \beta}
\label{eq:ssmall}
\end{array} \right),
\end{equation}
where $S_{\mathrm{ER}}$ is a $N_\mathrm{ER}\times N_\mathrm{ER}$ matrix.
$S_{\alpha\alpha}$ ($S_{\beta \beta})$ is a square matrix that includes all the $N_\alpha$ ($N_\beta$) orbitals within the EM on the left (right) of the ER, 
so that $N=N_\alpha+N_\beta+N_\mathrm{ER}$.
The dimensions of the off-diagonal matrix blocks are determined from the ones of the diagonal blocks, and are therefore not explicitly given. 
We subdivide the Hamiltonian matrix in an analogous way. 
Note that by construction the following important relations are satisfied: 
$S_{\alpha,\mathrm{ER}} U_{\mathrm{ER}}=0$, $S_{\beta,\mathrm{ER}} U_{\mathrm{ER}}=0$, $H_{\alpha,\mathrm{ER}} U_{\mathrm{ER}}=0$, 
and $H_{\beta,\mathrm{ER}} U_{\mathrm{ER}}=0$.

The Green's function (GF) matrix of the EM is then given by the standard form\cite{Smeagol}
\begin{equation}
G(z)=\left[z S-H-\Sigma_\mathrm{L}(z)-\Sigma_\mathrm{R}(z)\right]^{-1},
\label{eq:GF}
\end{equation}
which has the block-matrix structure analogous to the one of the $H$ and $S$ matrices in Eq.~(\ref{eq:ssmall}); $z$ is an arbitrary complex number, 
and $\Sigma_\mathrm{L}(z)$ and $\Sigma_\mathrm{R}(z)$ 
are the left and right leads' self-energies that describe 
the coupling of the EM to the leads. These are computed according the algorithm in Ref. \onlinecite{SelfEnergies}.
Importantly, $G(z)$ can be either the retarded GF if $z=E+i\delta$, with $E$ the real energy and $\delta$ a vanishingly small positive real number, 
or the Matsubara GF if $z=i\omega_n$, with $\omega_n=(2n+1)\pi/\beta$, for $n\in\mathbb{Z}$ and $\beta=1/k\theta$ the inverse temperature 
($k$ is the Boltzmann constant, $\theta$ is the temperature). 
Spin indices are omitted in above equations and in the following subsections to simplify the notation, 
but they will be explicitly re-introduced when required to emphasize spin-dependent relations.  

\subsection{Projection to the Anderson impurity subsystem}
\label{sec:projection}

Here we outline how to separate explicitly the AI sub-system from the rest of the system, which we refer to as the ``bath'', and that includes 
the orthogonal subspace to the AI within the EM as well as the semi-infinite electrodes.
To this aim we introduce a basis transformation matrix, $W$, which transforms the overlap, Hamiltonian, and self-energy matrices as
\begin{eqnarray}
\bar{S}=W^\dagger S W&,&\;\;\;\; \bar{H}=W^\dagger H W,\label{projection}\\
\bar{\Sigma}_\mathrm{L}(z)=W^\dagger \Sigma_\mathrm{L}(z)W&,&\;\;\;\; \bar{\Sigma}_\mathrm{R}(z)=W^\dagger \Sigma_\mathrm{R}(z)W,
\end{eqnarray}
where we denote the transformed matrices with a bar on top of the symbol.
This transformation is required to bring any general matrix extending over the orbitals of the EM, $M$, into a transformed form, $\bar{M}$, 
in which the top left corner describes the AI, the bottom right corner describes the part of the bath included in the EM (B), 
and the off-diagonal blocks describe the connection terms. We note that the full bath includes orbitals from both the EM and the semi-infinite electrodes, 
while the $N_\mathrm{B}\times N_\mathrm{B}$ matrix $\bar{M}_\mathrm{B}$ contains only the $N_\mathrm{B}=N-N_\mathrm{AI}$ bath orbitals within the EM.
In mathematical terms we therefore require $\bar{M}$ to have the following block-matrix structure:
\begin{equation}
\bar{M}=\left( \begin{array}{cc}
 \bar{M}_\mathrm{AI} &   \bar{M}_\mathrm{AI,B}           \\
 \bar{M}_\mathrm{B,AI} &   \bar{M}_\mathrm{B}           \\
\end{array} \right).
\label{eq:aibath}
\end{equation}
We also require the projection to lead to a zero overlap between the AI orbitals and the bath orbitals.

The form of $W$ to achieve this for the transport setup is derived in Appendix \ref{sec:appendixAfp}, and is given by
\begin{equation}
W=\left( \begin{array}{cccc}
0                        & 1_{N_\alpha}  & 0 & 0  \\
W_\mathrm{AI}          & 0             & W_\mathrm{NI}  & 0  \\
0                        &               0                  & 0 & 1_{N_\beta}
\end{array} \right).
\label{eq:wshort}
\end{equation}
Here and in the following we denote an identity matrix of dimensions $m\times m$ as $1_m$. 
We have introduced
\begin{eqnarray}
W_\mathrm{AI}&=& U_\mathrm{ER}   W_{2,\mathrm{AI}}, \label{eq:wai}
\end{eqnarray}
which is equal to the projection matrix $U_{\mathrm{ER}}$ [Eq. (\ref{eq:uer})], multiplied by the $N_\mathrm{AI}\times N_\mathrm{AI}$ matrix $W_{2,\mathrm{AI}}$, 
which in general is constructed in such a way to orthogonalize both the overlap and Hamiltonian matrices of the AI itself. 
In principle this second transformation is arbitrary and can also be omitted, depending on how the AI problem is solved. 
 The $N_\mathrm{ER}\times N_\mathrm{NI}$ matrix $W_\mathrm{NI}$ spans the orthogonal space to the AI inside the ER, so that $W_\mathrm{AI}^\dagger W_\mathrm{NI}=0$,
 and therefore projects onto the non-interacting part of the ER. 
We denote the number of non-interacting states within the ER as $N_\mathrm{NI}$, so that $N_\mathrm{NI}=N_\mathrm{ER}-N_\mathrm{AI}$.
There is some freedom in the construction of the matrices $W_{2,\mathrm{AI}}$ and $W_\mathrm{NI}$, since they are not uniquely defined. 
We construct them in such a way to leave the non-interacting part close to the original system, 
and the detailed relations to construct $W_{2,\mathrm{AI}}$ and $W_\mathrm{NI}$ are given in Appendix \ref{sec:appendixAfp}, Eqs. (\ref{eq:w2ai}) and (\ref{eq:wni}).

The explicit form of the final transformed overlap matrix is then evaluated to
\begin{equation}
\bar{S}=\left( \begin{array}{cccc}
1_{N_\mathrm{AI}}      & 0 & 0                  & 0            \\
0 & S_{\mathrm{\alpha\alpha}} & \bar{S}_{\mathrm{\alpha,\mathrm{NI}}}                    &  S_\mathrm{\alpha \beta}               \\
0 & \bar{S}_{\mathrm{\alpha,\mathrm{NI}}}^\dagger & \bar{S}_{\mathrm{NI}}                           &  \bar{S}_{\mathrm{\beta,\mathrm{NI}}}^\dagger                \\
0   & S_\mathrm{\alpha \beta}^\dagger                          & \bar{S}_{\mathrm{\beta,\mathrm{NI}}}     & S_\mathrm{\beta \beta}
\end{array} \right),
\label{eq:smatexp}
\end{equation}
which as required has zero overlap between AI and bath orbitals, and where
\begin{eqnarray}
\bar{S}_\mathrm{NI} &=& W_\mathrm{NI}^\dagger S_\mathrm{ER} W_\mathrm{NI},\\
\bar{S}_{\alpha,\mathrm{NI}} &=& S_{\alpha,\mathrm{ER}} W_\mathrm{NI},\\
\bar{S}_{\beta,\mathrm{NI}} &=& S_{\beta,\mathrm{ER}} W_\mathrm{NI}.
\end{eqnarray}
The final general form of the projected Hamiltonian matrix is
\begin{equation}
\bar{H}=\left( \begin{array}{cccc}
\epsilon_\mathrm{AI,D}      & 0 & \bar{H}_{\mathrm{AI,\mathrm{NI}}}                  & 0            \\
0 & H_{\mathrm{\alpha\alpha}} & \bar{H}_{\mathrm{\alpha,\mathrm{NI}}}                    &  H_\mathrm{\alpha \beta}               \\
\bar{H}_{\mathrm{AI,\mathrm{NI}}}^\dagger & \bar{H}_{\mathrm{\alpha,\mathrm{NI}}}^\dagger & \bar{H}_{\mathrm{NI}}        &  \bar{H}_{\mathrm{\beta,\mathrm{NI}}}^\dagger                \\
0   & H_\mathrm{\alpha \beta}^\dagger                          & \bar{H}_{\mathrm{\beta,\mathrm{NI}}}     & H_\mathrm{\beta \beta}
\end{array} \right),
\label{eq:hmatexp}
\end{equation}
where
\begin{eqnarray}
\epsilon_\mathrm{AI,D} &=& W_\mathrm{AI}^\dagger H_\mathrm{ER} W_\mathrm{AI},\\
\bar{H}_{\mathrm{AI},\mathrm{NI}} &=& W_\mathrm{AI}^\dagger H_\mathrm{ER} W_\mathrm{NI},\\
\bar{H}_\mathrm{NI} &=& W_\mathrm{NI}^\dagger H_\mathrm{ER} W_\mathrm{NI},\\
\bar{H}_{\alpha,\mathrm{NI}} &=& H_{\alpha,\mathrm{ER}} W_\mathrm{NI},\\
\bar{H}_{\beta,\mathrm{NI}} &=& H_{\beta,\mathrm{ER}} W_\mathrm{NI}.
\end{eqnarray}

The general structure of the resulting matrices has the required shape given in Eq. (\ref{eq:aibath}), with
\begin{eqnarray}
\bar{S}_\mathrm{B}&=&\left( \begin{array}{ccc}
 S_{\mathrm{\alpha\alpha}} & \bar{S}_{\mathrm{\alpha,\mathrm{NI}}}                    &  S_\mathrm{\alpha \beta}               \\
 \bar{S}_{\mathrm{\alpha,\mathrm{NI}}}^\dagger & \bar{S}_{\mathrm{NI}}                           &  \bar{S}_{\mathrm{\beta,\mathrm{NI}}}^\dagger                \\
 S_\mathrm{\alpha \beta}^\dagger                          & \bar{S}_{\mathrm{\beta,\mathrm{NI}}}     & S_\mathrm{\beta \beta}
\end{array} \right),
\label{eq:sb}\\
\bar{S}_\mathrm{AI,B}&=&\left( \begin{array}{ccc}
 0 & 0                  & 0           
\end{array} \right),
\end{eqnarray}
and 
\begin{eqnarray}
\bar{H}_\mathrm{B}&=&\left( \begin{array}{ccc}
 H_{\mathrm{\alpha\alpha}} & \bar{H}_{\mathrm{\alpha,\mathrm{NI}}}                    &  H_\mathrm{\alpha \beta}               \\
 \bar{H}_{\mathrm{\alpha,\mathrm{NI}}}^\dagger & \bar{H}_{\mathrm{NI}}                           &  \bar{H}_{\mathrm{\beta,\mathrm{NI}}}^\dagger                \\
 H_\mathrm{\alpha \beta}^\dagger                          & \bar{H}_{\mathrm{\beta,\mathrm{NI}}}     & H_\mathrm{\beta \beta}
\end{array} \right),\\
\bar{H}_\mathrm{AI,B}&=&\left( \begin{array}{ccc}
 0 & \bar{H}_{\mathrm{AI,\mathrm{NI}}}                  & 0           
\end{array} \right).\label{eq:hB}
\end{eqnarray}
Note that the transformed self-energy matrices extend only over the block of the bath inside the EM, and are zero for the other matrix blocks. 
This is ensured by the requirement that the orbitals of the left and right leads' unit cells included at the boundaries of the EM are not part of the IR.

The transformation for the GF is given by
\begin{eqnarray}
 \bar{G}(z)&=&W^{-1}G(z){W^{-1}}^\dagger,
\label{eq:gtran}
\end{eqnarray}
and can also be evaluated directly in the transformed system as
\begin{equation}
\bar{G}(z)=\left[z \bar{S}-\bar{H}-\bar{\Sigma}_\mathrm{L}(z)-\bar{\Sigma}_\mathrm{R}(z)\right]^{-1}.
\label{eq:GFbar}
\end{equation}

It is possible to evaluate the required inverse of $W$ by blocks (Appendix \ref{sec:appendixAip}), and the result is
\begin{equation}
W^{-1}=\left( \begin{array}{ccc}
0      &  W_\mathrm{iAI}  & 0   \\
1_{N_\alpha}    &  0   & 0   \\
0      & W_\mathrm{iNI}      & 0 \\
0      & 0      & 1_{N_\beta} 
\end{array} \right),
\label{eq:winvsimple}
\end{equation}
where
\begin{eqnarray}
W_\mathrm{iAI} &=& {W_{2,\mathrm{AI}}^{-1}} W_\mathrm{i,ER,\psi} S_{\mathrm{ER}},
\label{eq:wiabc}
\end{eqnarray}
with
\begin{equation}
W_\mathrm{i,ER,\psi}=\left({U_{\mathrm{ER}}}^\dagger S_\mathrm{ER} U_{\mathrm{ER}}\right)^{-1}{U_{\mathrm{ER}}}^\dagger.
\end{equation}
The form of the $N_\mathrm{NI}\times N_\mathrm{ER}$ block-matrix $W_\mathrm{iNI}$ is given in Appendix \ref{sec:appendixAip}, Eq. (\ref{eq:winvni}).

\subsection{Hybridization function}
\label{sec:hybridization}

By removing the AI orbitals from the system we can introduce the GF of only the bath orbitals within the EM, $\bar{g}(z)$, as
\begin{equation}
\bar{g}(z)=\left[z \bar{S}_\mathrm{B}-\bar{H}_\mathrm{B}-\left(\bar{\Sigma}_\mathrm{L}\right)_\mathrm{B}(z)-
\left(\bar{\Sigma}_\mathrm{R}\right)_\mathrm{B}(z)\right]^{-1},
\label{eq:barGB}
\end{equation}
which can be written in a block-matrix structure analogous to the one of $\bar{S}_\mathrm{B}$ [Eq. (\ref{eq:sb})] as
\begin{equation}
\bar{g}(z)= 
\left( \begin{array}{ccc}
 \bar{g}_{\mathrm{\alpha\alpha}}(z)  & \bar{g}_\mathrm{\alpha,NI} (z)      & \bar{g}_\mathrm{\alpha\beta}(z) \\
 \bar{g}_{\mathrm{NI,\alpha}} (z)  & \bar{g}_{\mathrm{NI}}  (z)          & \bar{g}_\mathrm{NI,\beta} (z)   \\
 \bar{g}_{\mathrm{\beta\alpha}}(z)   & \bar{g}_\mathrm{\beta,NI} (z)       & \bar{g}_\mathrm{\beta \beta}(z)
\end{array} \right).
\end{equation}

By using also Eqs. (\ref{eq:GFbar}), (\ref{eq:smatexp}) and (\ref{eq:hmatexp}) the GF on the AI is then obtained as
\begin{eqnarray}
\bar{G}_\mathrm{AI}(z)&=&\left[z - \epsilon_\mathrm{AI,D}-\bar\Delta_\mathrm{AI}(z)\right]^{-1}.
\label{eq:gbar2}
\end{eqnarray}
Here we have introduced the so-called hybridization function, $\bar\Delta_\mathrm{AI}(z)$, which is given by
\begin{equation}
\bar\Delta_\mathrm{AI}(z)=\bar{H}_\mathrm{AI,NI}\ \bar{g}_\mathrm{NI}(z) \bar{H}_\mathrm{AI,NI}^\dagger.
\label{eq:bardelta2}
\end{equation}
We remark that in general $\bar\Delta_\mathrm{AI}(z)$ is a dense $N_\mathrm{AI}\times N_\mathrm{AI}$ matrix. 
Therefore, optionally, as alternative possibility one can modify the transformation matrix $W_{2,\mathrm{AI}}$ 
to an energy-dependent form that diagonalizes $\bar\Delta_\mathrm{AI}(z)$ rather than $\bar H_\mathrm{AI}$. In Appendix \ref{sec:appendixhyb} 
we derive an equivalent, commonly used expression for the hybridization function. 
Furthermore, in Appendix \ref{sec:appendixlimitdelta} we show that $\bar\Delta_\mathrm{AI}(z)$ exhibits the correct physical decay for large $z$.

Once $\bar{G}_\mathrm{AI}(z)$ and $\bar{g}(z)$ are known, then the remaining block matrices of the GF within the EM [see Eq. (\ref{eq:aibath}) 
for the general matrix structure] can be evaluated to
\begin{eqnarray}
&\bar{G}_\mathrm{B}(z)= \bar{g}(z)+
\bar{g}(z)\bar{H}_{\mathrm{AI,B}}^\dagger\bar{G}_\mathrm{AI}(z)\bar{H}_{\mathrm{AI,B}}\bar{g}(z),\label{eq:gfexpand}\\
&\bar{G}_\mathrm{AI,B}(z)= - \bar{G}_\mathrm{AI}(z)\bar{H}_{\mathrm{AI,B}}\bar{g}(z),\\
&\bar{G}_\mathrm{B,AI}(z)= - \bar{g}(z)\bar{H}_{\mathrm{AI,B}}^\dagger\bar{G}_\mathrm{AI}(z).
\end{eqnarray}

\subsection{Effective Coulomb interaction and many-body self-energy}
\label{sec:interaction}

The AI is fully characterized at the KS-level through the ``on-site'' upper-block $\epsilon_\mathrm{AI,D}$ 
of the Hamiltonian matrix in Eq. (\ref{eq:hmatexp}), which is coupled  to the bath via the off-diagonal block $\bar{H}_\mathrm{AI,NI}$. 
The AIM Hamiltonian operator can be rewritten in its standard form as an operator in second quantization 
$\hat H_\mathrm{AIM}=\hat {\bar H}_\mathrm{AI,D}+\hat{ \bar{H}}_\mathrm{TB}+\hat {\bar H}_\mathrm{AI,NI}$, with
\begin{eqnarray}
&\hat {\bar H}_\mathrm{AI,D} =\sum_{i,j=1}^{N_\mathrm{AI}}(\epsilon_\mathrm{AI,D})_{ij}\sum_\sigma \hat d^\dagger_{i\sigma}\hat d_{j\sigma},  \\
&\hat {\bar{H}}_\mathrm{TB}=\sum_{p,q=1}^{\infty}(\bar{H}_\mathrm{TB})_{pq}\hat c^\dagger_{p\sigma}\hat c_{q\sigma},  \\
& \hat {\bar H}_\mathrm{AI,NI}= \sum_{i=1}^{N_\mathrm{AI}}\sum_{p=1}^{N_\mathrm{NI}}\Big[(\bar{H}_\mathrm{AI,NI})_{ip}\hat d^\dagger_{i\sigma}\hat c_{p\sigma}+c.c\Big].
\end{eqnarray}
 Here 
$\hat d^{(\dagger)}_{i\sigma}$ is the annihilation (creation) operator for an electron of spin $\sigma$ in the orbital $i$ of the AI, while
$\hat c^{(\dagger)}_{p\sigma}$ is the annihilation (creation) operator for an electron of spin $\sigma$ in an orbital $p$ of the total bath,
whose Hamiltonian matrix is formally written as $\bar{H}_\mathrm{TB}$. We note that $\bar{H}_\mathrm{TB}$   
includes the elements of $\bar{H}_\mathrm{B}$ obtained from the projection of the EM Hamiltonian with the Eq.~(\ref{projection}),
as well as the elements of the semi-infinite electrodes. 
The effect of the leads is accounted for in the bath Green's function [Eq.~(\ref{eq:barGB})], and therefore in the hybridization function [Eq.~\ref{eq:bardelta2}]
through the projected self-energies $\left(\bar{\Sigma}_\mathrm{L}\right)_\mathrm{B}(z)$ and
$\left(\bar{\Sigma}_\mathrm{R}\right)_\mathrm{B}(z)$, so that the full $\bar{H}_\mathrm{TB}$ never needs to be explicitly computed.

In order to account for many-body correlation effects, we supplement the AI  
with an effective Coulomb interaction, expressed by the operator
\begin{equation}
\hat H_{\mathrm{I}}=\hat H_{\mathrm{C}}-\hat H_{\mathrm{dc}}.
\end{equation}
Then the interacting AIM Hamiltonian operator is defined as
\begin{equation}
\hat H_\mathrm{IAIM}=\hat H_\mathrm{AIM}+ \hat H_{\mathrm{I}}.\label{IAIM}
\end{equation}
$\hat H_{\mathrm{C}}$ is typically chosen to have the form of a generalized Hubbard-like interaction, while $\hat H_{\mathrm{dc}}$ is the double-counting correction\cite{Kotliar}. 
This double-counting correction is required in order to subtract the correlation effects in the AI that 
are already included at the KS level. 
The exact form of $\hat H_{\mathrm{dc}}$ is unfortunately not known, and several approximations have been introduced, the most common one being the 
 so-called ``fully localized limit''\cite{Kotliar}. This is the one that we also employ.

For a single impurity one-orbital Anderson model (SIAM) one has
\begin{equation}
\hat H_{\mathrm {C}}=U \hat n_\uparrow \hat n_\downarrow,\label{Hubbard} 
\end{equation}
with the Hubbard $U$ being a real number, and $\hat n_\uparrow$ ($\hat n_\downarrow$) the occupation operator for up-spin (down-spin) electrons on the AI.
The double-counting correction in the fully localized limit has the simple expression
\begin{equation}
\hat H_{\mathrm{dc}}= U (n-\frac 1 2)\sum_{\sigma}\hat n_{\sigma},
\label{Hdc}
\end{equation}
where $n$ is the DFT total occupation of the impurity.\\

The solution of the interacting AIM leads to the many-body GF on the AI, $\bar{G}_\mathrm{AI}^\mathrm{MB}(z)$. 
One can then define the many-body self-energy on the AI, $\bar{\Sigma}^\mathrm{MB}_\mathrm{AI}(z)$, by means of the Dyson equation
\begin{equation}
\bar{\Sigma}^\mathrm{MB}_\mathrm{AI}(z)=\bar{G}_\mathrm{AI}^{-1}(z)-\left(\bar{G}_\mathrm{AI}^\mathrm{MB}\right)^{-1}(z).
\label{eq:sai}
\end{equation}
Since we are considering the case of a single interacting region, the full many-body self-energy matrix of the EM is zero for all elements, except for those of the AI
\begin{equation}
\bar{\Sigma}^\mathrm{MB}(z)=\left( \begin{array}{cc}
\bar{\Sigma}^\mathrm{MB}_\mathrm{AI}(z)                       & 0 \\
 0 & 0_{N_\mathrm{B}}
\end{array} \right).
\label{eq:sigmamb}
\end{equation}
Here we use the notation $0_m$ to denote a $m\times m$ matrix with all zeroes; the sizes of the off-diagonal blocks are determined by the ones of the diagonal blocks. 
If $\bar{\Sigma}^\mathrm{MB}(z)$ is known, it can be used to directly calculate the many-body GF as
\begin{equation}
\bar{G}^\mathrm{MB}(z)=\left[z \bar{S}-\bar{H}-\bar{\Sigma}_L(z)-\bar{\Sigma}_R(z)-\bar{\Sigma}^\mathrm{MB}(z)\right]^{-1}.
\label{eq:gmbbar}
\end{equation}

In cases when the many-body self-energy is required in the original basis, this can be obtained by applying the inverse transformation as
\begin{equation}
\Sigma^\mathrm{MB}(z)={W^{-1}}^\dagger\;\bar{\Sigma}^\mathrm{MB}(z)\;W^{-1}.
\end{equation}
By using Eqs. (\ref{eq:winvsimple}) and (\ref{eq:sigmamb}) the explicit structure of $\Sigma^\mathrm{MB}(z)$ becomes
\begin{equation}
\Sigma^\mathrm{MB}(z)=\left( \begin{array}{ccc}
0_{N_\alpha}        & 0                                   & 0                \\
0                       &\Sigma^\mathrm{MB}_{\mathrm{ER}}(z)                 & 0                \\
0                       & 0                         & 0_{N_\beta}
\end{array} \right).
\end{equation}
The non-zero block of $\Sigma^\mathrm{MB}$ extends over the ER, and is given by
\begin{eqnarray}
\Sigma^\mathrm{MB}_\mathrm{ER}(z)=W_\mathrm{iAI}^\dagger \bar{\Sigma}^\mathrm{MB}_\mathrm{AI}(z) W_\mathrm{iAI},
\label{eq:sigmambinv}
\end{eqnarray}
with $W_\mathrm{iAI}$ given in Eq. (\ref{eq:wiabc}).
After calculating $\Sigma^\mathrm{MB}_\mathrm{ER}(z)$ one can then obtain the full many-body GF in the original basis as
\begin{equation}
G^\mathrm{MB}(z)=\left[z S-H-\Sigma_L(z)-\Sigma_R(z)-\Sigma^\mathrm{MB}(z)\right]^{-1}.
\label{eq:gmb}
\end{equation}

\subsection{Current and transmission}
\label{sec:teq}
The current flowing out of the EM into the right electrode, $I_\mathrm{R}$, can be written as sum of a component transmitted into the right electrode from the left electrode, $I_\mathrm{R,L}$, and a component flowing from the AI into the right electrode, $I_\mathrm{R,AI}$,\cite{Meir,Ness}
\begin{equation}
I_\mathrm{R}=I_\mathrm{R,L}+I_\mathrm{R,AI}.
\label{eq:ir}
\end{equation}
A similar expression holds for the current flowing from the left electrode into the EM, $I_\mathrm{L}$, as outlined in Appendix \ref{sec:appendixcurrent}. 
At steady state the current conservation condition implies that $I_\mathrm{L}=I_\mathrm{R}$.\\
The value of $I_\mathrm{R,L}$ can be evaluated using the left to right energy dependent many-body transmission coefficient, $T_\mathrm{R,L}(E)$, 
which we will simply denote as transmission, $T(E)$, from now on  in order to simplify the notation. $I_\mathrm{R,L}$ is then given by
\begin{eqnarray}
I_\mathrm{R,L}&=&\frac{e}{h}\int dE \;\left(f_\mathrm{L}(E)-f_\mathrm{R}(E)\right)T(E),
\label{eq:ielastic}
\end{eqnarray}
where $e$ is the electron charge, $h$ is the Planck constant, $f_L$ ($f_R$) is the Fermi Dirac distribution at the chemical potential of the left (right) electrode, and
\begin{eqnarray}
T(E)&=&\mathrm{Tr}\left[\Gamma_\mathrm{L}(E) {G^\mathrm{MB}}^\dagger(E) \Gamma_\mathrm{R}(E) G^\mathrm{MB}(E)\right]\nonumber\\
&=&\mathrm{Tr}\left[\bar{\Gamma}_\mathrm{L}(E) {\bar{G}^{\mathrm{MB}\dagger}}(E) \bar{\Gamma}_\mathrm{R}(E) \bar{G}^\mathrm{MB}(E)\right].\label{eq:tre}
\end{eqnarray}
$T$ includes all interference effects between the possible transport channels across the system, 
including the effects of interactions on the AI.  
The so called coupling matrices $\Gamma_{\mathrm{L,R}}(E)$ are defined as
\begin{eqnarray}
\Gamma_\mathrm{L}(E)&=& i\left[\Sigma_\mathrm{L}(E)-\Sigma_\mathrm{L}^\dagger(E)\right],\label{eq:gammal}\\
\Gamma_\mathrm{R}(E)&=& i\left[\Sigma_\mathrm{R}(E)-\Sigma_\mathrm{R}^\dagger(E)\right],\label{eq:gammar}
\end{eqnarray}
and $\bar{\Gamma}_\mathrm{\{L,R\}}(E)=W^\dagger \Gamma_\mathrm{\{L,R\}}(E) W$. 
Note that if the Matsubara GF is computed by using Eqs. (\ref{eq:GF}) or (\ref{eq:gmbbar}) with $z=i\omega_n$,
one has to perform the analytic continuation from the complex Matsubara energies to the real axis in order to obtain the retarded GF prior to the calculation of the transmission. 
The procedure that we have chosen to carry out such operation is described in Sec.\ref{sec:analytic}.
Note also that we can equally express the transmission and current in terms of the quantities in the original basis and in the transformed one. 
In what follows we work in the transformed basis.\\
The transport properties of a molecular device in DFT+NEGF are usually expressed in terms of the transmission coefficient 
for the KS system\cite{Smeagol}, denoted as $T_0(E)$, which is given by\cite{Smeagol}
\begin{eqnarray}
T_0(E)&=&\mathrm{Tr}\left[\bar{\Gamma}_\mathrm{L}(E) \bar{G}^\dagger(E) \bar{\Gamma}_\mathrm{R}(E) \bar{G}(E)\right].\label{eq:tr0}
\end{eqnarray}
It has the analogous structure to $T(E)$, but with the many-body GF replaced by the KS one. Therefore $T(E)$ accounts for the renormalization of the 
coherent transport properties via many-body effects not included at the KS DFT level, and in the following, we refer to $I_\mathrm{R,L}$ as the coherent, or elastic, component of
the current. In contrast, $I_\mathrm{R,AI}$ represents the incoherent component of the current. \\
The incoherent component of the current flowing from the AI to the right electrode is given by (see Appendix \ref{sec:appendixcurrent})
\begin{eqnarray}
I_\mathrm{R,AI}&=&\frac{e}{h}\int dE \;\mathrm{Tr}\left[
\left(\bar{F}^\mathrm{MB}(E)-f_\mathrm{R}(E)\right)
\right.\nonumber\\
&&\left.\left(\bar{\Gamma}^\mathrm{MB}(E) \bar{G}^{\mathrm{MB}\supi{\dagger}}(E)\bar{\Gamma}_\mathrm{R}(E) \bar{G}^\mathrm{MB}(E)\right)\right],
\label{eq:irai}
\end{eqnarray}
where we have introduced the occupation matrix of the AI, $\bar{F}_\mathrm{MB}(E)$,\cite{Ness} and 
\begin{equation}
\bar{\Gamma}^\mathrm{MB}(E)=i\left[\bar{\Sigma}^\mathrm{MB}(E)-\left(\bar{\Sigma}^\mathrm{MB}(E)\right)^\dagger\right],
\end{equation}
which is defined in a similar way to the coupling matrices in Eqs.~(\ref{eq:gammal}) and (\ref{eq:gammar}), but with the many-body self-energy replacing the the leads' self-energies.
 The general structure of the occupation matrix is analogous to the one of $\bar{\Sigma}^\mathrm{MB}$ in Eq. (\ref{eq:sigmamb}), and can be written as
\begin{equation}
\bar{F}^\mathrm{MB}(E)=\left( \begin{array}{cc}
\bar{F}^\mathrm{MB}_\mathrm{AI}(E)                       & 0 \\
 0 & 0_{N_\mathrm{B}}
\end{array} \right).
\label{eq:fmat}
\end{equation}
The quantity inside the trace of Eq. (\ref{eq:irai}) can be interpreted as a transmission matrix,
given by $\left(\bar{\Gamma}^\mathrm{MB}(E) \bar{G}^{\mathrm{MB}\supi{\dagger}}(E)\bar{\Gamma}_\mathrm{R}(E) \bar{G}^\mathrm{MB}(E)\right)$, 
times the difference in the distribution matrices, given by $\left(\bar{F}^\mathrm{MB}(E)-f_\mathrm{R}(E)\right)$. 
It therefore has the analogous structure to $I_\mathrm{R,L}$ in Eq. (\ref{eq:ielastic}), 
with the only difference that since $\bar{F}^\mathrm{MB}$ is a matrix it cannot be moved outside the trace. 
The analogous equations for the current from the left lead into the EM, $I_\mathrm{L}$, are given in Appendix \ref{sec:appendixcurrent}.

If we assume that the matrices $H_{\alpha\beta}$ and $S_{\alpha\beta}$ are zero, 
or more generally that their values are small enough that they can be neglected, and that $\Gamma_\mathrm{L}(E)$ ($\Gamma_\mathrm{R}(E)$) 
extends only over the region $\alpha$ ($\beta$), 
then we can obtain the transmission function from the GF block matrix of the NI as
\begin{equation}
T(E)=\mathrm{Tr}\left[\bar{\Gamma}_\mathrm{L,NI}(E) {\bar{G}}_\mathrm{NI}^{\mathrm{MB}\dagger}(E) \bar{\Gamma}_\mathrm{R,NI} (E)
\bar{G}_\mathrm{NI}^\mathrm{MB}(E)\right],
\end{equation}
where $\bar{\Gamma}_\mathrm{L,NI}(E)=i\left[\bar\Sigma_\mathrm{L,NI}(E)-\bar\Sigma_\mathrm{L,NI}^\dagger(E)\right]$ 
and $\bar{\Gamma}_\mathrm{R,NI}(E)=i\left[\bar\Sigma_\mathrm{R,NI}(E)-\bar\Sigma_\mathrm{R,NI}^\dagger(E)\right]$. 
The matrices $\bar\Sigma_\mathrm{\{L,R\},NI}(E)$ are given by
\begin{eqnarray} 
\bar{\Sigma}_\mathrm{L,NI}(E)&=\bar{K}_{\mathrm{NI},\alpha}\left[\bar{K}_{\alpha\alpha}-
\left(\bar{\Sigma}_{\mathrm{L}}\right)_{\alpha\alpha}(E)\right]^{-1}\bar{K}_{\alpha,\mathrm{NI}},\\
\bar{\Sigma}_\mathrm{R,NI}(E)&=\bar{K}_{\mathrm{NI},\beta}\left[\bar{K}_{\beta\beta}-
\left(\bar{\Sigma}_{\mathrm{R}}\right)_{\beta\beta}(E)\right]^{-1}\bar{K}_{\beta,\mathrm{NI}},
\end{eqnarray}
with $\bar{K}=E\bar{S}-\bar{H}$ having the analogous block-matrix structure to $\bar{H}$ in Eq. (\ref{eq:hmatexp}), 
and for real energies $\bar{K}_{\mathrm{NI},\beta}=\bar{K}_{\beta,\mathrm{NI}}^\dagger$ and $\bar{K}_{\mathrm{NI},\alpha}=\bar{K}_{\alpha,\mathrm{NI}}^\dagger$. 
Note that $\bar{K}_{\alpha\alpha}=K_{\alpha\alpha}$,  $\bar{K}_{\beta\beta}=K_{\beta\beta}$, 
and with the assumptions made in this paragraph we have $\left(\bar{\Sigma}_{\mathrm{L}}\right)_{\alpha\alpha}=\left(\Sigma_{\mathrm{L}}\right)_{\alpha\alpha}$ 
and $\left(\bar{\Sigma}_{\mathrm{R}}\right)_{\beta\beta}=\left(\Sigma_{\mathrm{R}}\right)_{\beta\beta}$.\\
With Eq. (\ref{eq:gfexpand}) the GF block matrix of the NI can be obtained from the one of the AI as
\begin{equation}
\bar{G}^\mathrm{MB}_\mathrm{NI}(E)= \bar{g}_\mathrm{NI}(E)+\Delta\bar{G}^\mathrm{MB}_\mathrm{NI}(E),
\end{equation}
with
\begin{equation}
\Delta\bar{G}^\mathrm{MB}_\mathrm{NI}(E)= \bar{g}_\mathrm{NI}(E)\bar{H}_{\mathrm{AI,NI}}^\dagger\bar{G}^\mathrm{MB}_\mathrm{AI}(E)\bar{H}_{\mathrm{AI,NI}}\bar{g}_\mathrm{NI}(E).
\end{equation}
The transmission can then be decomposed in three components
\begin{equation}
T(E)=T_\mathrm{B}(E)+T_\mathrm{AI}(E)+T_\mathrm{I}(E),
\label{eq:tedecomp}
\end{equation}
where we have introduced the background or bath transmission
\begin{equation}
T_\mathrm{B}(E)=\mathrm{Tr}\left[\bar{\Gamma}_\mathrm{L,NI}(E) {\bar{g}}_\mathrm{NI}^{\dagger}(E) \bar{\Gamma}_\mathrm{R,NI} (E)
\bar{g}_\mathrm{NI}(E)\right],
\label{eq:tbath}
\end{equation}
the transmission through the AI
\begin{equation}
T_\mathrm{AI}(E)=\mathrm{Tr}\left[\bar{\Gamma}_\mathrm{L,NI}(E) {\Delta\bar{G}}_\mathrm{NI}^{\mathrm{MB}\dagger}(E) \bar{\Gamma}_\mathrm{R,NI} (E)
\Delta\bar{G}_\mathrm{NI}^\mathrm{MB}(E)\right],
\label{eq:tAI1}
\end{equation}
and the interference term of the transmission
\begin{eqnarray}
T_\mathrm{I}(E&)&=\mathrm{Tr}\left[\bar{\Gamma}_\mathrm{L,NI}(E) {\Delta\bar{G}}_\mathrm{NI}^{\mathrm{MB}\dagger}(E) \bar{\Gamma}_\mathrm{R,NI} (E)
\bar{g}_\mathrm{NI}(E)\right]\nonumber\\
&+&
\mathrm{Tr}\left[\bar{\Gamma}_\mathrm{L,NI}(E) {\bar{g}}_\mathrm{NI}^{\dagger}(E) \bar{\Gamma}_\mathrm{R,NI} (E)
\Delta\bar{G}_\mathrm{NI}^\mathrm{MB}(E)\right].
\label{eq:tinterference}
\end{eqnarray}
Often the background and AI transmission do not interfere significantly, so that it is useful to give estimates for their separate values. 
In that case typically the transmission is composed of a background value, onto which the peaks due to transport through the AI are added. 
The transmission through the AI can be rewritten as
\begin{equation}
T_\mathrm{AI}(E)=\mathrm{Tr}\left[\bar{\gamma}_\mathrm{L,AI}(E) {\bar{G}}_\mathrm{AI}^{\mathrm{MB}\dagger}(E) \bar{\gamma}_\mathrm{R,AI} (E)
\bar{G}_\mathrm{AI}^\mathrm{MB}(E)\right],
\label{eq:tAI2}
\end{equation}
which depends only on the GF of the AI, and on the hybridization matrices of the AI, given by
\begin{eqnarray}
\bar{\gamma}_\mathrm{L,AI}(E)=&\bar{H}_\mathrm{AI,NI}\ \bar{g}_\mathrm{NI}(E) \bar{\Gamma}_\mathrm{L,NI}(E) \bar{g}_\mathrm{NI}^\dagger(E)\bar{H}_\mathrm{AI,NI}^\dagger,\\
\bar{\gamma}_\mathrm{R,AI}(E)=&\bar{H}_\mathrm{AI,NI}\ \bar{g}_\mathrm{NI}^\dagger(E) \bar{\Gamma}_\mathrm{R,NI}(E) \bar{g}_\mathrm{NI}(E)\bar{H}_\mathrm{AI,NI}^\dagger.
\end{eqnarray}
We can perform the analogous transformations also for $I_\mathrm{R,AI}$ from Eq. (\ref{eq:irai}), and obtain
\begin{eqnarray}
I_\mathrm{R,AI}&=&\frac{e}{h}\int dE \;\mathrm{Tr}\left[
\left(\bar{F}^\mathrm{MB}_\mathrm{AI}(E)-f_\mathrm{R}(E)\right)
\right.\nonumber\\
&&\left.
\bar{\Gamma}^\mathrm{MB}_\mathrm{AI}(E) \bar{G}^{\mathrm{MB}\supi{\dagger}}_\mathrm{AI}(E)
\bar{\gamma}_\mathrm{R,AI}(E) \bar{G}^\mathrm{MB}_\mathrm{AI}(E)\right],
\end{eqnarray}
where we have defined
\begin{equation}
\bar{\Gamma}^\mathrm{MB}_\mathrm{AI}(E)=i\left[\bar{\Sigma}^\mathrm{MB}_\mathrm{AI}(E)-\bar{\Sigma}^{\mathrm{MB}\supi{\dagger}}_\mathrm{AI}(E)\right].
\label{eq:gammamb}
\end{equation}
In general therefore the calculation of the current requires the knowledge of the impurity occupation matrix $F^\mathrm{MB}(E)$, 
which in turn requires the solution of the non-equilibrium problem. 
In Refs.~\cite{Ng,Ferretti} a specific shape is implicitly assumed for this matrix\cite{Ness}, which allows simplified estimates of the currents. 
The approach proposed in those references requires the inversion of the coupling matrices $\Gamma_\mathrm{L}$ and $\Gamma_\mathrm{R}$, 
which is however not defined in the general case \cite{SelfEnergies}.

We now consider the special case where $\bar{\gamma}_\mathrm{L,AI}(E)=\lambda\bar{\gamma}_\mathrm{R,AI}(E)$, with $\lambda$ a constant\cite{Meir}. 
This rather strict condition can usually only be fulfilled for a SIAM, so that in the remainder of this section we consider only the SIAM. 
In this case it is possible to avoid the calculation of $\bar{F}^\mathrm{MB}(E)$ by using the current conservation condition $I_\mathrm{L}=I_\mathrm{R}$\cite{Meir}, and one obtains
\begin{eqnarray}
I_\mathrm{R,AI}=\frac{e}{h}\int dE \;\left(f_\mathrm{L}(E)-f_\mathrm{R}(E)\right)T_\mathrm{R,AI}(E),
\label{eq:iraispecial}
\end{eqnarray}
with
\begin{eqnarray}
&&T_\mathrm{R,AI}=\nonumber\\
&&
\bar{\Gamma}^\mathrm{MB}_\mathrm{AI}(E) \bar{G}^{\mathrm{MB}\supi{\dagger}}_\mathrm{AI}(E)
\frac{\bar{\gamma}_\mathrm{L,AI}(E)\bar{\gamma}_\mathrm{R,AI}(E)}{\bar{\gamma}_\mathrm{L,AI}(E)+
\bar{\gamma}_\mathrm{R,AI}(E)}
\bar{G}^\mathrm{MB}_\mathrm{AI}(E),
\end{eqnarray}
where we have used the fact that for the SIAM all quantities in the equation are just numbers instead of matrices.

Using Eqs. (\ref{eq:ielastic}) and (\ref{eq:iraispecial}) for the special case of a SIAM, and $\bar{\gamma}_\mathrm{L,AI}(E)=\lambda\bar{\gamma}_\mathrm{R,AI}(E)$, 
we can write the total current as
\begin{eqnarray}
I_\mathrm{R}=\frac{e}{h}\int dE \;\left(f_\mathrm{L}(E)-f_\mathrm{R}(E)\right)T_\mathrm{t}(E),
\label{eq:itotal}
\end{eqnarray}
with the total effective transmission, which includes elastic and incoherent terms, given by
\begin{equation}
T_\mathrm{t}(E)=T_\mathrm{B}(E)+T_\mathrm{AI}(E)+T_\mathrm{I}(E)+T_\mathrm{R,AI}(E).
\label{eq:tedecomptotal}
\end{equation}
We can collect the terms that describe the effective total transmission across the AI, $T_\mathrm{t,AI}(E)$, as\cite{Meir}
\begin{eqnarray}
T_\mathrm{t,AI}(E)&=&T_\mathrm{AI}(E)+T_\mathrm{R,AI}(E)=\nonumber\\
&=&\frac{\bar{\gamma}_\mathrm{L,AI}(E)\bar{\gamma}_\mathrm{R,AI}(E)}{\bar{\gamma}_\mathrm{L,AI}(E)+
\bar{\gamma}_\mathrm{R,AI}(E)}\mathrm{Im}\left[-\bar{G}_\mathrm{AI}^\mathrm{MB}(E)\right].
\label{eq:ttai}
\end{eqnarray}
Eqs. (\ref{eq:itotal}-\ref{eq:ttai}) extend the results of Ref.~\cite{Meir} to the more general case including background transmission and interference terms, 
as typically found for STM experiments of molecules on surfaces.\\
In this article we only consider applications to linear-response transport, so that the conductance $\mathcal{G}=dI/dV$ is given by
\begin{equation}
 \mathcal{G}= \frac {e^2}{h}\int dE \bigg(-\frac {df}{dE}\bigg)~T_\mathrm{t,AI}(E).
\end{equation}
This equation generalizes the Landauer formula used in DFT+NEGF to the case of an interacting EM, with the only difference that $T_0(E)$ (used in the Landauer formula) is replaced by  $T_\mathrm{t,AI}(E)$.
Therefore, following the standard practice used in DFT+NEGF that consists in analyzing 
the zero-bias transport properties by means of the transmission at equilibrium $T_0(E)$, 
here we will present the results for zero-bias transport in presence of many-body effects by plotting $T_\mathrm{t,AI}(E)$.

\subsection{CTQMC impurity solver}\label{sec:ai}
\label{sec:ctqmc}

In the present work the AIM is solved by using CTQMC for quantum systems in thermodynamic equilibrium\cite{Gull}, since we only address linear-response transport. 
In this case the method is well-established. We note however that recently
there have been a number of developments towards the extension of CTQMC to out-of-equilibrium problems, both in steady-state and time-dependent frameworks\cite{Dirks,Schiro,Werner2,Werner3,Cohen,Profumo}.\\
In this work, two different algorithms have been considered:
the weak-coupling approach, called continuous-time auxiliary field (CT-AUX)\cite{Gull2}, and the hybridization expansion, strong coupling approach (CT-HYB)\cite{Werner}. 
CT-AUX scales as the product of the interaction $U$ and of the inverse temperature,
so that its application turns out too computationally demanding for Kondo systems at temperatures of the order of only a few Kelvin. 
In contrast, we found CT-HYB able to provide quite accurate results at 
a reasonable computational cost for the specific Au/TOV system considered in the following. 
Since this system is described as a SIAM, here we present the method only for this case.
However, we remark that our implementation can treat multi-orbital systems as well, although only for density-density interaction terms.

CTQMC is restricted to finite-temperatures, where the partition function is given by $Z=  \mathrm{Tr}\big[e^{-\beta \hat H_{\mathrm{IAIM}}}\big]$.
The starting step, which is common to all algorithms, is to separate the interacting AIM Hamiltonian [Eq. (\ref{IAIM})] in two parts, a reference Hamiltonian $\hat H_1$ 
and a perturbation Hamiltonian $\hat H_2$, so that
$\hat H_{\mathrm{IAIM}}=\hat H_1+ \hat H_2$.
Each CTQMC algorithm differs in the exact definition of $\hat H_1$ and $\hat H_2$\cite{Gull}. While in CT-AUX and other weak-coupling approaches 
$\hat H_2$ is set equal to the effective Coulomb interaction term $\hat H_{\mathrm {C}}$ [Eq.~(\ref{Hubbard})],
in CT-HYB one imposes\cite{Werner}
\begin{eqnarray}
\hat H_{1}&=&\hat {\bar H}_\mathrm{AI,D} +\hat {\bar{H}}_\mathrm{TB}  +\hat H_{\mathrm {I}}\label{h1}\\
\hat H_2&=&\hat {\bar H}_\mathrm{AI,NI}\label{h2}
\end{eqnarray}
with $\hat {\bar H}_\mathrm{AI,D}$, $\hat{\bar{H}}_\mathrm{TB}$, $\hat H_{\mathrm {I}}$ and $\hat {\bar H}_\mathrm{AI,NI}$ defined in subsection \ref{sec:interaction}. 
Therefore the perturbation Hamiltonian in CT-HYB
is represented by the bath-AI coupling, while the reference Hamiltonian is that of the decoupled atomic-like correlated AI
 and of the isolated bath.

 Having divided the Hamiltonian in two parts, one is able to introduce the interaction picture, where an operator $\hat O$ depends on the imaginary time $\tau$ as
$\hat O(\tau)= e^{\tau \hat H_1} \hat O e^{-\tau \hat H_1}$, with $0<\tau<\beta$, and 
the partition function is written as the standard time-ordered exponential
\begin{eqnarray}
 Z &=  \mathrm{Tr}\big[e^{-\beta \hat H_1} T_{\tau}e^{-\int_0^\beta d\tau \hat H_2(\tau)}\big]\nonumber\\
&=\sum_{n=0}^\infty\int_0^\beta d\tau_1...\int_{\tau_{n-1}}^\beta d{\tau}_{n}\,  w_n, \label{part_f}
\end{eqnarray}
with $T_\tau$ the time-ordering operator, and
\begin{equation}
 w_n=\mathrm{Tr}\big[ e^{-(\beta-\tau_n)\hat H_1}(-\hat{H}_2)...e^{-(\tau_2-\tau_1)\hat H_1}(-\hat H_2)e^{-\tau_1 \hat H_1}\big].\label{w_part_f}
\end{equation}
The partition function has the form of an integral over a configuration space. In such space,
any particular configuration is specified by the expansion order $n$, the times $\{\tau_1,..., \tau_n\}$ 
and a set of discrete variables, for instance the spin, and it is characterized by the probability distribution $p_n=w_n \prod_{k=1}^n d\tau_k$. 
It is this integral, which is ultimately evaluated by Monte Carlo techniques.\\
By using the definitions of $\hat H_1$ and $\hat H_2$ in Eqs (\ref{h1}) and (\ref{h2}), $w_n$ in Eq.~(\ref{w_part_f}) becomes\cite{Werner, Gull}
\begin{multline}
w_n=\\Z_{\mathrm{B}}\mathrm{Tr}\bigg[ e^{-\beta\hat H_{loc}} T_{\tau} \prod_\sigma  \hat d_\sigma(\tau_{n_\sigma}^\sigma)\hat d^\dagger_\sigma({\tau'}_{n_\sigma}^\sigma)...
\hat  d_\sigma(\tau_{1}^\sigma)\hat d^\dagger_\sigma({\tau'}_{1}^\sigma)\bigg]\\
 \times\prod_\sigma \det{D_\sigma^{-1}}(\tau_{1}^\sigma,...,\tau_{n_\sigma}^\sigma;{\tau'}_{1}^\sigma,...,{\tau'}_{n_\sigma}^\sigma).\label{wc}
\end{multline}
Here $Z_{\mathrm{B}}$ is the bath partition function, $\hat H_{loc}=\hat {\bar H}_\mathrm{AI,D}+\hat{ H}_{\mathrm {I}}$,
$\hat d^{(\dagger)}_\sigma(\tau)= e^{\tau\hat H_{loc}} \hat d^{(\dagger)}_\sigma e^{-\tau \hat H_{loc}}$ and the matrix $D_\sigma^{-1}$ has elements 
\begin{equation}
(D_\sigma^{-1})_{ij}=\tilde\Delta^f_{\mathrm{AI},\sigma}(\tau_i^\sigma-{\tau'}_j^\sigma), \label{delta_tau}
\end{equation}
which are the Fourier transforms of the hybridization function
\begin{equation}
\tilde\Delta^f_{\mathrm{AI},\sigma}(\tau)=\frac 1 \beta \sum_{n=-\infty}^\infty e^{-i\omega_n\tau}\bar\Delta_{\mathrm{AI},\sigma}(i\omega_n),\label{Fourier}
\end{equation}
evaluated for the imaginary-time interval $\tau_i^\sigma-{\tau'}_j^\sigma$, 
which separate pairs of operators $\hat d_\sigma(\tau_i^\sigma)$ and $\hat d^\dagger_\sigma(\tau_j^\sigma)$
in Eq. (\ref{wc}). In the function $w_n$ 
the trace accounts for the impurity that fluctuates between different quantum states as electrons jump in and out,
while the determinants resum the bath evolutions which are compatible with the sequence of quantum fluctuations in the impurity.
Note that here we have explicitly re-introduced the spin-index that was dropped from the equations in the previous subsections. However,
if the bath is non-magnetic like in most cases, the hybridization function is the same for spin-up and down, and can therefore be obtained 
from non spin-polarized calculations.

In the specific case of a one-orbital Hubbard interaction, CT-HYB can be efficiently implemented by using the so-called ``segment representation''\cite{Werner}.
This means that each configuration is depicted by segments, 
which represent time intervals $\tau^\sigma-{\tau'}^\sigma$ 
during which an electron of a given spin resides on the impurity. Notably, with such representation the trace in $w_n$ can be evaluated in polynomial time.
New configurations are then obtained by either adding or removing segments. 
This is enough to ensure ergodicity, although other operations 
(such as shifting the segments' end-points) and global updates must be implemented to ensure an efficient sampling. 
Each update is accepted or rejected according to the Metropolis algorithm. The acceptance probability is efficiently 
computed with standard fast matrix update methods\cite{Gull}. 

As seen in Eq.~(\ref{delta_tau}), CT-HYB requires the Fourier transform of the hybridization function, which in principle is calculated through a summation 
over an infinite number of frequencies.
However, in practice only a finite number of frequencies smaller than a certain cutoff $N_\omega$ can be inevitably  summed up, 
although the high-frequency limit of $\bar \Delta_{\mathrm{AI}}(i\omega_n)$ determines $\tilde\Delta_{\mathrm{AI}}^f(\tau)$ close to $\tau=0$.
Therefore, in order to accurately calculate $\tilde\Delta^f_{\mathrm{AI}}(\tau)$ for any arbitrary  $\tau$, 
we use a standard approach that consists in adding and removing from Eq.~(\ref{Fourier})
the high frequency limit of $\bar\Delta_{\mathrm{AI}}(i\omega_n)$. This limit is given by $-i M_1/\omega_n$, with $M_1$ given in Eq.~(\ref{eq:m1}) of the Appendix \ref{sec:appendixlimitdelta}.
Hence we evaluate $\tilde\Delta^f_{\mathrm{AI}}(\tau)$ as
\begin{multline}
\tilde\Delta^f_{\mathrm{AI},\sigma}(\tau)=\frac{2 M_1}{\beta}
\sum_{n=0}^{N_\omega}
\Big[ \Re\Big\{\Delta_{\mathrm{AI}}(i\omega_n)\Big\}\cos{\omega_n\tau}+\\
+\Big(\Im\Big\{\Delta_{\mathrm{AI}}(i\omega_n)\Big\}-\frac{M_1}{\omega_n}\Big)\sin{\omega_n\tau}\Big]+\frac{ M_1}2,
\end{multline}
where the summation is restricted to positive frequencies, since $\Re\Big\{\Delta_{\mathrm{AI}}(i\omega_n)\Big\}=\Re\Big\{\Delta_{\mathrm{AI}}(-i\omega_n)\Big\}$ 
and $\Im\Big\{\Delta_{\mathrm{AI}}(i\omega_n)\Big\}=-\Im\Big\{\Delta_{\mathrm{AI}}(-i\omega_n)\Big\}$.\\
During Monte Carlo sampling the many-body Matsubara GF can be directly estimated.
However, following Boehnke {\it et al.} \cite{Boehnke}, we use an expansion of the $\bar G^\mathrm{MB}_{\mathrm{AI},\sigma}(i\omega_n)$ 
in terms of Legendre polynomials, $P_l[(2\tau/\beta) -1]$,
\begin{multline}
\bar G_{\mathrm{AI},\sigma}^\mathrm{MB}(i\omega_n)=\\
\sum_{l \geq 0} \bar G_{\mathrm{AI},\sigma}^\mathrm{MB}(l)
\frac{\sqrt{2l+1}}{\beta}\int_0^\beta d\tau e^{i\omega_n\tau} P_l(2\tau/\beta -1),\label{Gl}
\end{multline}
and we estimate the expansion coefficients $\bar G_{\mathrm{AI},\sigma}^\mathrm{MB}(l)$. 
The advantage of this choice is twofold. First, the transformation (\ref{Gl}) 
can be written as a unitary transformation. 
Second, only few Legendre coefficients are needed to express the Matsubara GF. This is because the Legendre coefficients for a Matsubara GF 
decay faster than any power of the Legendre expansion index $l$\cite{Boehnke}.
A valuable effect of this is that the statistical noise is filtered out due to the cutoff at a certain expansion order. 
After extensive tests for the specific system studied in this work, we found that an appropriate choice for the 
cutoff of the Legendre polynomials is around 100. However, such cutoff must generally be determined for each specific case.\\

The many-body self-energy is obtained either from the Dyson Eq.~(\ref{eq:sai}), or by using the expression\cite{Bulla2}
\begin{equation}
 \bar\Sigma^\mathrm{MB}_{\mathrm{AI},\sigma}(i\omega_n)=U\frac {\bar F_{\mathrm{AI},\sigma}^\mathrm{MB}(i\omega_n)}{\bar G^\mathrm{MB}_{\mathrm{AI},\sigma}(i\omega_n)},\label{FG}
\end{equation}
where $\bar F_{\mathrm{AI},\sigma}^\mathrm{MB}(i\omega_n)$ is the Matsubara representation of the correlation function
$\tilde F_{\mathrm{AI},\sigma}^\mathrm{MB}(\tau-\tau')=-\langle T_\tau\hat d_{-\sigma}(\tau)\hat d_{-\sigma}^\dagger(\tau')\hat n_\sigma(\tau')\rangle$,
which can be easily computed in the Legendre polynomial basis at no extra cost\cite{Hafermann}. 
This last approach usually provides much  more accurate results than the calculation through the Dyson equation. 
The inversion of the GF in Eq. (\ref{eq:sai})
amplifies the statistical noise, in particular at high Matsubara frequencies, when the difference between 
interacting and non-interacting GF is very small. In contrast, such problem is not present when using Eq. (\ref{FG}).\\
Finally, we note that, although in this subsection we have made explicit the dependence of the GF on the spin-index for notation completeness, 
the spin up and down GFs are equal as long as the substrate is non-magnetic, and there is no Zeeman-like term in the AIM Hamiltonian.
Since this is the case for the application presented in this work, we will once again drop this index in the following.

\subsection{Analytic continuation}\label{sec:analytic}
\label{sec:analyticCont}

In order to compute the transport properties, as outlined in subsection \ref{sec:teq}, we need 
the retarded many-body GF on the AI, $\bar G_\mathrm{AI}^\mathrm{MB}(E)$, whose 
imaginary part defines the spectral function
\begin{equation}
 A_\mathrm{AI}^\mathrm{MB}(E)=-\frac 1 \pi \Im \bar G_\mathrm{AI}^\mathrm{MB}(E),\label{A_MB}
\end{equation} 
 normalized as
\begin{equation}
  \int_{-\infty}^{\infty}d E~A_\mathrm{AI}^\mathrm{MB}(E)=1,\label{normalization}
\end{equation}
and which is related to the real part via the Kramers-Kronig relation
\begin{equation}
\Re \bar G_\mathrm{AI}^\mathrm{MB}(E)=- \mathrm{P}\int_{-\infty}^{\infty} d E' \frac{A_\mathrm{AI}^\mathrm{MB}(E)}{E'-E}. 
\end{equation}
However, CTQMC returns the Matsubara GF $\bar G_\mathrm{AI}^\mathrm{MB}(i\omega_n)$ and not $\bar G_\mathrm{AI}^\mathrm{MB}(E)$.
The relation between the spectral function, Eq. (\ref{A_MB}), and the Matsubara GF is determined by the Hilbert transformation
\begin{equation}
\bar G_\mathrm{AI}^\mathrm{MB}(i\omega_n)=-\int_{-\infty}^{\infty}d E \frac{A_\mathrm{AI}^\mathrm{MB}(E)}{i\omega_n-E}, \label{analytic}
\end{equation}
which must therefore be inverted to find the unknown $A_\mathrm{AI}^\mathrm{MB}(E)$ from $\bar G_\mathrm{AI}^\mathrm{MB}(i\omega_n)$.
Unfortunately, the inversion of the Hilbert transformation belongs to the class of ill-posed problems, for which there is no unique solution in a mathematical sense. 
Despite that, several numerical methods have been proposed during the last few decades to deal with this problem in the context of quantum Monte Carlo methods\cite{Vidberg,Jarrell,Sandvik, Beach, Fuchs}. 
Here we employ the stochastic optimization (SO), and our implementation is based on the original proposal by Mishchenko {\it et al.}\cite{Mishchenko}, 
with some modifications. 
We have verified that there are only
negligible quantitative differences between the results obtained with our code and the original version by Mishchenko\cite{Igor}.
Other methods, such as the Pade approximation\cite{Vidberg} and the maximum entropy\cite{Jarrell} methods, 
which are routinely employed by the DMFT practitioners, have been tested as well. 
However, the results are found to depend critically on the precision of the Monte Carlo data, so that they result in overall unsatisfactory performances. 

The SO relies on the parametrization of the 
spectral function $A_\mathrm{AI,t}^\mathrm{MB}(E)$ as a sum of $K$ rectangles determined by the height $h_t$, the length $l_t$ and the center $c_t$,
and which is normalized as in Eq. (\ref{normalization}). This means that
\begin{equation}
A_{\mathrm{AI},t}^\mathrm{MB}(E)=\sum_{t=1}^K f(c_t,l_t,h_t, E)\,\,,\,\,\,\sum_{t=1}^K l_t h_t=1,
\end{equation}
with
\begin{equation}
 f(c_t,l_t,h_t,E)=\begin{cases} h_t \mbox{    if   } E\in[c_t-l_t/2,c_t+l_t/2]\\ 0 \mbox{    otherwise}\end{cases}.
\end{equation}
Accordingly, the GF obtained by using this $A_{\mathrm{AI},t}^\mathrm{MB}(E)$ in Eq.~\ref{analytic} reads
\begin{equation}
\bar G_{\mathrm{AI},t}^\mathrm{MB}(i\omega_n)=-\sum_{t=1}^K h_t \ln\bigg[\frac{c_t-l_t/2-i\omega_n}{c_t+l_t/2-i\omega_n}\bigg].  \label{Gt_analytic}
\end{equation}
The optimization algorithm proceeds through a series of global stochastic updates that change the number, the size and the location of the rectangles
in order to minimize a given deviation function 
between $G_{\mathrm{AI},t}^\mathrm{MB}(i\omega_n)$ and the original set of GFs obtained by CTQMC. 
Importantly, the fact that $\bar G_{\mathrm{AI},t}^\mathrm{MB}(i\omega_n)$ has an analytic expression [Eq.~(\ref{Gt_analytic})]
largely improves the performances of the method.

Once the many-body retarded GF is computed for real energies, the many-body self-energy on the real energy axis can be obtained by using the Dyson Eq. (\ref{eq:sai}).
Alternatively, the analytic continuation of the self-energy $\bar \Sigma_{\mathrm{AI},t}^\mathrm{MB}(i\omega_n)$ can be directly carried out
\cite{Fuchs2,Goth}.

\section{Computational details}
\label{sec:comput}

\subsection{DFT}
The DFT calculations for the TOV molecule in gas phase are performed by using the Siesta code\cite{Siesta}.
Norm-conserving Troullier-Martin pseudopotentials are used together with a double-$\zeta$  plus polarization quality basis set. 
The Perdew-Burke-Ernzerhof (PBE) generalized gradient approximation (GGA)\cite{PBE1,PBE2} for the exchange-correlation density functional is considered. 
Additional all-electron calculations are carried out with the FHI-AIMS package\cite{Blum,Ren,Caruso}.
In particular, FHI-AIMS is used to compare the results of PBE with those obtained with the PBE0 hybrid functional\cite{PBE0}.
Furthermore, the G$_0$W$_0$ method\cite{Ferdi} of the many-body perturbation theory is also employed. 
G$_0$W$_0$ calculations are carried out by using the
(generalized)-KS orbitals and eigenvalues obtained by either PBE or PBE0 (the notation
G$_0$W$_0$@PBE and G$_0$W$_0$@PBE0 is used in order to distinguish between the two cases). 
The computational details are the same as those described in Ref.~\cite{Droghetti}. \\ 
FHI-AIMS is also employed to optimize the geometry of the TOV molecule on Au within a supercell approach. 
The Au(111) surface is experimentally found to present FCC and HCP domains separated by ridge regions, and with the molecules that are adsorbed on both domains\cite{JLiu}.
Here we consider the FCC surface, which is modeled as a 4-layer slab with a (5x5) square unit cell, and each slab is separated by 60 \AA~of vacuum from its periodic 
image.
Only the molecule and the first two Au layers are allowed to relax until forces are smaller than $0.01$ eV/$\AA$. 
The PBE+vdW$^{surf}$ method\cite{Tkatchenko,VdW_surf,VdW_surf2} is employed 
in order to include the effect of the van der Waals (vdW) interactions. 
PBE+vdW$^{surf}$ has been extensively tested for molecules both physisorbed and chemisorbed
on metallic substrates, and the results are generally more accurate than those obtained
with other common vdW-corrected GGA functionals\cite{WLiu, WLiu2}. The values of the 
screened $C_6$ coefficient, the vdW radius and the polarizability for Au are provided by Ruiz {\it et al.} in the original article about  PBE+vdW$^{surf}$\cite{VdW_surf}.
The standard numerical atom-centered orbitals basis set ``tier 1'' and ``tier 2'' are considered for Au and H, C, N, O, respectively. 
The used $k$-points mesh is equal to $4\times4\times1$.
All DFT calculations for the gas phase molecule and for the molecule on Au are spin-polarized in order to account for the magnetism.

\subsection{DFT+NEGF}
\label{sec:tdft}
\textit{Smeagol} electron transport calculations are performed with the relaxed geometries described in the previous subsection, where 
further Au layers are added below the molecule, and the top electrode with the Au STM tip is included [Fig. \ref{fig:structtransport}(a)].
The real space mesh is set by an equivalent energy cutoff of 300 Ry, and we use the local density approximation (LDA) for the exchange correlation potential. 
We verified that the transport properties are essentially the same when using the LDA or GGA functionals. 
We employ a double-$\zeta$ plus polarization basis set for the all atoms of the molecule, a double-$\zeta$ basis for the Au atoms, 
and a double-$\zeta$ plus polarization basis with extended orbital cutoffs for the Au atoms of the tip.
This ensures that there is an appropriate electronic coupling between tip and molecule also at extended tip-molecule separations.
We verified that due to the large size of the supercell in the plane we can evaluate the electronic structure at the $\Gamma$-point in the Brillouin zone perpendicular 
to the transport direction. 

All calculations of the DFT+NEGF transport properties without the inclusion of the many-body self-energy are spin-polarized. 
In contrast, non-spin-polarized calculations are used to extract the hybridization function 
$\bar\Delta_\mathrm{AI}(z)$ and on-site Hamiltonian $\epsilon_\mathrm{AI,D}$ of the AIM, since  
the magnetism of the TOV is accounted for within the AIM. Spin-polarized calculations of the hybridization function are only required if the substrate is magnetic. 

\section{Results}
\label{sec:results}
\begin{figure}[t!]
\centerline{
\includegraphics[width=0.26\textwidth,clip=true]{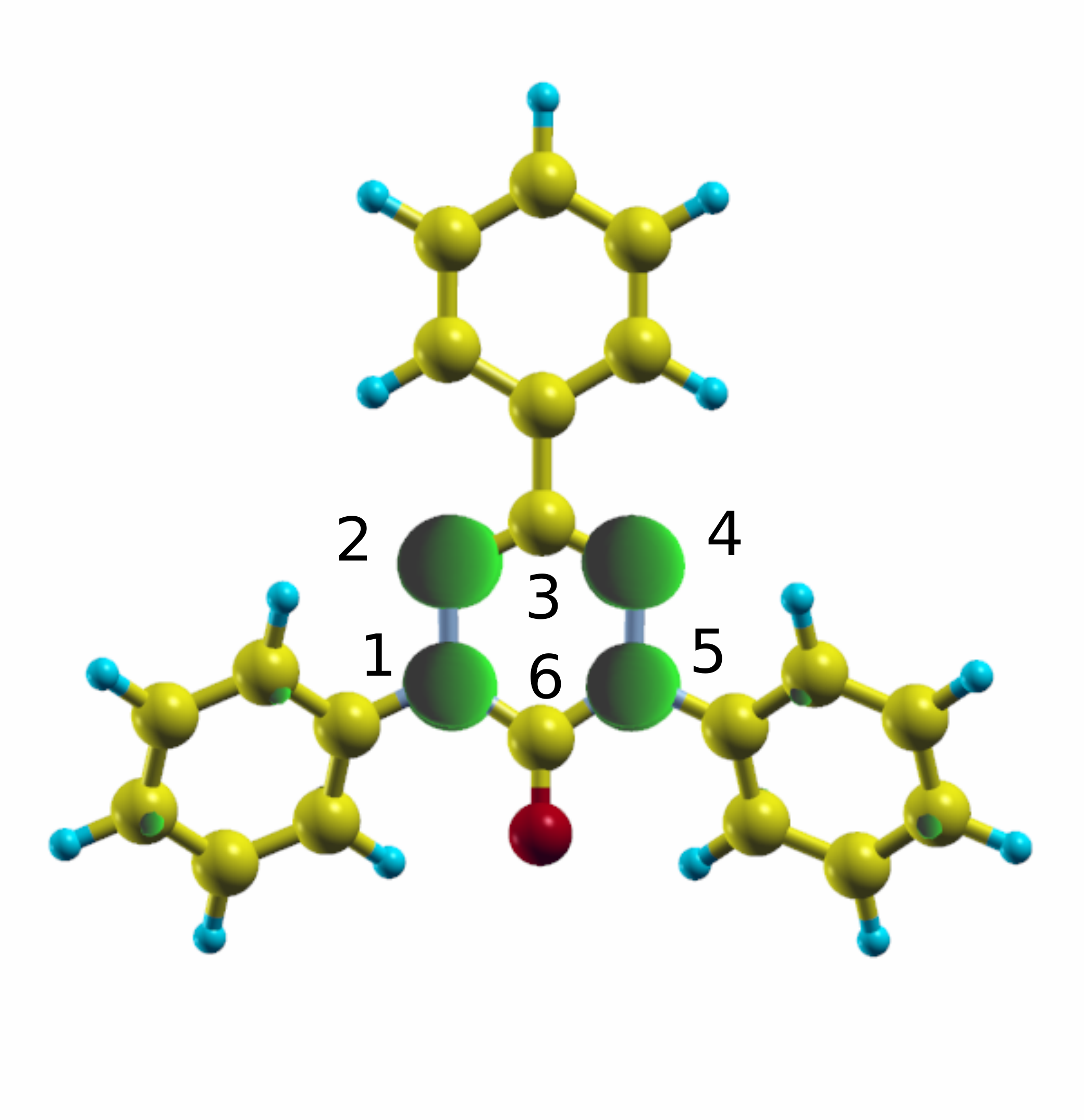}
\includegraphics[width=0.18\textwidth,clip=true]{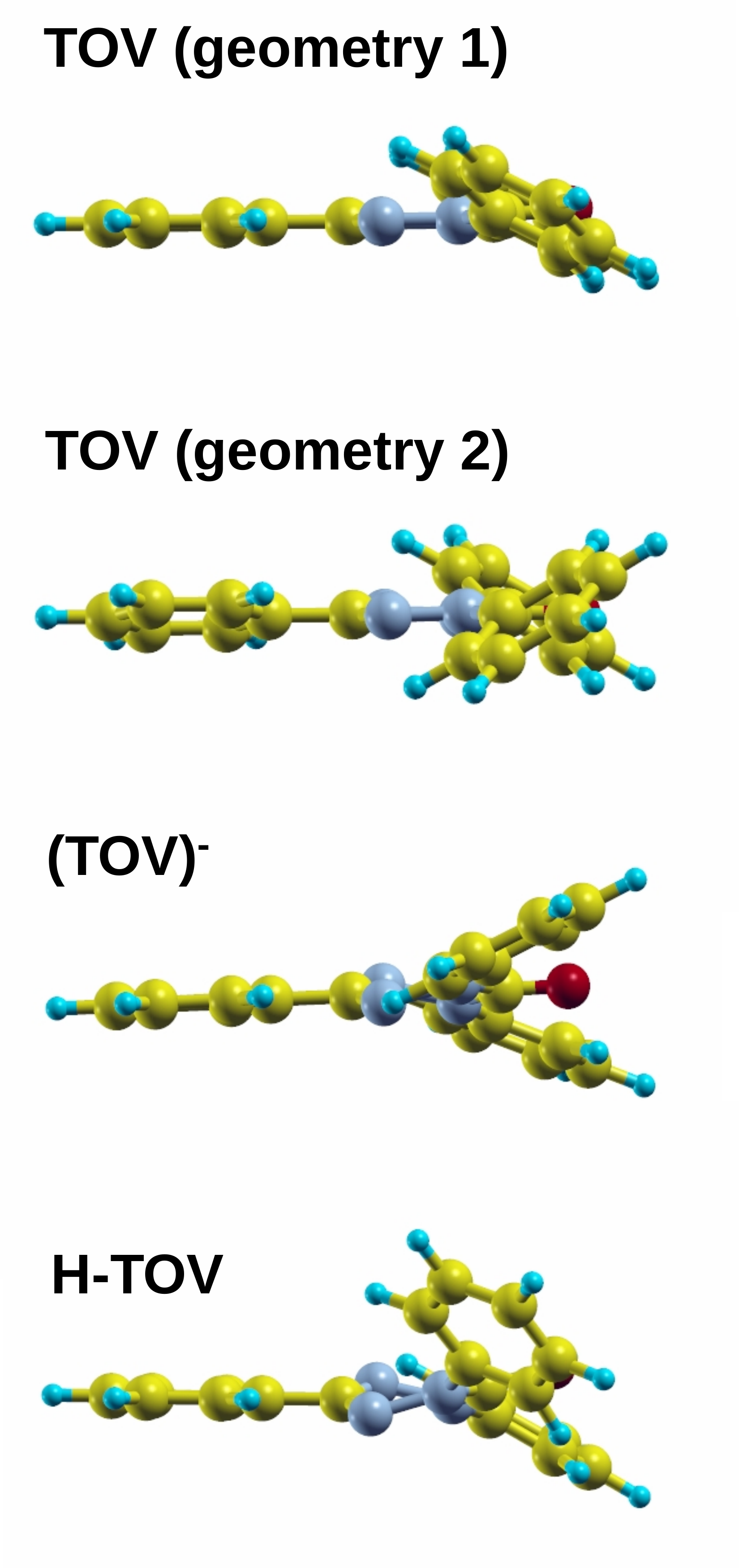}
}
\caption{(Color online) Left: top view of the TOV molecule, indicating also the electron density isosurface of the TOV SOMO (green bubbles) extending over the nitrogen atoms; 
the atoms N1, N2, C3, N4, N5 and C6 are explicitly indicated.
Right: side view of the optimized geometry for TOV with the two energetically-equal conformations, the negatively-charged TOV and H-TOV.
 Color code: C atoms - yellow, H atoms - cyan, N atoms - gray, O atom -red. Note that the N-atoms are not visible in the top view as they are surrounded by the green isosurface.}
\label{fig.LDOS}
\end{figure}
\begin{figure}[t!]
\centering\includegraphics[width=0.5\textwidth,clip=true]{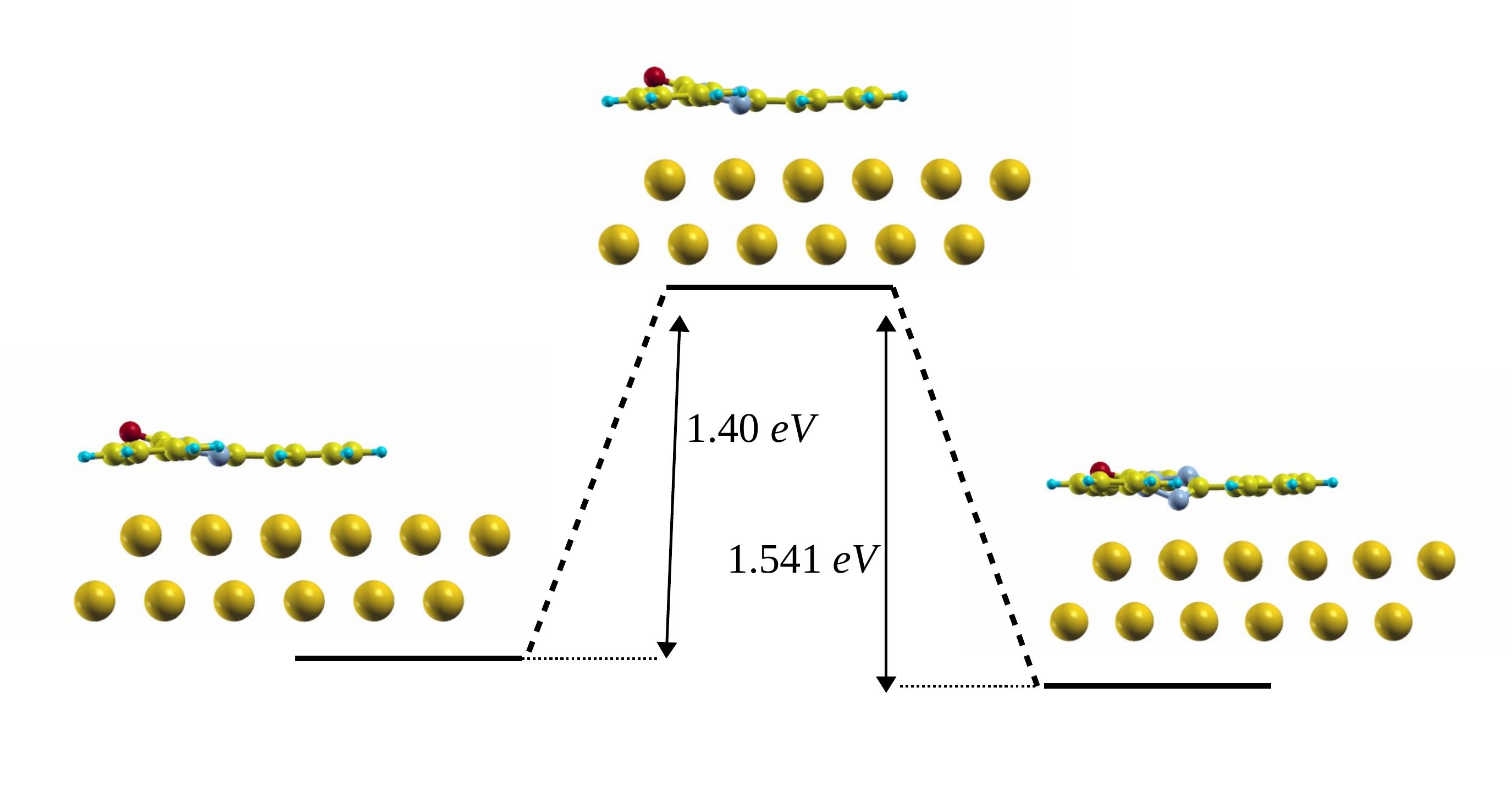}
\caption{(Color online) Energy barrier separating the configurations CFG$_1$ (left) and CFG$_2$ (right).}
\label{fig.barrier}
\end{figure}
\begin{table}
{%
\begin{tabular}{lcccc}\hline\hline
 Method  & $E_{\mathrm{SOMO}}$ (eV) &  $E_{\mathrm{SUMO}}$(eV)& $E_\mathrm{gap}$ (eV)   \\\hline\hline
PBE (KS eigenvalues) & $-4.54$  & $-3.84$ & $0.7$   \\
PBE0 (KS eigenvalues) & $-5.77$ &$-2.97$ & $2.8$ \\
$\Delta$SCF (PBE)& $-6.53$ & $-1.87$ & $3.56$ \\
G$_0$W$_0$@PBE & $-6.26$ & $-2.39$ & $3.87$\\
G$_0$W$_0$@PBE0 & $-6.86$ & $-2.01$ & $4.85$ \\
\hline
\end{tabular}}
\caption{SOMO and SUMO energies, $E_{\mathrm{SOMO}}$ and $E_\mathrm{SUMO}$, and SOMO-SUMO energy gap $E_\mathrm{gap}=E_\mathrm{SUMO}-E_\mathrm{SOMO}$ 
computed with several methods: PBE, PBE0, $\Delta$SCF, and 
the G$_0$W$_0$ perturbative correction to the (generalized-)KS spectrum computed with PBE(0). 
$\Delta$SCF provides the SOMO and SUMO energies 
for an isolated molecule through finite total energy differences
$E_{\mathrm{SOMO}} =E(N)-E(N-1)$ and $E_\mathrm{SUMO}=E(N+1)-E(N)$, where $E(N)$ is the PBE energy of the neutral molecule and $E(N-1)$ [$E(N+1)$] 
is the energy of the corresponding cation (anion)\cite{Jones}. 
}\label{SOMO_SUMO}
\end{table}
\subsection{TOV molecule and adsorption geometry}
\label{sec:geo}
The TOV molecule has spin 1/2 due to an unpaired electron, delocalized mainly over the nitrogen $\pi$ orbitals of the planar verdazyl heterocycle (Fig. \ref{fig.LDOS}). 
The atoms N2 and N4 carry slightly more spin than N1 and N5, and
the difference of the Mulliken populations for spin up and down is equal to $0.31$ for  N2 and N4, and to $0.16$ for N1 and N5. The spin delocalization over the verdazyl heterocycle
reflects the character of the SOMO, and of the corresponding singly unoccupied molecular orbital (SUMO) (Fig.~\ref{fig.LDOS}, left panel).
In contrast, the highest doubly-occupied molecular orbital (HOMO) and the lowest unoccupied molecular orbital (LUMO) are mainly located on the phenyl rings.
The results are overall consistent with the established picture for the TOV electronic structure\cite{Koivisto,Hicks}. \\
Although the four N atoms in the verdazyl heterocycle are in the same plane, the molecule does not have a flat conformation. In fact, we find that the 
 the two phenyl groups
attached to N1 and N5 are twisted out of the verdazyl heterocycle plane, either in the same direction by about $40$ degrees (geometry 1 in right panel of Fig.~\ref{fig.LDOS}), 
or in opposite directions by about $40 $ and $140$ degrees (geometry 2 in the right panel of Fig.~\ref{fig.LDOS}).  
Only the latter conformation (geometry 2) has been previously reported in the literature\cite{Koivisto}, 
but we find that that the energy difference with the geometry 1 is below $1$ meV,
which is the numerical accuracy of our calculations set by the used basis set\cite{Blum}. In contrast,  
the energy difference with the geometry obtained by constraining all three phenyl groups in the same plane as the verdazyl heterocycle is about $0.15$ eV. \\
The molecule's spin can be fully compensated either by charging or by attaching a hydrogen to the oxygen atom, thus forming the H-TOV molecule. 
In both cases, we find a considerable geometrical rearrangement involving also the N atoms, so that the verdazyl core appears largely distorted, 
and the phenyl groups are further moved out-of-plane in an asymmetric way. 
The optimized geometries for the negatively-charged TOV and H-TOV are displayed in Fig.~\ref{fig.LDOS}, right panel. 
Negligible differences are found between the results obtained with PBE and PBE0, and with Siesta and FHI-AIMS.\\  
The SOMO and the SUMO are well separated in energy from all other molecular states, and the
SOMO-SUMO gaps obtained with different methods are reported in Tab.~\ref{SOMO_SUMO}.
Several works have demonstrated that G$_0$W$_0$ yields excellent results for gaps when the generalized KS orbitals from hybrid DFT are used as a starting point for the
perturbative calculation\cite{Droghetti,Blase,Marom,Marom2,Marom3}. For this reason, and due to the lack of experimental data, in the following we
consider the G$_0$W$_0$@PBE0 estimate of $4.85$ eV as reference value. 
We note that that no difference is found within our numerical accuracy for the computed SOMO-SUMO gap in case of the two different TOV conformations (geometries 1 and 2 in in the right panel of Fig.~\ref{fig.LDOS}),
so that the results in Tab.~\ref{SOMO_SUMO} apply to both.\\
The geometry optimization of the TOV/Au system reveals that the molecule can be adsorbed in two different configurations, labelled CFG$_1$ and CFG$_2$. 
In configuration CFG$_1$, TOV is physisorbed and assumes an almost flat conformation on the surface. The distances of N2/N4 and N1/N3 from the top Au layer are
$2.94$ \AA~ and $3.13$ \AA, respectively. The average adsorption height of the phenyl rings attached to N1 and N3 is $3.09$ \AA, while that of the phenyl ring attached to C3 is $2.95$\AA. 
The adsorption energy is equal to $-2.564$ eV, a value that is comparable to that of typical molecules of similar size 
(such as perylenetetracarboxylic dianhydride, PTCDA) physisorbed on Au\cite{VdW_surf}. 
By analyzing the Mulliken populations, we note that a small charge transfer of about $0.2$ electrons 
from the Au to the molecule occurs. This leads to an overall decrease of the SOMO-SUMO exchange splitting and to a partial compensation 
of the molecule magnetic moment, which reduces to $0.3~\mu_B$. 
Configuration CFG$_2$ corresponds to a qualitatively different physisorbed geometry, which has an adsorption energy of $-2.706$ eV. 
For CFG$_2$ one full electron is transferred from the surface to the TOV,
so that the spin on the molecule completely disappears. Although the adsorption energy for CFG$_1$ and CFG$_2$ differs by only about $150$ meV,
the energy barrier separating the two states is quite large ($\approx 1.4$-$1.5$ eV). Indeed we observe that the increased charge transfer is accompanied 
by a substantial distortion of the 
TOV verdazyl heterocycle, with one of the N atoms (N4) being displaced out of the molecular plane towards the Au surface (see Fig.~\ref{fig.barrier}). 
The distance between N4 and the surface decreases to $2.3$\AA, while the distance between N2 and the surface increases to $3.23$\AA.
This distortion of the verdazyl core is similar to that observed in the gas phase
for the negatively-charged molecule (Fig.~\ref{fig.LDOS}), taking into account that now the phenyl groups are kept parallel to the surface because of the vdW interaction. 
We note that our calculations for physisorbed H-TOV reveal a similar deformation of the molecule.
We point out that the charge transfer and the consequent spin reduction or compensation are likely overestimated 
by the calculations, because PBE underestimates the SOMO-SUMO gap. This problem is commonly found in DFT-based electronic transport simulations,
and it has been cured in practice by using scissor operator schemes\cite{Amaury1,Quek,Suarez,Strange1,Strange2, Cehovin} (see the following section). 
The overestimated charge transfer may have important implications for the geometry optimization, since it may artificially stabilize the configuration CFG$_2$ over the configuration CFG$_1$. 
The calculation of the correct charge transfer for structural relaxations can currently not be addressed on a fully quantitative level even with state-of-art methods. 
DFT energy functionals that increase the SOMO-SUMO gap, such as hybrid functionals, do not provide a good description of the metal electronic structure.
For configuration CFG$_1$ the charge transfer is overall rather small, and in cases of small or negligible charge-transfer PBE+vdW$^{surf}$ has been shown to perform remarkably well 
when compared to experimental data\cite{Burker}. 
We therefore expect that the final geometry for CFG$_1$, and in particular the molecule-surface average distance, is accurately predicted.
In experiments two types of molecules are observed on the Au surface, denoted as type-A and type-B\cite{JLiu}. 
While type-B molecules display a Kondo resonance in the low bias conductance, such resonance is absent in type-A molecules. 
Type-A molecules have been identified as H-TOV, while type-B molecules correspond to configuration CFG$_1$\cite{JLiu}. 
We note that based on our results the type-A molecules could also correspond to molecules adsorbed with geometry CFG$_2$. 
However, as discussed in the previous paragraph, it is likely that the calculations overestimate the stability of CFG$_2$,
and in experiments most type-A molecules may indeed be H-TOV. 
In the remaining part of the manuscript we will address only type-B molecules, where the Kondo state has been measured experimentally.

\subsection{Transport properties from DFT+NEGF}
\label{sec:transportDFT}
\begin{figure}[t!]
\centering\includegraphics[width=0.45\textwidth,clip=true]{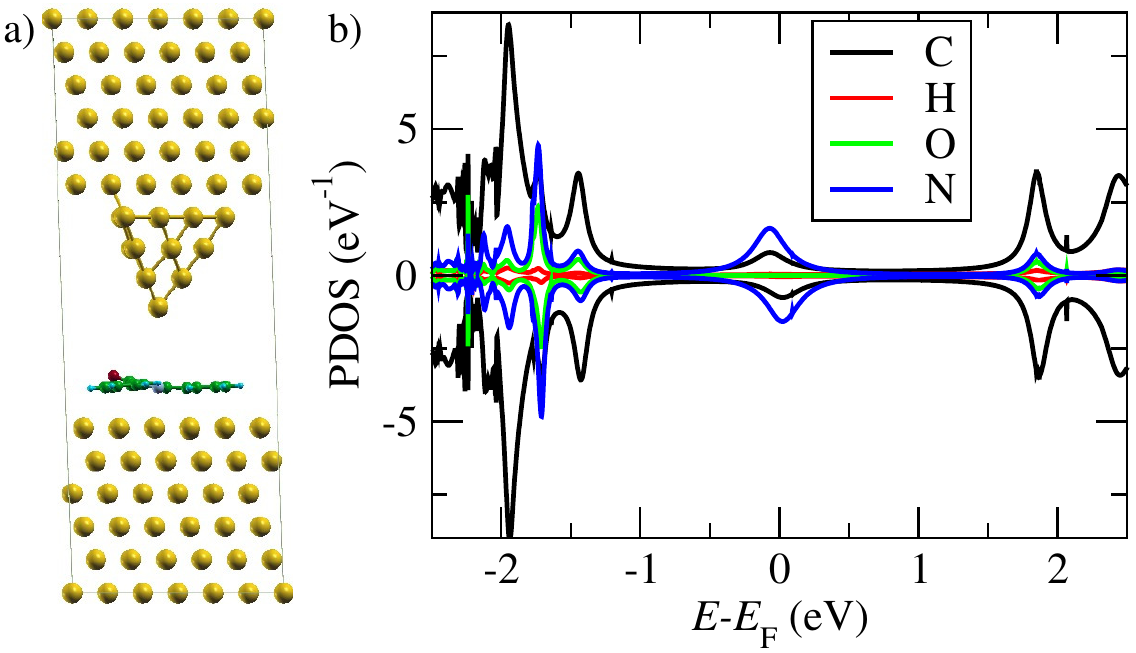}
\caption{(Color online) (a) Scanning tunneling microscope (STM) setup used in the electron transport simulations,
and (b) projected density of states (PDOS) on the atoms of the molecule on the Au surface (positive values are for spin-up states, 
negative values for spin-down states).}
\label{fig:structtransport}
\end{figure}

The spin-polarized projected density of states (PDOS) on the molecule in the transport setup for configuration CFG$_1$ 
is shown in Fig. \ref{fig:structtransport}(b). 
After adsorption, the former gas-phase SOMO and SUMO appear as Lorentzian-like peaks 
that extend around $E_\mathrm{F}$. The combined effects of the hybridization of the SOMO and SUMO states with the Au substrate, 
and of the small surface-to-molecule charge transfer, induce a reduction 
of the exchange splitting compared to the gas phase, and consequently of the magnetic moment. 
The full width at half maximum of the SOMO and SUMO Lorentzian peaks gives their electronic coupling to the substrate, $\Gamma\approx-2\Im\bar\Delta_{\mathrm{AI}}(E=E_\mathrm{F})$, which we calculate to be $\Gamma\approx$ 290 meV. This 
system therefore presents a rather strong electronic coupling to the substrate.
In contrast, the electronic coupling to the tip is usually much smaller, and it can be neglected compared to $\Gamma$.

The analysis of the transport properties is carried out by looking at the transmission coefficient, $T$ (Sec. \ref{sec:teq}). 
This decays exponentially with the tip height and depends sensitively on the in-plane position of the STM tip. 
The transmission through the SOMO and SUMO is largest when the tip is on top of the N atoms of the molecule, 
so that we place it above atom N2 in the subsequent evaluation of the transport properties, with a tip height of about 5.7 \AA~above the plane of the molecule.

\begin{figure}[t!]
\centering\includegraphics[width=0.43\textwidth,clip=true]{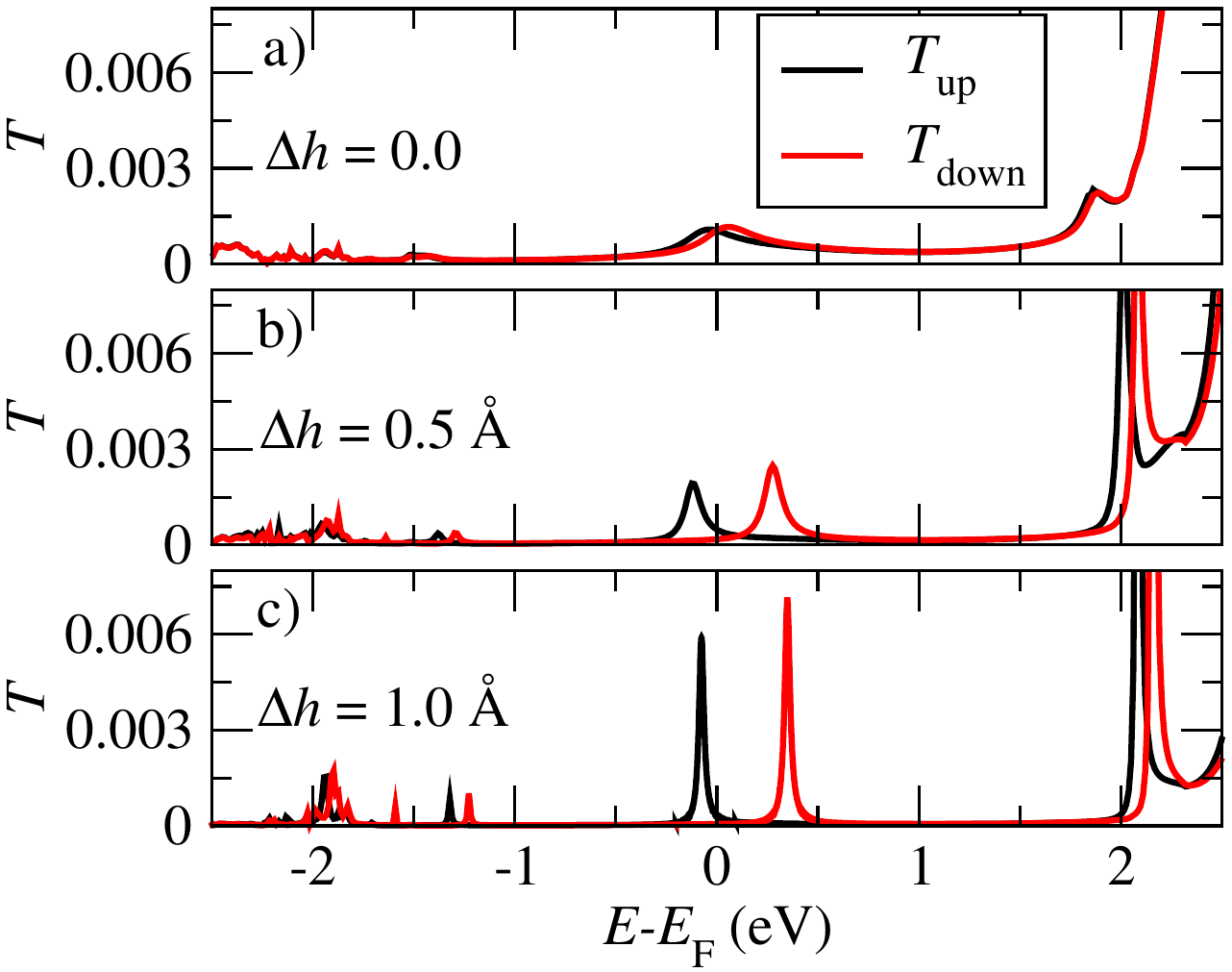}
\caption{(Color online) Transmission for different molecule-substrate separations, given by about 3.1 \AA + $\Delta h$.}
\label{fig:tdistance}
\end{figure}

The spin-polarized transmission as function of energy is shown in Fig. \ref{fig:tdistance}(a),
and it can be seen that, like the PDOS, also $T$ is only slightly spin-split. 
We investigate the change of $T$ for reduced coupling by rigidly shifting the molecule off from the surface by a distance, $\Delta h$, 
of 0.5 \AA, and 1.0 \AA (Fig. \ref{fig:tdistance}(b-c)). The results demonstrate that if the coupling is reduced by increasing the molecule-surface separation, 
then the peak of the radical state is less broad, 
and the spin splitting increases. For $\Delta h$ of 0.5 \AA~the radical is fully spin-split, while it is only partly spin-split for $\Delta h=$ 0.0 \AA. 
This shows that in quantitative predictions of hybridizations of molecules to surfaces the correct evaluation of the contact geometry is of central importance. 
Experimentally such a change of coupling might be achievable by systematically varying the groups attached to the TOV molecule. 

\begin{figure}[t!]
\centering\includegraphics[width=0.43\textwidth,clip=true]{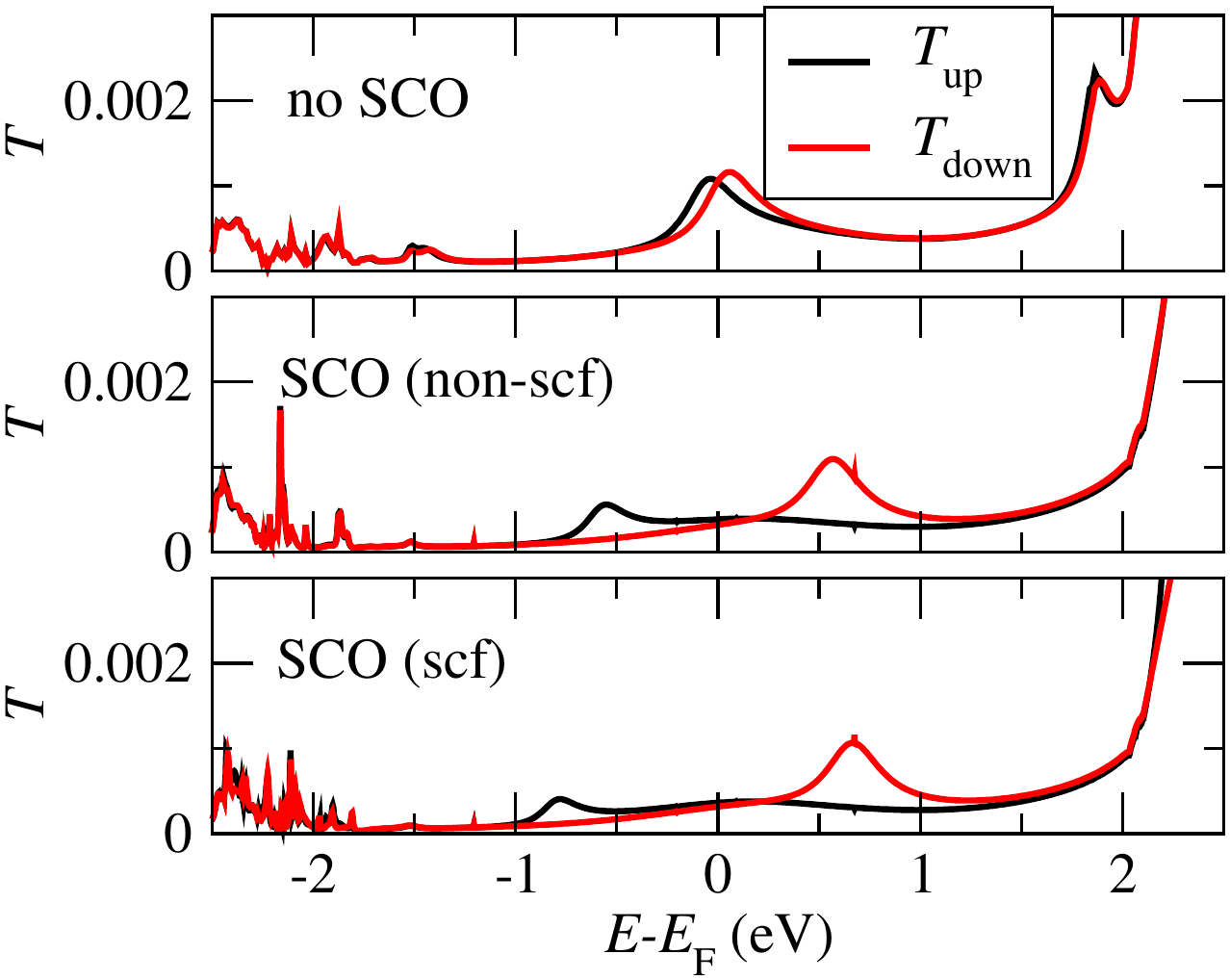}
\caption{(Color online) Transmission for no applied scissor operator (SCO), 
and with a SCO applied non selfconsistently for the converged LDA charge density [SCO (non-scf)], and with a SCO applied selfconsistently [SCO (scf)].}
\label{fig:T_DFT}
\end{figure}

As discussed in the previous section, the LDA gas phase KS SOMO-SUMO gap is underestimated when compared to experiment.
On a surface however the gap is reduced due to the image charge effect, which is not captured by the LDA KS eigenvalues\cite{Amaury1}.
A fortuitous error cancellation between the gap underestimation and the neglect of image-charge effects 
 may sometimes happen, but this is not generally the case. 
Here we evaluate the effect that a larger gap has on the transport properties by using a scissor operator (SCO) approach\cite{Amaury1,Quek,Suarez,Strange1, Cehovin}, 
where we increase the SOMO-SUMO gap by about 1 eV to illustrate the general trend (Fig. \ref{fig:T_DFT}). 
We apply the SCO correction non-self-consistently as a postprocessing step on top of the LDA converged charge density as well as self-consistently.  
The opening of the SOMO-SUMO gap leads to a fully spin-split state, where one can clearly identify SOMO and SUMO peaks despite their large broadening. 
Self-consistency does not qualitatively change the results, demonstrating that applying the correction as a post-processing step to 
the converged LDA solution is a good approximation. The exact value of the SOMO-SUMO gap, which corresponds to the interaction energy $U$ [Eq. (\ref{Hubbard})], 
for molecules on surfaces can only be estimated. We will address this issue for the TOV molecule deposited on Au in Sec.~\ref{sec_Kondo}.

Finally, we note that the DFT+NEGF PDOS and transmission are spin-split and a magnetic moment appears due to the unequal occupation of spin up and down states. 
This is a characteristic broken-symmetry picture, which is only valid for applied magnetic fields\cite{Kepenekian}. 
In absence of magnetic field the symmetry is not broken, 
and the DFT symmetry breaking is an artifact caused by the effective exchange-correlation magnetic field. 

\begin{figure}[t!]
\centering\includegraphics[width=0.43\textwidth,clip=true]{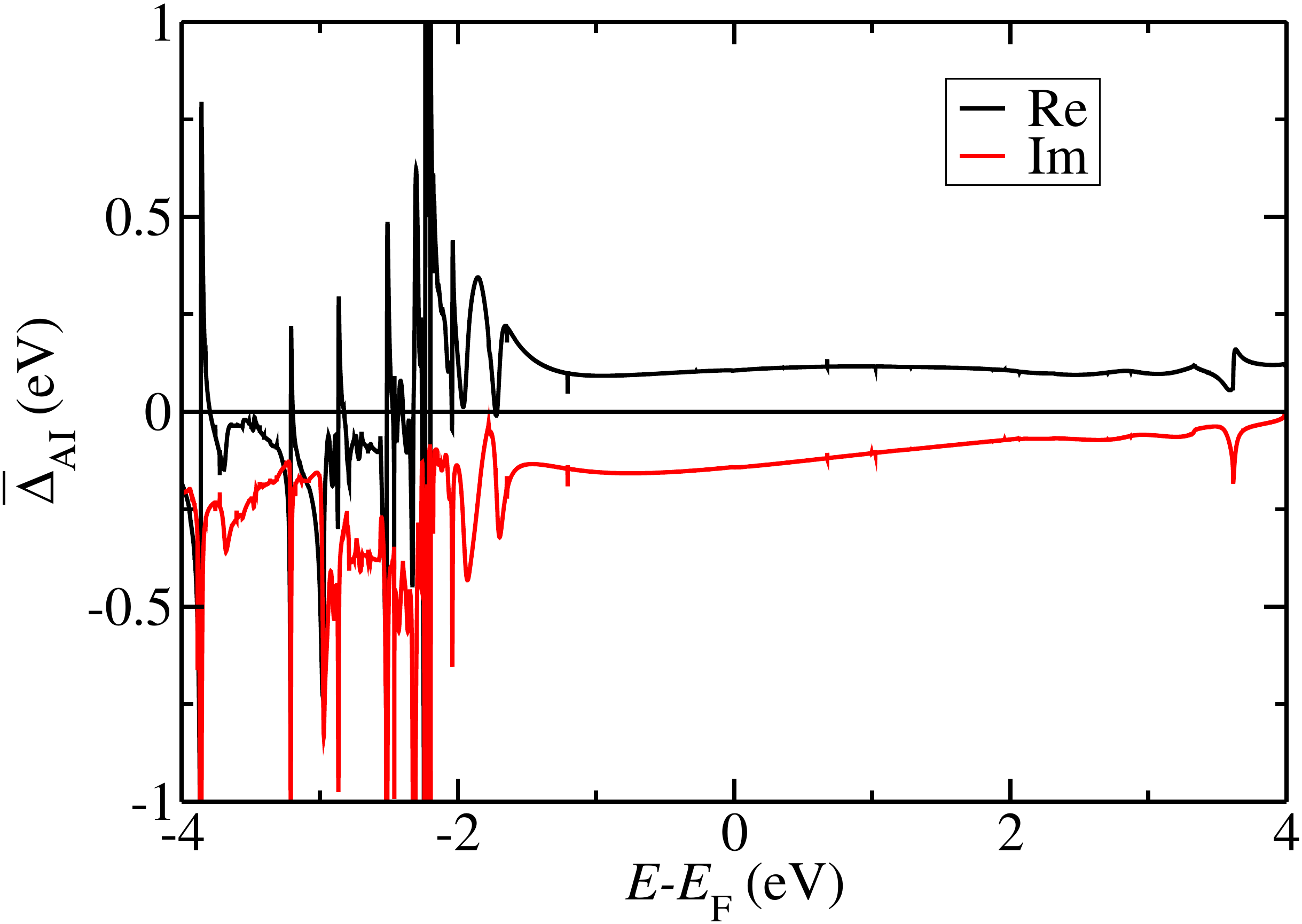}
\caption{(Color online) Real and imaginary parts of the hybridization function, $\bar\Delta_\mathrm{AI}$ (Eq. (\ref{eq:bardelta2})).}
\label{fig.Delta}
\end{figure}
\subsection{Mean-field Anderson impurity model}
\label{sec_MF}

The TOV adsorbed on Au is represented as a SIAM 
through the projection described in Sec. \ref{sec:method}. We remind that here non-spin-polarized DFT calculations are used, since magnetism is 
 accounted for in the SIAM calculation.
The hybridization function, calculated with Eq. (\ref{eq:bardelta2}), is shown in Fig.~\ref{fig.Delta}, and 
mainly reflects the Au DOS. Both the real and imaginary parts are rather featureless over a wide energy range, where only the $4s$ states contribute to the Au DOS,
while very sharp peaks are present at energies below about $-2$ eV, where the fully filled $3d$ bands are located.

\begin{figure}[t!]
\centering\includegraphics[width=0.4\textwidth]{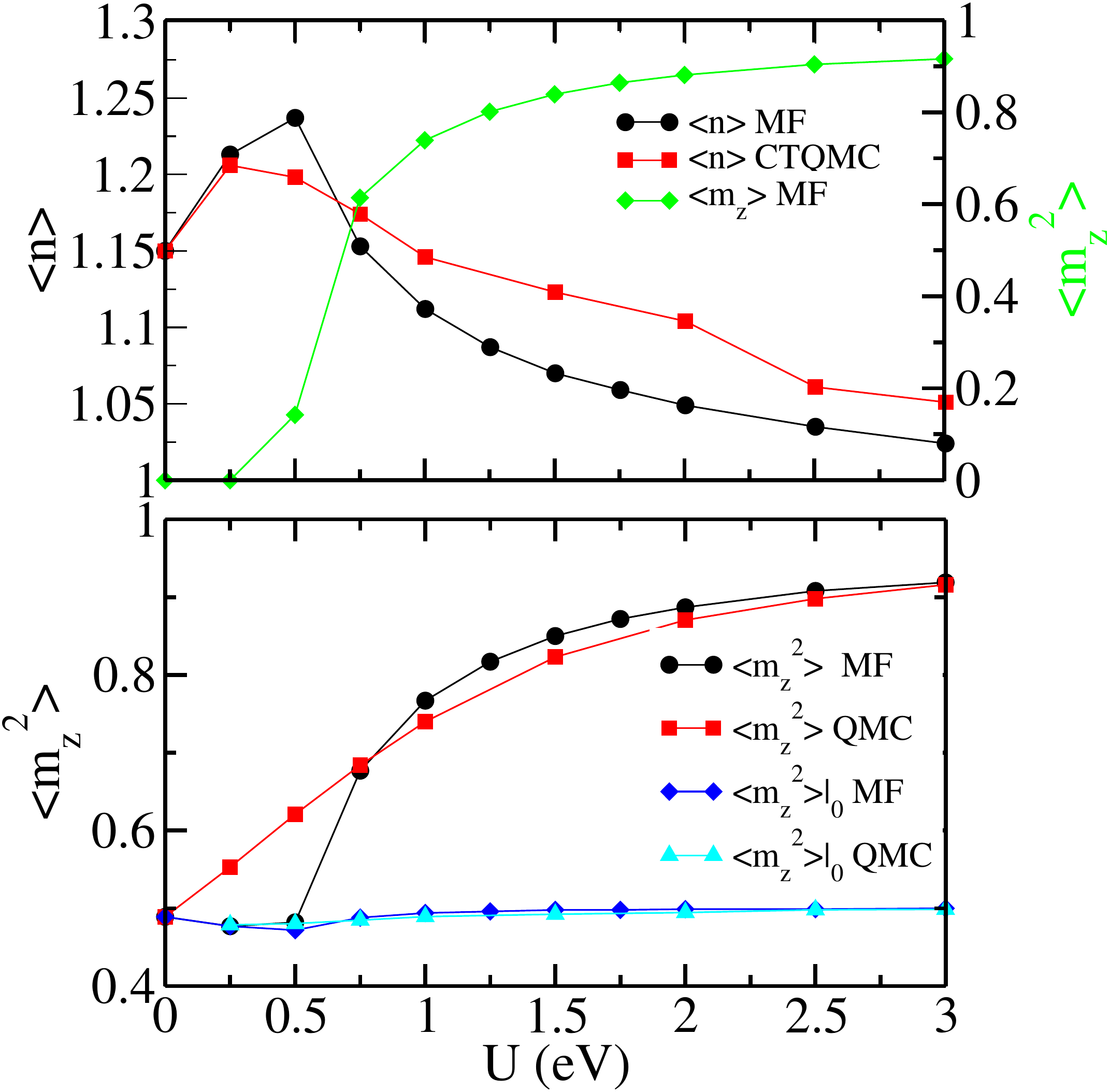}
\caption{(Color online) Top panel: mean-field (MF) and CTQMC on-site occupation
$\langle \hat n\rangle=\langle \hat n_\uparrow\rangle +\langle \hat n_\downarrow\rangle$ and mean-field local magnetic moment 
$\langle \hat  m_z\rangle=\langle n_\uparrow\rangle -\langle \hat n_\downarrow\rangle$ as function of $U$.
Bottom panel: MF and CTQMC squared local magnetic moment $\langle \hat m_z^2\rangle=\langle (\hat n_\uparrow -\hat n_\downarrow)^2\rangle$ and
the corresponding square local magnetic moment in the limit of no correlations $\langle \hat m_z^2\rangle\vert_{0}=\langle \hat n\rangle-\langle \hat n\rangle^2/2$. 
Results are for $\theta=20 K$. Error bars in the CTQMC estimates are smaller than the symbols.
}
\label{fig.n_m}
\end{figure}
\begin{figure}[t!]
\centering\includegraphics[width=0.4\textwidth]{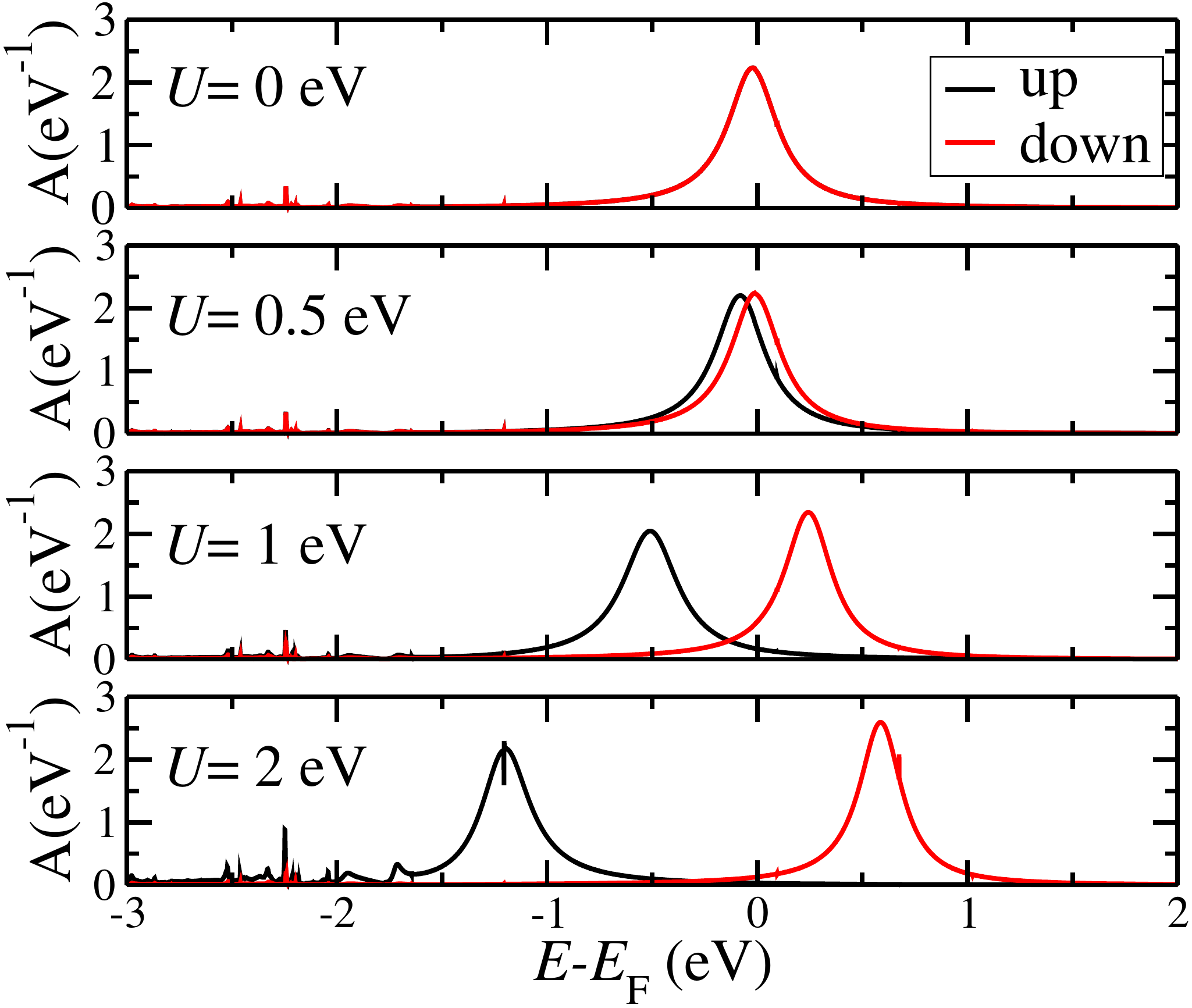}
\caption{(Color online) Mean-field spectral function at zero temperature and for $U= 0$, $0.5$, $1$ and $2$ eV. For $U=0$ eV 
the spectral function for spin up and down are superimposed as 
the systems is spin-degenerate.}
\label{fig.Dos}
\end{figure}
Before solving the SIAM by using CTQMC, we address it within the mean-field approximation,
where the Hamiltonian $\hat{ H}_\mathrm{C}$ in Eq. (\ref{Hubbard}) is replaced by 
 $\hat{ H}_\mathrm{C}^{\mathrm{MF}}=U\sum_\sigma \hat n_\sigma\langle \hat n_{-\sigma}\rangle$, and the double-counting correction 
is the same as in CTQMC [Eq. (\ref{Hdc})]. 
 The mean-field solution 
 can be directly compared to the broken-symmetry DFT+NEGF results, where
$\hat H_\mathrm{C}^{\mathrm{MF}}$ plays the same role as the KS potential including the SCO, and where $U$ determines the SOMO-SUMO gap. 

At $U=0$, 
the total AI occupation  $\langle \hat n\rangle=\langle \hat n_\uparrow\rangle +\langle \hat n_\downarrow\rangle$ is equal to the LDA occupation of the AI, 
obtained with the projection described in Sec. \ref{sec:method}. 
This is slightly larger than $1$, reflecting the small metal-to-molecule charge transfer predicted by DFT, 
and discussed in Sec. \ref{sec:geo}. The magnetic moment $\langle \hat m_z\rangle=\langle \hat n_\uparrow\rangle -\langle  \hat n_\downarrow\rangle$ is equal to zero, 
because the AIM is mapped from a non-spin-polarized calculation and $\hat{H}_\mathrm{C}^{\mathrm{MF}}=0$.
For $U\neq 0$ eV, we see in Fig.\ref{fig.n_m} (top panel) that the total occupation $\langle\hat n\rangle$ initially increases as function of $U$, until 
the system undergoes a transition to the magnetic state. At this point (between $U=0.5 $ and $U=0.75 $ eV) 
the formation of the magnetic moment is accompanied by the reduction of $\langle\hat n \rangle$, 
which then keeps decreasing monotonically for increasing $U$ towards the half-filling limit.
This behavior is qualitatively the same that we find when we apply the SCO in the DFT+NEGF (Sec. \ref{sec:transportDFT}).

The spectral function of the AI is a Lorentzian at $U=0$ eV (Fig. \ref{fig.Dos}), 
and gradually splits into two (almost) Lorentzian functions as $U$ is increased,
one for spin up and one for spin down. Importantly, due to the small charge transfer from the Au substrate, these functions are not symmetric around $E_\mathrm{F}$. 
Moreover, the spectral function for spin down is slightly sharper than that for spin up, 
which is due to the decrease of absolute value of the imaginary part of the hybridization function with increasing energy (Fig. \ref{fig.Delta}). 
This is overall similar to the DFT+NEGF transmission peaks displayed in Fig.~\ref{fig:T_DFT}.

\begin{figure}[t!]
\centering\includegraphics[width=0.4\textwidth]{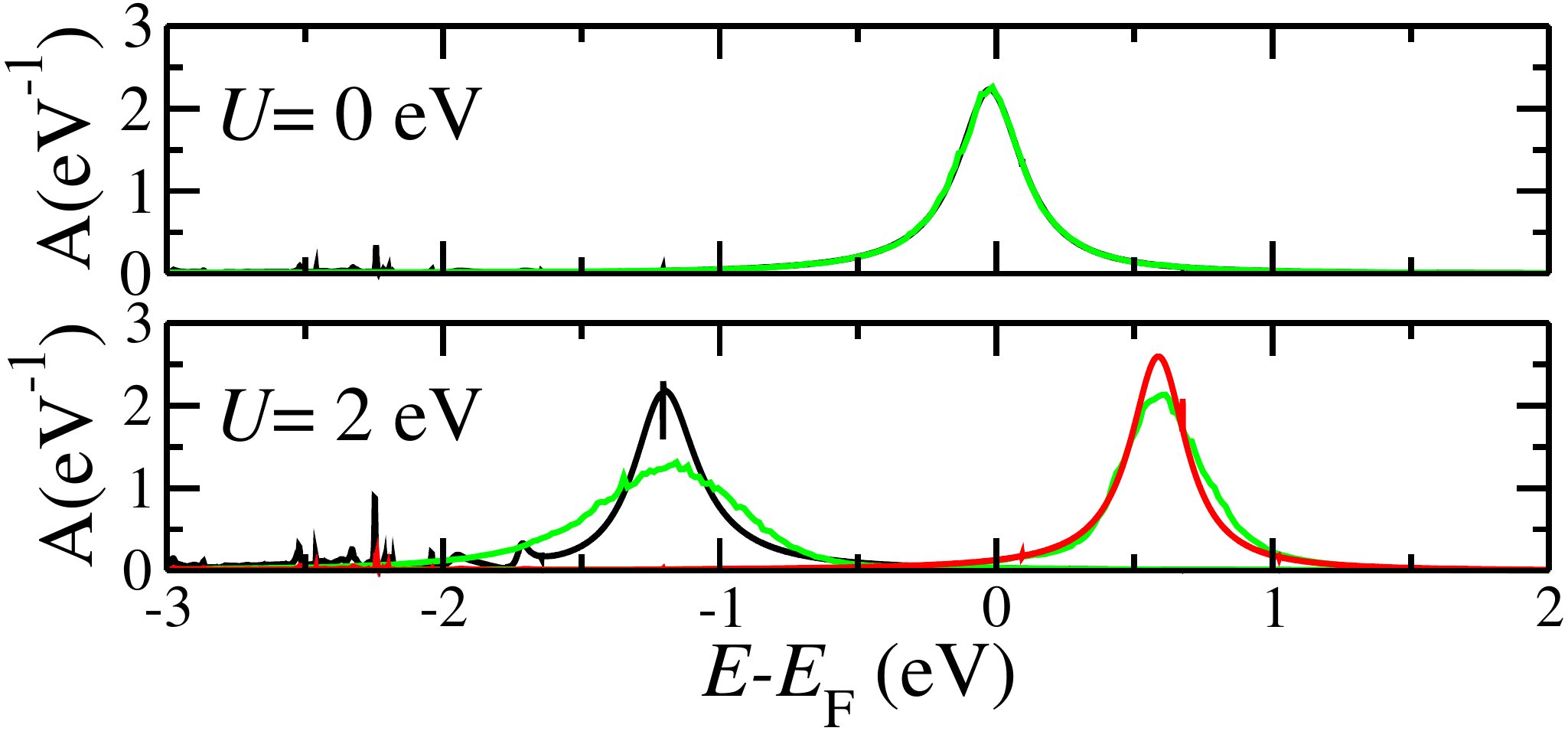}
\caption{(Color online) Comparison between the zero-temperature mean-field spectral function (black and red lines for spin up and down, respectively), 
and the spectral function obtained by performing the analytic continuation of the mean-field Matsubara Green's function at $T=20$ K (green line). }
\label{fig.aMF}
\end{figure}

To conclude this section, we take advantage of the known exact mean-field spectral function for real energies in order to assess the accuracy of the analytic continuation (Sec. \ref{sec:analytic}). 
This is shown in Fig.~\ref{fig.aMF}, which compares the exact spectral function evaluated directly for real energies, to the
one ``analytically continued'' from complex Matsubara frequencies to real energies, and which should in principle be identical to one another.
The accuracy is very good for energies around $E_\mathrm{F}$, where the two spectral functions are almost indistinguishable,
while the agreement becomes progressively worse for much larger (in absolute values) energies.
In particular, the analytically continued spectral function appears broader than the exact one the further one moves away from $E_\mathrm{F}$. 
This additional broadening is an artifact of the analytic continuation, and it is caused by the limited accuracy of the SO (Sec. \ref{sec:analytic}).

\begin{figure}[t!]
\centering\includegraphics[width=0.43\textwidth]{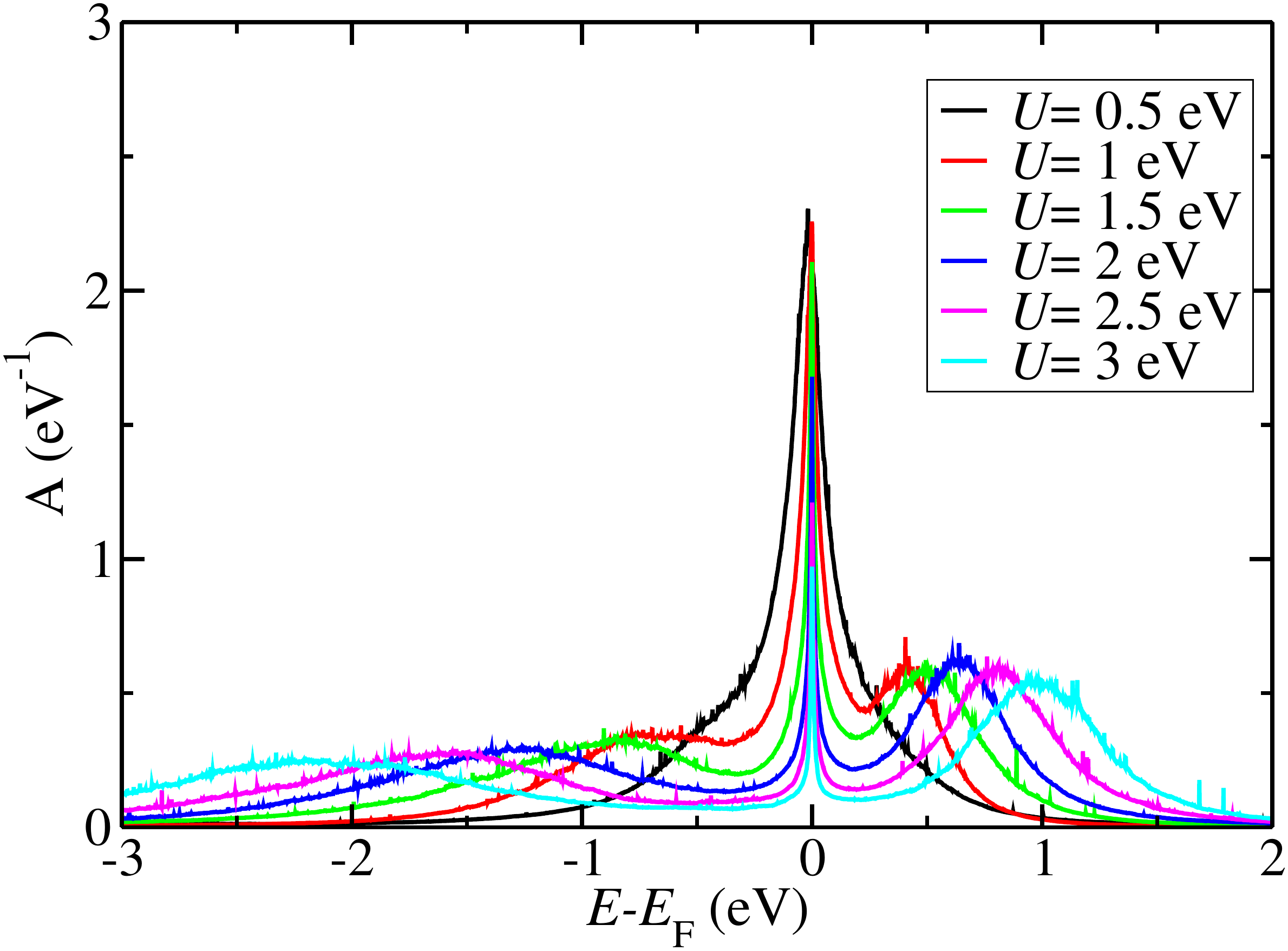}
\caption{(Color online) Spectral function for several $U$ and at $\theta=20$ K.
The noise is a well-know drawback of the SO method\cite{Mishchenko}. 
In principle, it can be greatly reduced by averaging the results of many independent optimizations for the same data set, in practice however it cannot be completely removed. 
The presented results are obtained by averaging over 250 independent runs.}
\label{fig.spectral_U}
\end{figure}
\begin{figure}[t!]
\centering\includegraphics[width=0.43\textwidth]{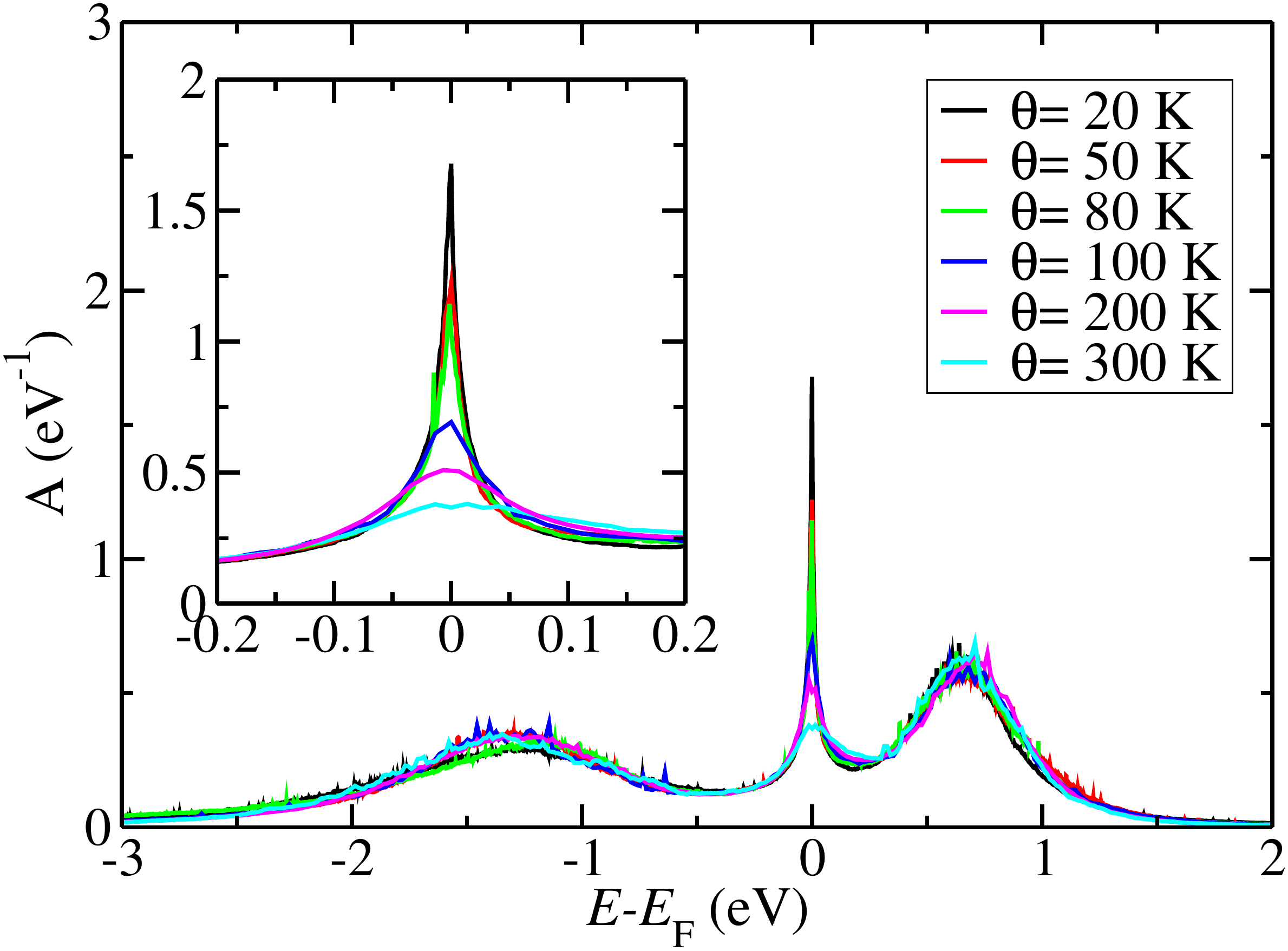}
\caption{(Color online) Spectral function for $U=2$ eV and at several different temperatures, $\theta$. The inset shows a zoom for the $0.4$ eV-range around the Fermi energy. }
\label{fig.spectral_T}
\end{figure}
\subsection{CTQMC simulations and Kondo effect}
\label{sec_Kondo}

The CTQMC impurity occupation $\langle\hat  n\rangle $ and squared local magnetic moment $\langle \hat m_z^2\rangle$ 
 for several $U$-values at $\theta=20$ K are presented in Fig.~\ref{fig.n_m}. 
 We remark that  
$\langle \hat  m_z^2\rangle$ and not $\langle \hat m_z\rangle$ is the appropriate quantity to look at in the analysis of the magnetization for the CTQMC results,
since the occupation for spin up and down is equal in absence of any Zeeman-like term in the Hamiltonian, even in the local moment regime.
This shows that the exact CTQMC solution does not have the unphysical symmetry breaking found for the DFT and mean field solutions. 
The CTQMC $\langle \hat m_z^2\rangle$ gradually increases as function of $U$, 
and there is no evidence of an abrupt magnetic transition at $U\approx 0.5$ eV.
Nevertheless, when mean-field predicts a stable local magnetic moment for $U> 0.5$ eV, the CTQMC $\langle \hat m_z^2\rangle$ 
is smaller than the corresponding mean-field value, while $\langle \hat n\rangle $ is slightly larger.
In this region of parameters the effect of electron correlation is therefore to favor the AI double occupancy compared to the mean-field picture,
so that the magnetic moment is partly quenched. This is a typical behavior expected for a Kondo system. 
For large $U$, after the transition from the Kondo screened to the local magnetic moment regime, the mean-field 
and the CTQMC $\langle\hat  m_z^2\rangle$ converge to the same value,  and $\langle\hat n\rangle $ approaches the half-filling limit.\\
Important insights into the dominant physics at different energy scales are provided by the AI spectral function $A(E)$. 
In Fig.~\ref{fig.spectral_U} we observe that the Lorentzian peak found for $U=0$ eV
 gradually shrinks into a narrow and sharp peak for increasing values of $U$, thus forming the Kondo resonance centered around $E_\mathrm{F}$. 
At the same time, two broad satellite peaks, separated approximately by $U$, develop, and they correspond to the SOMO and SUMO states found in the DFT+NEGF and in the mean-field DOS.
Note that, unlike in the mean-field case of Fig.~\ref{fig.Dos}, here the SOMO and SUMO are not formed by breaking the spin-symmetry through an unequal spin-occupation. 
 Furthermore, they are much broader. Even though the broadening is partly an artifact caused by the analytic continuation (see Sec. \ref{sec_MF}), 
we can still appreciate the real intrinsic  broadening due to electron-electron interaction by comparing Fig.~\ref{fig.spectral_U} to Fig.~\ref{fig.aMF}. \\
In Fig.~\ref{fig.spectral_T} the spectral function is plotted at fixed $U=2$ eV, but at different temperatures, ranging from $20$ K to $300$ K. 
The SOMO and SUMO-related peaks do not change significantly as function of $\theta$ within the stochastic and numerical precision of the results, since $U$ is constant.
The Kondo peak on the other hand progressively shrinks with increasing $\theta$, in analogy to what has been reported in experiments.
We note that the energy scale for the Kondo effect is set by the ratio between $U$ and the electronic coupling of the molecule to the substrate $\Gamma$ (see below). 
Therefore, for a fixed $\Gamma$, increasing $\theta$ at fixed $U$ has the analogous effect on the Kondo feature
as fixing $\theta$ and increasing $U$.\\
The quality of the computed spectral function at $E_F$ 
can be checked by means of  
the generalized Friedel sum rule\cite{Hewson}
\begin{equation}
 A_\mathrm{AI}^\mathrm{MB}(E_F=0)=-\frac{\sin^2(\pi \langle\hat  n\rangle/2)}{\pi \Im\bar\Delta_\mathrm{AI}(0)}.\label{Friedel}
\end{equation}
We remind that in this work we set $E_F$ to $0$ as explained in subsection \ref{sec:negf_setup}.
Although in principle the generalized Friedel sum rule is valid only at $\theta=0$ K,
we find that it is satisfied within $2\%$ whenever the temperature is low enough that the system can be considered deep in the Kondo regime. 
This turns out to be the case for $U\leqslant 1.5$ eV at $\theta=20$ K. Deviations for larger $U$ are mostly due to the finite temperature, since the system gets closer to its Kondo temperature (see below). \\
 We have also assessed the large energy behaviour of the spectral functions
by computing their moments.
The normalization condition defined in Eq.~(\ref{normalization})
fixes the value of the zero moment, and is enforced in the analytic continuation (Sec. \ref{sec:analyticCont}).
The sum rule involving the first moment\cite{Potthoff}
\begin{equation}
 \int_{-\infty}^{\infty}d E~E~A_\mathrm{AI}^\mathrm{MB}(E)=\epsilon_\mathrm{AI,D}- \epsilon_{\mathrm{dc}}+U\langle\hat  n\rangle/2
\end{equation}
is fulfilled within $5\%$ to $10\%$ for every considered value of $U$ and $\theta$
[$\epsilon_{\mathrm{dc}}=U (n- 1/2)$ is the contribution to the on-site energy deriving from the double-counting correction
in Eq.~(\ref{Hdc})]. Much larger errors are found for higher moments. Overall these results and the verification of the generalized Friedel sum rule confirm the observations in the previous sub-section 
about the expected accuracy of the computed spectral functions at different energies. 
While the accuracy is good for energies around $E_\mathrm{F}$, there is a loss of accuracy at high energies, reflected by the error in the moments beyond the first.  \\
In order to quantitatively relate our simulations to the experiment\cite{JLiu} 
we compute the Kondo temperature $\theta_\mathrm{K}$. This is defined as\cite{Hewson1,Hewson2} 
\begin{equation}
 k\theta_\mathrm{K}=-\frac{\pi}{4} Z \Im\bar\Delta_{\mathrm{AI}}(E_F=0),\label{T_k1}
\end{equation}
where $Z$ is the quasiparticle weight, also known as the quasiparticle mass-renormalization factor, 
\begin{equation}
 Z= \bigg[1-\frac{\partial }{\partial E}\Re\bar\Sigma^\mathrm{MB}_\mathrm{AI}(E)\bigg|_{E=0}\bigg]^{-1}.\label{Z_Real}
\end{equation}
Since CTQMC simulations yield the self-energy at the Matsubara frequencies, $Z$ is approximated in the discrete form\cite{Blumer} 
\begin{equation}
Z\approx  \bigg[ 1- \frac{\Im\bar\Sigma^\mathrm{MB}_\mathrm{AI}(i\omega_0)}{\omega_0}\bigg]^{-1}.\label{Z_Mats}
\end{equation}
with $\omega_0=\pi k\theta$. We note that the definition of $\theta_\mathrm{K}$ in Eq.~\ref{T_k1} was originally derived for the symmetric AIM in the Kondo limit, where the occupation is $n=1$, and where
the charge susceptibility is zero and the magnetic susceptibility becomes equal to that of the $sd$ model, also called Kondo model, with $S=1/2$ for $\theta\rightarrow 0$\cite{Hewson}. 
Despite that, in recent works\cite{Surer,Dang} it has been employed to study general AIMs with the parameters obtained from DFT, and it was shown able to provide
reliable estimates of $\theta_\mathrm{K}$, so that its use has become justified in practice.
Moreover, in our case the symmetric AIM is a good approximation to the studied AIM,
since the investigated system is close to half-filling and its hybridization function appears 
rather featureless over a wide energy range around the Fermi energy (Fig.~\ref{fig.Delta}).

In experiments\cite{JLiu} $\theta_\mathrm{K}$ is extracted from the measurement of the full width at half maximum of the Kondo peak, 
$\Gamma_\mathrm{K}(\theta)$, 
as a function of temperature by means of the approximate relation\cite{Nagaoka}
\begin{equation}
\theta_\mathrm{K}=\frac{1}{k\sqrt{2}}\sqrt{\left(\frac{\Gamma_\mathrm{K}(\theta)}{2}\right)^2-\left(\pi k\theta\right)^2}.\label{T_k2}
\end{equation}
Therefore, additionally to Eq.~\ref{T_k1}, we have also considered this equation, with $\theta$ equal to our simulation temperature, 
and $\Gamma_\mathrm{K}$ as the Kondo-peak width of the spectral functions in Fig.~\ref{fig.spectral_U}.\\

\begin{figure}[t!]
\centering\includegraphics[width=0.45\textwidth]{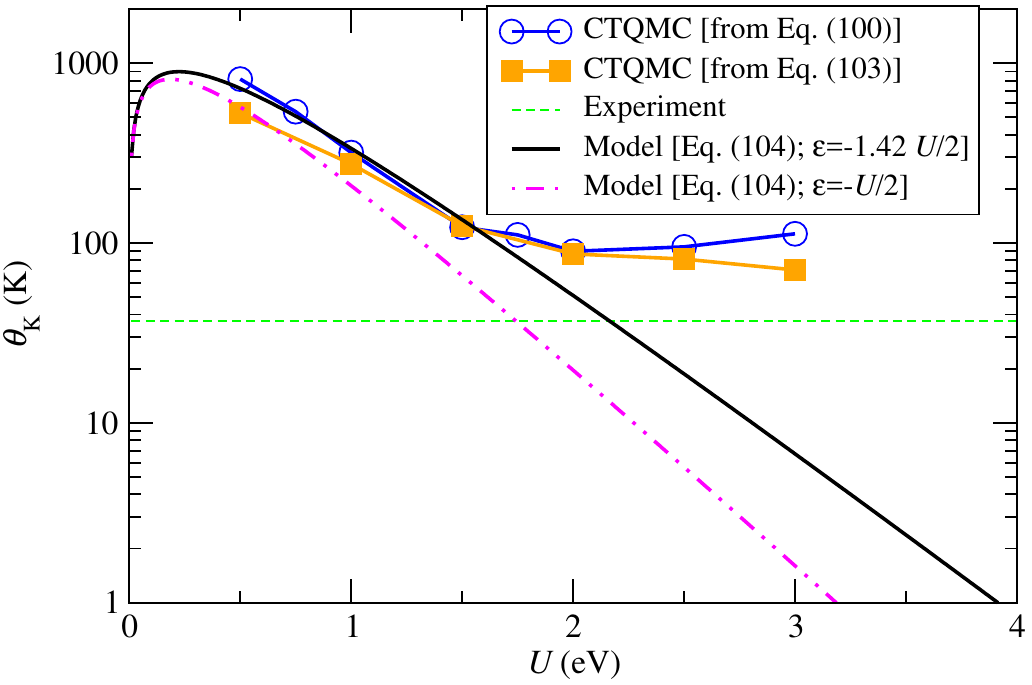}
\caption{(Color online)Computed Kondo temperature as function of $U$ for $\theta=20$ K. 
The CTQMC results obtained with Eqs.~(\ref{T_k1}) (blue circles) and (\ref{T_k2}) (orange squares) 
are compared to the experimental value (horizontal green dashed line). 
The black curve represents the fit of the CTQMC data from Eq. (\ref{T_k1}) in the range of $U\leqslant 1.5$ 
by using the model Eq. (\ref{gamma_u_t_asym}) for the asymmetric SIAM (with $\epsilon=-1.42~U/2$). 
This fit allows the extrapolation of the calculated $\theta_\mathrm{K}$ to larger $U$. 
For comparison, the pink dash-dotted line shows the expected $\theta_\mathrm{K}$ for the symmetric SIAM ($\epsilon=-U/2$).}
\label{fig.Kondo_T}
\end{figure}
The estimated $\theta_\mathrm{K}$ as a function of $U$ for simulations at $\theta=20$ K is plotted in Fig.~\ref{fig.Kondo_T} on a logarithmic scale. 
The results obtained by using the two alternative definitions in Eqs.~(\ref{T_k1}) and~(\ref{T_k2}) are in good agreement, 
showing that both methods are largely equivalent for the evaluation of $\theta_\mathrm{K}$.
Note that the numerical and stochastic precision of the computed $Z$, and consequently of $\theta_\mathrm{K}$ through Eq. \ref{T_k1}, degrade for $U$ larger than 2 eV
as $Z$ becomes small, although its error bar stays of the order of 0.1.
In this case the system is approaching the transition point between the Kondo and the local moment regime, 
where $Z$ is not a well-defined quantity anymore. A precise evaluation of the Kondo temperature via Eq. \ref{T_k1} for $U>2$ would require much lower simulation
 temperatures than $\theta=20$ K, where the system would be deep in the Kondo regime, but these simulations turned out too computationally demanding. 

The results can then be compared to the Kondo temperature for the asymmetric SIAM, given by (see page 168 in Ref.~[\onlinecite{Hewson}])
\begin{equation} 
\theta_\mathrm{K}=\frac{1}{2}\sqrt{\Gamma U} e^{-\frac{\pi | \epsilon | | \epsilon+U |}{U \Gamma}},\label{gamma_u_t_asym}
\end{equation} 
which is valid for $U/\pi\Gamma>1$. Here $\Gamma=-2\Im\bar\Delta_{\mathrm{AI}}(E_\mathrm{F}=0)$, 
and $\epsilon$ is the on-site energy of the impurity. The asymmetric SIAM reduces to the symmetric SIAM for $\epsilon=-U/2$.
Eq.~\ref{gamma_u_t_asym} shows that for large $U$ the value of $\theta_\mathrm{K}$ decreases exponentially with $U$. 
In our calculations $\epsilon$ is used a parameter that is obtained 
by fitting with Eq.~\ref{gamma_u_t_asym} the $\theta_\mathrm{K}$, which is calculated via Eq. (\ref{T_k1}) in Fig. \ref{fig.Kondo_T},
for small $U$ ($\pi\Gamma<U \leqslant 1.5$ eV). 
We then obtain $\epsilon=-1.42~U/2$. This value deviates from the symmetric value $\epsilon=-U/2$ because, as we pointed out above, 
there is a small substrate-to-molecule charge-transfer.
The fitted $\epsilon$ approximately agrees with $\epsilon_\mathrm{AI,D}- \epsilon_{\mathrm{dc}}\approx - \epsilon_{\mathrm{dc}}$ 
($\epsilon_\mathrm{AI,D}\ll \epsilon_{\mathrm{dc}}$ for $\pi\Gamma<U \leqslant 1.5$), and accordingly also to
the position of the SOMO peak with respect to $E_\mathrm{F}$ in the spectral functions in Figs. \ref{fig.spectral_U} and \ref{fig.spectral_T}.
This agreement confirms the applicability of the model, and shows that for $U\leqslant 1.5$ eV our DFT+NEGF+CTQMC calculations provide reliable estimates of the Kondo temperature. 
Then, by using Eq. (\ref{gamma_u_t_asym}) with the fitted $\epsilon$, we can extrapolate the calculated values of $\theta_\mathrm{K}$ 
to larger values of $U$, where the $\theta_\mathrm{K}$ computed directly from the CTQMC data becomes inaccurate.
We note that for the special case of the symmetric SIAM 
with $\epsilon=-U/2$ in Eq. (\ref{gamma_u_t_asym}), we obtain 
a steeper decay of the predicted Kondo temperature as function $U$ than for the fitted asymmetric case (see Fig. \ref{fig.Kondo_T}).

%*****************************************************************
\begin{figure*}[t!]
\center
\includegraphics[width=17.8cm,clip=true]{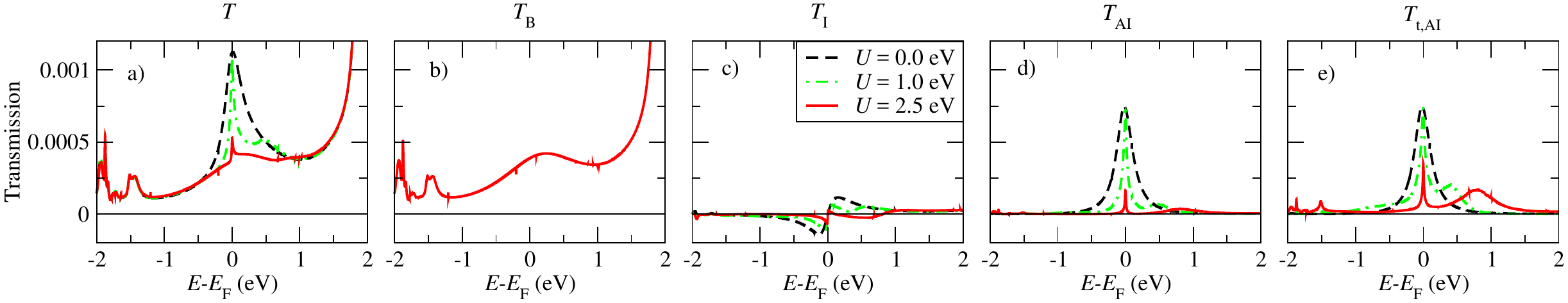}
\caption{(Color online) Transmission including correlations for different values of $U$: 
a) total elastic transmission ($T$), b) background transmission ($T_\mathrm{B}$),
 c) interference terms of the transmission ($T_\mathrm{I}$),
 d) elastic transmission of the Anderson impurity ($T_\mathrm{AI}$), 
and e) total transmission of the Anderson impurity ($T_\mathrm{t,AI}$). }
\label{fig:T_MB}
\end{figure*}
%*****************************************************************
The TOV/Au system has an experimental $\theta_\mathrm{K}$ of $37$ K\cite{JLiu}, and the
estimated value of $U$ that gives such $\theta_\mathrm{K}$ in our calculation for the fitted asymmetric SIAM equation is $U\approx2.2$ eV (Fig. \ref{fig.Kondo_T}). Consistently, for this $U$ the value of
 $Z$ is calculated to be approximately zero, and the squared local magnetic moment $\langle \hat m_z^2\rangle$ approaches the mean-field value (see Fig~\ref{fig.n_m}). 
The estimated $U$ that reproduces the experimental Kondo temperature can then be compared to
the SOMO-SUMO gap obtained for the gas phase molecule (Tab. \ref{SOMO_SUMO}), which is computed to be $4.85$ with G$_0$W$_0$@PBE0.
When the molecule is put on the Au surface at a height of about 3.1 \AA~(Sec. \ref{sec:geo}), metal-induced image-charge corrections reduce the static gap\cite{Perrin, Heimel}.
In order to estimate this static gap reduction, a classical image-charge model can be used, with an image plane placed at 1 \AA~above the Au surface \cite{Quek2,Souza_cdft,Amaury1,Amaury2}. 
With this model the reduction of the gap results to be about 3.4 eV, which then leads to an approximate static $U$ for the molecule on the surface that is $1.5$ eV. 
This static $U$ corresponds to the lower limit for $U$, since it implies full screening of any charge fluctuation on the molecule by the Au surface. 
The value of $U\approx2.2$ eV that best matches experiment lies somewhat above this lower limit, but well below the upper limit of the gas phase molecule values. 
We explain this result by observing that, since the fluctuations responsible for the Kondo physics happen on a rather fast time-scale, 
they will generally not be fully screened. 
While much effort has been dedicated to compute fully-screened $U$ values in both molecular and solid state systems 
with several alternative methods, such as constrained Random Phase Approximation (cRPA)\cite{Springer,Aryasetiawan,Aryasetiawan2}, 
constrained DFT (cDFT)\cite{Gunnarsson,McMahan,Hybertsen,Imada,Sau,Amaury1,andrea_cdft}, screened hybrid functionals\cite{Sharifzadeh,Egger} and (self-consistent) linear response\cite{Cococcioni,Kulik},
the intermediate situation that we suggest here has not been analyzed yet. 
This represents therefore an important aspect for future theoretical studies about Kondo physics.

We point out that an accurate estimate of the Kondo temperature is a great challenge. 
As seen in Eq.~(\ref{gamma_u_t_asym}), the Kondo temperature is an exponential function of the ratio between $U$ and $\Gamma$.
This latter quantity is determined by the molecule adsorption distance and position, which means that the quantitative modeling of Kondo systems critically 
depends on the accuracy of the DFT structural relaxation for molecules on surfaces.
 This in turn requires a correct description of both covalent and vdW interactions, which is an open direction of research in itself\cite{acc_chem},
despite notable recent advancements, such as the introduction of the PBE+vdW$^{surf}$ functional\cite{Tkatchenko,VdW_surf,VdW_surf2} that we use here.

\subsection{Transport properties from DFT+NEGF+CTQMC}
\label{sec:tctqmc}

%*****************************************************************
\begin{figure*}
\center
\includegraphics[width=17.8cm,clip=true]{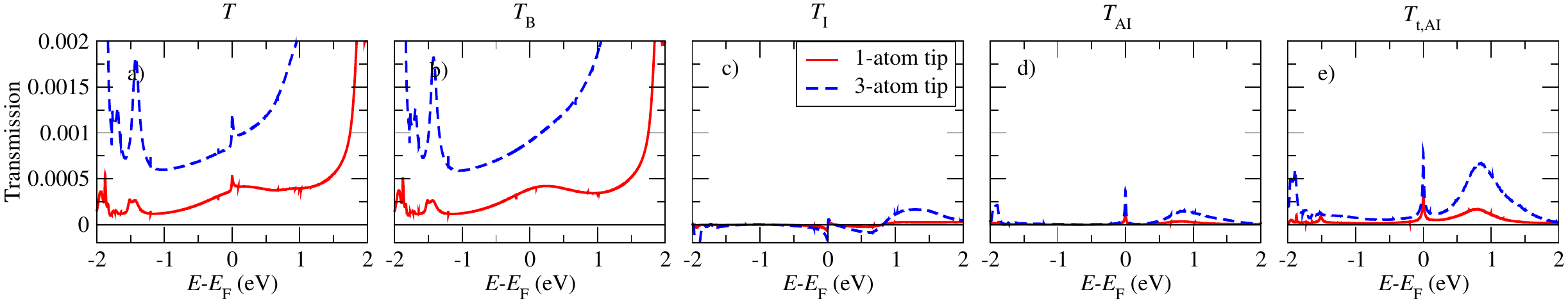}
\caption{(Color online) Transmission including correlations for $U=2.5$ eV for a tip with a single atom at the top (1-atom tip), and a tip with three atoms at the top (3-atom tip). 
The different sub-figures correspond to the ones described in Fig. \ref{fig:T_MB}}
\label{fig:T_MB_BT}
\end{figure*}
%*****************************************************************
 
The total elastic transmission in presence of many-body correlations, $T$, 
of the system is calculated with Eq. (\ref{eq:tre}). Importantly, we
have verified that the same result is obtained by using the original or the projected matrices, so that 
the evaluation of the inverse projection of the many-body self-energy [Eq. (\ref{eq:sigmambinv})] 
is in principle not required for the calculation of the transmission. The results for the setup outlined in Sec. \ref{sec:tdft}, 
and for three different values of $U$, are given in Fig. \ref{fig:T_MB}(a). 
The non-spin-polarized transmission without correlations, $T_\mathrm{0}$, corresponds to $T$ for $U=0$, 
and has a broad Lorentzian peak around $E_\mathrm{F}$. When increasing $U$ this peak gets narrower and becomes the Kondo peak, 
with two broad satellite peaks separated by about $U$, following the evolution of the DOS in Fig. \ref{fig.spectral_U}. 
The peak below $E_\mathrm{F}$ is less pronounced than the one above $E_\mathrm{F}$, which indicates an energy dependent 
coupling to the electrodes, and is in qualitative agreement with the results obtained 
using a scissor operator (Fig. \ref{fig:T_DFT}). The breakdown of $T$ into its components, 
as derived in Sec. \ref{sec:teq}, is shown in Figs. \ref{fig:T_MB}(b-d). 
By construction the background transmission in Fig. \ref{fig:T_MB}(b) [Eq. (\ref{eq:tbath})] is independent of $U$, 
and reflects the DOS and coupling of the tip to the substrate mediated by non-interacting states of the molecule. 
For the considered values of $U$ the transmission of the AI itself, defined in Eq. (\ref{eq:tAI2}) and 
shown in Fig. \ref{fig:T_MB}(d), decays to zero for energies $2$ eV below and above $E_\mathrm{F}$, 
where the AI DOS vanishes (Fig. \ref{fig.spectral_U}). 
The interference terms are shown in Fig. \ref{fig:T_MB}(c) [Eq. (\ref{eq:tinterference})], 
and are rather small, so that in this case $T$ 
largely corresponds to the sum of $T_\mathrm{B}$ and $T_\mathrm{AI}$. In general however  $T_\mathrm{I}$ 
can be large and lead to a change of shape of the Kondo peak from a symmetric Lorentzian-type to an asymmetric Fano-type, 
and our method allows to evaluate the contributions individually. 
For the contribution of the AI we evaluate also the total transmission including incoherent components, $T_\mathrm{t,AI}$, 
and the results are shown in Fig. \ref{fig:T_MB}(d).
 Qualitatively $T_\mathrm{t,AI}$ and $T_\mathrm{AI}$ are similar, 
but $T_\mathrm{AI}$ becomes significantly smaller than $T_\mathrm{t,AI}$ as $U$ increases, since it neglects incoherent components of the electron transport. 
Note that at zero temperature the many-body self-energy at $E_\mathrm{F}$ vanishes in the Kondo regime, so that one should have $T(E_\mathrm{F})=T_\mathrm{0}(E_\mathrm{F})$ and $T_\mathrm{t,AI}(E_\mathrm{F})=T_\mathrm{AI}(E_\mathrm{F})$.
However, at finite temperature the many-body self-energy is small, but not exactly 0, 
which results in the Kondo transmission peak shrinking when compared to the non-interacting case and 
to $T_\mathrm{t,AI}(E_\mathrm{F})\ne T_\mathrm{AI}(E_\mathrm{F})$ for $\theta>0$.

We can now evaluate how the transmission depends on the tip shape. Since the tip-molecule coupling is rather weak at our considered tip height, 
we can assume that the tip does not significantly influence the electron correlation effects in the molecule, 
so that we use the same many-body self-energy for the different tip shapes. 
To evaluate the transmission for a blunt tip, we remove the single-tip Au atom, so that we have a tip with 3 Au atoms in the bottom-most plane closest to the molecule. 
We then shift the height of these atoms from the molecule to be equal to the one of the single-atom tip considered in Fig. \ref{fig:T_MB}. 
The comparison of the transmission for the two tip shapes, and for $U=2.5$ eV, is shown in Fig. \ref{fig:T_MB_BT}. 
The main difference is that the 3-atom tip has a stronger coupling to the molecule, which increases all contributions to the transmission. 
The background transmission becomes less featured, since the increased DOS around $E_\mathrm{F}$ for the 1-atom tip is generally broadened for a 3-atom tip \cite{Quek2}. 
With the used value of $U=2.5$ eV, which is close to the value obtained by comparing the Kondo temperature to experiment, 
our transport results are in good agreement with experimental conductance measurements in Ref. \cite{JLiu}, 
which show a Kondo resonance at small bias on top of a background conductance, which is approximately constant in the considered low bias range. 
In general the relative height of background transmission to Kondo peak depends on tip shape, size and position, 
and can therefore be one of the tools to infer a nanoscale junction geometry.

\section{Conclusions}
\label{sec:conclusions}
We have presented a rigorous way to extract an effective AIM from DFT+NEGF calculations for molecules adsorbed on a substrate.
This effective AIM is then solved at finite-temperature with our implementation of CTQMC. 
With the developed method we can describe the linear-response conductance of systems in the Kondo regime, 
and evaluate the separate contributions to the overall conductance originating from the transmission through the molecule itself, 
from the background transmission, and from interference terms. 
Furthermore, for the SIAM we can compare the elastic contributions to the total conductance including incoherent effects. 
The results for the specific case of the TOV radical adsorbed on Au demonstrate 
that the method can potentially return accurate values for the Kondo temperature, 
provided that DFT gives reliable
adsorption geometries, and that one can estimate the value of $U$ for the effective electron-electron interaction. 
We propose that a partially screened $U$ has to be considered to describe the Kondo physics for such molecules on metallic surfaces. \\
Our results show that the developed method is a promising tool to evaluate the linear-response 
conductive properties of strongly correlated nanoscale systems. We note that the extension beyond linear response to finite-bias problems is in principle straightforward, 
and the current can be calculated as outlined in Sec. \ref{sec:teq}. The main developments required for such finite bias calculations are accurate and computationally efficient solvers for the out-of-equilibrium AIM in the steady-state.
Such developments are still an open research direction, and important progresses have been recently made\cite{Dirks,Kurth2,Wingreen,Aliga,Fujii}.

\section{Acknowledgments}
The authors are grateful to L. Chioncel, A.R. Rocha and S. Kurth for useful discussions. 
A.D thanks I. De Marco for some interactions about the problem of the analytic continuation.
The work was sponsored by the European Union through 
the project 618082 ACMOL. AD received additional support from the ``Ministerio de Economia y Competitividad'' (Mineco) of Spain.

\appendix
\section{Projection Scheme}
\label{sec:appendixA}

\subsection{Forward projection}
\label{sec:appendixAfp}

As first step in the projection scheme we introduce a new basis set $\{\tilde{\phi}_\mu\}$, which is identical to the original set of 
$\left\{\phi_\mu\right\}$, with the exception that the $N_\mathrm{AI}$ basis orbitals, which have largest overlap with the set
$\left\{\psi_{1,\mathrm{ER}}, \psi_{2,\mathrm{ER}}, \psi_{3,\mathrm{ER}}, \dots , \psi_{N_\mathrm{AI},\mathrm{ER}}\right\}$
are replaced with the corresponding $\psi_{i,\mathrm{ER}}$ of the AI. Note that by construction the replaced basis orbitals are always part of the IR, 
since $\psi_{i,\mathrm{ER}}$ has non-zero elements only within IR basis orbitals. This transformation is achieved with the projection matrix
\begin{equation}
W_0=\left( \begin{array}{cccc}
0                        & 1_{N_\alpha} & 0    & 0   \\
U_\mathrm{ER}    & 0  & U_\mathrm{NI}  & 0  \\
0                        & 0  & 0              & 1_{N_\beta}
\end{array} \right).
\end{equation}
The matrix $U_\mathrm{NI}$ has the dimension $N_\mathrm{ER}\times N_\mathrm{NI}$, and is an identity matrix with the $N_\mathrm{AI}$ columns removed, 
which correspond to the ones of the replaced basis functions. 
We collect all the unity vectors removed when forming $U_\mathrm{NI}$
 into a $N_\mathrm{ER}\times N_\mathrm{AI}$ matrix $U_\perp$, so that ${U_\mathrm{NI}}^\dagger U_\mathrm{NI}=1_{N_\mathrm{NI}}$ and ${U_\perp}^\dagger U_\mathrm{NI}=0$. 

The overlap and Hamiltonian matrices in the new basis, $\tilde{S}$ and $\tilde{H}$, are then obtained as
\begin{equation}
\tilde{S}=W_0^\dagger S W_0,\ \ \ \ \tilde{H}=W_0^\dagger H_\mathrm{EM} W_0.
\end{equation}
The general structure of the resulting $\tilde{S}$ is
\begin{equation}
\tilde{S}=\left( \begin{array}{cccc}
\tilde{S}_\mathrm{AI}      & 0 & \tilde{S}_\mathrm{AI,NI}                          & 0            \\
0 & S_{\mathrm{\alpha\alpha}} & \tilde{S}_{\mathrm{\alpha,\mathrm{NI}}}                    &  S_\mathrm{\alpha \beta}               \\
\tilde{S}_\mathrm{AI,NI}^\dagger & \tilde{S}_{\mathrm{\alpha,\mathrm{NI}}}^\dagger & \tilde{S}_{\mathrm{NI}}   &  \tilde{S}_{\mathrm{\beta,\mathrm{NI}}}^\dagger                \\
0   & S_\mathrm{\alpha \beta}^\dagger                          & \tilde{S}_{\mathrm{\beta,\mathrm{NI}}}     & S_\mathrm{\beta \beta}
\end{array} \right),
\end{equation}
where $\tilde{S}_\mathrm{AI}$ is a $N_\mathrm{AI}\times N_\mathrm{AI}$ matrix, $\tilde{S}_\mathrm{NI}$ is an $N_\mathrm{NI}\times N_\mathrm{NI}$ matrix, 
and the dimensions of the off-diagonal blocks are determined from the ones of the diagonal blocks. Importantly, the sub-blocks for the overlap 
between the AI and the $\alpha$ and $\beta$ subsystem are 0, since the ER is chosen to be large enough to guarantee this by construction.

Here we have introduced
\begin{eqnarray}
\tilde{S}_\mathrm{AI}&=& {U_\mathrm{ER}}^\dagger S_\mathrm{ER} {U_\mathrm{ER}},\label{eq:saitilde} \\
\tilde{S}_\mathrm{AI,NI}&=& {U_\mathrm{ER}}^\dagger S_{\mathrm{ER}} U_{\mathrm{NI}}, \\
\tilde{S}_{\mathrm{NI}}&=& U_\mathrm{NI}^\dagger S_{\mathrm{ER}} U_{\mathrm{NI}}, \\
\tilde{S}_{\mathrm{\alpha,\mathrm{NI}}} &=& S_\mathrm{\alpha,ER}U_{\mathrm{NI}}, \\
\tilde{S}_{\mathrm{\beta,\mathrm{NI}}} &=& S_\mathrm{\beta,ER}U_{\mathrm{NI}}.
\end{eqnarray}
The transformed $\tilde{H}$ has an analogous structure.

The basis functions of the AI are generally not orthogonal to the other ER orbitals, so that $\tilde{S}_\mathrm{AI,NI}\ne 0$. 
If the Anderson impurity solver can deal with such non-orthogonal systems, then one can pass this matrix directly to such a solver. 
Most AI solvers however require orthogonality between the AI and all other orbitals in the system. We therefore perform a second transformation, 
which orthogonalizes all the orbitals to the ones of the AI. Importantly, we require such transformation to keep the orbitals of the AI, 
and of the $\alpha$ and $\beta$ regions, unchanged. One can achieve this with the following transformation matrix:
\begin{equation}
W_1=\left( \begin{array}{cccc}
1_{N_\mathrm{AI}}        & 0                       & -W_\mathrm{SB}   & 0\\
0                        & 1_{N_\alpha}            &          0 & 0  \\
0                        & 0                  & 1_{N_\mathrm{NI}}   & 0  \\
0                        & 0     & 0 & 1_{N_\beta}
\end{array} \right),
\end{equation}
with 
\begin{eqnarray}
W_\mathrm{SB} &=& \tilde{S}_\mathrm{AI}^{-1} \tilde{S}_{\mathrm{AI,NI}}.
\end{eqnarray}
This leads to the transformed Hamiltonian, $\dtilde{H}=W_1^\dagger \tilde{H} W_1$, and overlap, $\dtilde{S}=W_1^\dagger \tilde{S} W_1$, matrices.  
$\dtilde{S}$ has the new structure
\begin{equation}
\dtilde{S}=\left( \begin{array}{cccc}
\tilde{S}_\mathrm{AI}      & 0 & 0                  & 0            \\
0 & S_{\mathrm{\alpha\alpha}} & \tilde{S}_{\mathrm{\alpha,\mathrm{NI}}}                    &  S_\mathrm{\alpha \beta}               \\
0 & \tilde{S}_{\mathrm{\alpha,\mathrm{NI}}}^\dagger & \dtilde{S}_{\mathrm{NI}}                           &  \tilde{S}_{\mathrm{\beta,\mathrm{NI}}}^\dagger                \\
0   & S_\mathrm{\alpha \beta}^\dagger                          & \tilde{S}_{\mathrm{\beta,\mathrm{NI}}}     & S_\mathrm{\beta \beta}
\end{array} \right),
\end{equation}
with
\begin{equation}
\dtilde{S}_{\mathrm{NI}}=\tilde{S}_{\mathrm{NI}}- \tilde{S}_{\mathrm{AI,NI}}^\dagger \tilde{S}_\mathrm{AI}^{-1} \tilde{S}_{\mathrm{AI,NI}}.
\end{equation}
We therefore now have the AI orthogonal to all bath basis orbitals.

As a further optional step we can diagonalize $\tilde{S}_\mathrm{AI}$  and $\tilde{H}_\mathrm{AI}$.
To this aim we first orthogonalize the orbitals in the AI via a L\"owdin transformation 
by calculating the square root of the inverse of the AI overlap matrix, $\tilde{S}_\mathrm{AI}^{-\frac{1}{2}}$,
and then obtain the eigenvalues and the matrix of all eigenvectors, $U_{\mathrm{AI}\psi}$, of the resulting eigenvalue system 
$\left(\tilde{S}_\mathrm{AI}^{-\frac{1}{2}}\tilde{H}_\mathrm{AI}\tilde{S}_\mathrm{AI}^{-\frac{1}{2}}\right)U_{\mathrm{AI}\psi}=U_{\mathrm{AI}\psi}\epsilon_{\mathrm{AI},D}$. 
Here $U_{\mathrm{AI}\psi}$ is a $N_\mathrm{AI}\times N_\mathrm{AI}$ matrix, with each column equal to an eigenvector of the eigenvalues system, and $\epsilon_{AI,D}$ 
is a $N_\mathrm{AI}\times N_\mathrm{AI}$ diagonal matrix, 
where the diagonal elements are the corresponding eigenvalues.
The resulting transformation matrix to the new basis acts only on the orbitals of the AI, and is 
\begin{equation}
W_2=\left( \begin{array}{cccc}
W_{2,\mathrm{AI}}        & 0   & 0 & 0\\
0                        & 1_{N_\alpha} & 0 & 0  \\
0                        &  0  & 1_{N_\mathrm{NI}}   & 0  \\
0                        &  0  & 0      & 1_{N_\beta}
\end{array} \right),
\end{equation}
with
\begin{equation}
W_{2,\mathrm{AI}}=\tilde{S}_\mathrm{AI}^{-\frac{1}{2}}U_{\mathrm{AI}\psi}.
\label{eq:w2ai}
\end{equation}
The transformed matrices are $\bar{H}=W_2^\dagger \dtilde{H} W_2$, and $\bar{S}=W_2^\dagger \dtilde{S} W_2$, 
and keep the non-zero structure of $\dtilde{H}$ and $\dtilde{S}$, except that the top-left $N_\mathrm{AI}\times N_\mathrm{AI}$ elements become diagonal matrices.

The total transformation matrix, $W$, then is
\begin{equation}
W=W_0 W_1 W_2,
\label{eq:w123}
\end{equation}
which when multiplied out gives the result in Eq. (\ref{eq:wshort}), repeated in the following for convenience
\begin{equation}
W=\left( \begin{array}{cccc}
0                        & 1_{N_\alpha}  & 0 & 0  \\
W_\mathrm{AI}          & 0             & W_\mathrm{NI}  & 0  \\
0                        &               0                  & 0 & 1_{N_\beta}
\end{array} \right).
\end{equation}
Here we have introduced
\begin{eqnarray}
W_\mathrm{AI}&=& {U_\mathrm{ER}}   W_{2,\mathrm{AI}}, \label{eq:waiapp} \\
W_\mathrm{NI}&=& \left(1_{N_\mathrm{NI}} -W_{\mathrm{ER},\psi} S_{ER}\right) U_\mathrm{NI},\label{eq:wni}
\end{eqnarray}
with
\begin{eqnarray}
W_{\mathrm{ER},\psi}&=&{U_\mathrm{ER}}\left({U_\mathrm{ER}}^\dagger S_\mathrm{ER} {U_\mathrm{ER}}\right)^{-1}{U_\mathrm{ER}}^\dagger.
\end{eqnarray}

\subsection{Inverse projection}
\label{sec:appendixAip}

In this section we evaluate the inverse of $W$ [Eq. (\ref{eq:wshort})]. 
By using Eq. (\ref{eq:w123}) we obtain its general form as
\begin{equation}
W^{-1}=W_2^{-1} W_1^{-1} W_0^{-1}.
\end{equation}
The inverse of $W_0$ is
\begin{equation}
W_0^{-1}=\left( \begin{array}{ccc}
0               & A    & 0   \\
1_{N_\alpha}    & 0  & 0   \\
0               & B      & 0 \\
0               & 0      & 1_{N_\beta} 
\end{array} \right),
\end{equation}
where $A$ is a $N_\mathrm{AI}\times N_\mathrm{ER}$ matrix, and $B$ is a $N_\mathrm{NI}\times N_\mathrm{ER}$ matrix. They are obtained by evaluating
\begin{eqnarray}
\left( \begin{array}{c}
A \\
B 
\end{array} \right)=
\left( \begin{array}{cc}
{U_\mathrm{ER}} & U_\mathrm{NI}
\end{array} \right)^{-1}.
\end{eqnarray}
If we recollect that all the unity vectors removed when forming $U_\mathrm{NI}$ are combined into a matrix $U_\perp$ ($U_\perp^\dagger U_\mathrm{NI}=0$), 
then we can give analytic expressions for $A$ and $B$ as
\begin{eqnarray}
A&=&\left( U_\perp^\dagger {U_\mathrm{ER}} \right)^{-1} U_\perp^\dagger,\nonumber\\
B&=&U_\mathrm{NI}^\dagger\left(1_{N_\mathrm{IR}} - {U_\mathrm{ER}} \left(U_\perp^\dagger {U_\mathrm{ER}} \right)^{-1} U_\perp^\dagger\right).
\label{eq:abu}
\end{eqnarray}
Note that since for large systems the matrices $U_\mathrm{NI}$ and $U_\perp$ consist mainly of zeroes, 
and the matrix $\left( U_\perp^\dagger {U_\mathrm{ER}} \right)$ is only of dimensions $N_\mathrm{AI} \times N_\mathrm{AI}$, 
the calculation of the non-zero elements of $W_0^{-1}$ by using Eq. (\ref{eq:abu}) is a fast Order($N$) operation.

The inverse of $W_1$ is
\begin{equation}
W_1^{-1}=\left( \begin{array}{cccc}
1_{N_\mathrm{AI}}        & 0                       &  W_\mathrm{SB}   & 0\\
0                        & 1_{N_\alpha}            &          0 & 0  \\
0                        & 0                  & 1_{N_\mathrm{NI}}   & 0  \\
0                        & 0     & 0 & 1_{N_\beta}
\end{array} \right),
\end{equation}
and the inverse of $W_2$ is
\begin{equation}
W_2^{-1}=\left( \begin{array}{cccc}
W_{2,\mathrm{AI}}^{-1}        & 0   & 0 & 0\\
0                        & 1_{N_\alpha} & 0 & 0  \\
0                        &  0  & 1_{N_\mathrm{NI}}   & 0  \\
0                        &  0  & 0      & 1_{N_\beta}
\end{array} \right),
\end{equation}
Multiplying all components gives the result in Eq. (\ref{eq:winvsimple}), with
\begin{equation}
W_\mathrm{iNI}=B=U_\mathrm{NI}^\dagger\left(1_{N_\mathrm{IR}} - {U_\mathrm{ER}} \left(U_\perp^\dagger {U_\mathrm{ER}} \right)^{-1} U_\perp^\dagger\right).
\label{eq:winvni}
\end{equation}

\subsection{Alternative expression for the hybridization function}
\label{sec:appendixhyb}
We write the dense $N\times N$ matrix $G$ in the block matrix form analogous to the one of the overlap matrix [Eq. (\ref{eq:ssmall})] as
\begin{equation}
G(z)=\left( \begin{array}{ccc}
G_{\mathrm{\alpha\alpha}}(z)         & G_{\mathrm{\alpha, ER}}(z)  & G_{\alpha\beta} (z)          \\
G_{\mathrm{ER, \alpha}}(z)  & G_{\mathrm{ER}} (z)         & G_{\mathrm{ER, \beta}} (z)   \\
G_{\beta\alpha}  (z)                & G_{\mathrm{\beta, ER}} (z)  & G_\mathrm{\beta \beta}(z) 
\label{eq:gsmall}
\end{array} \right).
\end{equation}
By using Eqs. (\ref{eq:gtran}) and (\ref{eq:winvsimple}) we obtain the following expression for the Green's function projected on the AI:
\begin{eqnarray}
\bar{G}_\mathrm{AI}(z) =W_\mathrm{iAI}G_\mathrm{ER}(z)  W_\mathrm{iAI}^\dagger.
\label{eq:gbar1}
\end{eqnarray}
which is an equivalent expression to the one given in Eq. (\ref{eq:gbar2}). If the task is to calculate $\bar\Delta_\mathrm{AI}(z) $, 
one can therefore also first use Eq. (\ref{eq:gbar1}) to obtain $\bar G_\mathrm{AI}(z) $, and then Eq. (\ref{eq:gbar2}) to evaluate $\bar\Delta_\mathrm{AI}(z) $ as
\begin{eqnarray}
\bar{\Delta}_\mathrm{AI}(z) &=&z-\epsilon_\mathrm{AI,D}-\bar G_\mathrm{AI}^{-1}(z) .
\label{eq:bardelta1}
\end{eqnarray}
However, especially for large $z$, this indirect way of obtaining $\bar{\Delta}_\mathrm{AI}(z) $ is significantly less accurate 
than the direct way through Eq. (\ref{eq:bardelta2}). 
The reason is that for large $z$ the quantities $z$ and $\bar G_\mathrm{AI}^{-1}(z)$ in Eq. (\ref{eq:bardelta1}) are both very large,
 while the resulting difference is small, with a leading term proportional to $1/z$ (see Sec. \ref{sec:appendixlimitdelta}).
 In a practical calculation we therefore always use Eq. (\ref{eq:bardelta2}) to calculate $\bar{\Delta}_\mathrm{AI}(z)$.

\subsection{Asymptotic limits of the hybridization function for large energies}
\label{sec:appendixlimitdelta}

By means of Eq. (\ref{eq:bardelta2}) we can evaluate the limit of $\bar{\Delta}_\mathrm{AI}(z)$ as $z$ goes to infinity. 
In Eq. (\ref{eq:bardelta2}) $\bar{\Delta}_\mathrm{AI}(z)$ is given by a multiplication between energy independent matrices, 
$\bar{H}_\mathrm{AI,NE}$ and $\bar{H}_\mathrm{AI,NE}^\dagger$, with the energy dependent matrix $\bar{g}_\mathrm{NI}(z)$. 
We therefore need to evaluate the asymptotic limit of $\bar{g}_\mathrm{NI}(z)$, defined in Eq. (\ref{eq:barGB}), for large $z$. 

To this aim we first evaluate the limit for large $z$ of $\Sigma_\mathrm{L}(z)$ and $\Sigma_\mathrm{R}(z)$. 
The left self-energy satisfies the following recursive relation\cite{SelfEnergies}
\begin{eqnarray}
\Sigma_\mathrm{L}(z)=\left(z S_{\mathrm{L},1}^\dagger-H_{\mathrm{L},1}^\dagger\right)
&&\left(z S_{\mathrm{L},0}-H_{\mathrm{L},0}-\Sigma_\mathrm{L}(z)\right)^{-1}\nonumber\\
&&\ \ \ \ \ \ \ \ \ \left(z S_{\mathrm{L},1}-H_{\mathrm{L},1}\right),
\label{eq:senerecursive}
\end{eqnarray}
where $S_{\mathrm{L},0}$ ($H_{\mathrm{L},0}$) is the left lead's onsite overlap (Hamiltonian) matrix, and 
$S_{\mathrm{L},1}$ ($H_{\mathrm{L},1}$) is the left lead's coupling overlap (Hamiltonian) matrix between nearest neighbor cells. 
If we expand this relation for large $z$, we can see that there is a dominant term proportional to $z$, 
a constant term, plus terms proportional to $1/z$ and smaller. For large $z$ we can therefore write
\begin{equation}
\Sigma_\mathrm{L}=z\ \Sigma_{\mathrm{L};S} + \Sigma_{\mathrm{L};0} + O(\frac{1}{z}),
\end{equation}
with $\Sigma_{\mathrm{L};S}$ and $\Sigma_{\mathrm{L};0}$ constant matrices to be evaluated. 
If we insert this relation in Eq. (\ref{eq:senerecursive}) and evaluate it for large energies, we obtain the recursive relation
\begin{eqnarray}
\Sigma_{\mathrm{L;S}}&=&S_{\mathrm{L},1}^\dagger
\left(S_{\mathrm{L},0}-\Sigma_\mathrm{L;S}\right)^{-1}
S_{\mathrm{L},1},
\end{eqnarray}
which has the analogous form of Eq. (\ref{eq:senerecursive}) when the Hamiltonian is set to 0, multiplied by $1/z$. $\Sigma_{\mathrm{L;S}}$ can therefore
be evaluated by using the standard method of Ref. \onlinecite{SelfEnergies} with the Hamiltonian set to 0, and can be interpreted as the ``self-energy'' matrix 
in the calculation of the inverse of the bare left semi-infinite overlap matrix. 

When we perform the perturbation expansion of Eq. (\ref{eq:senerecursive})to the next lower order we obtain the following recursive relation for $\Sigma_{\mathrm{L};0}$:
\begin{eqnarray}
\Sigma_{\mathrm{L;0}}&=&
S_{\mathrm{L},1}^\dagger \left(S_{\mathrm{L},0}-\Sigma_\mathrm{L;S}\right)^{-1}   \left(H_{\mathrm{L},0}+\Sigma_{\mathrm{L;0}}\right) \nonumber\\
&&\left(S_{\mathrm{L},0}-\Sigma_\mathrm{L;S}\right)^{-1} S_{\mathrm{L},1}\nonumber\\
&&-H_{\mathrm{L},1}^\dagger \left(S_{\mathrm{L},0}-\Sigma_\mathrm{L;S}\right)^{-1} S_{\mathrm{L},1}\nonumber\\
&&-S_{\mathrm{L},1}^\dagger \left(S_{\mathrm{L},0}-\Sigma_\mathrm{L;S}\right)^{-1} H_{\mathrm{L},1}.
\end{eqnarray}
Since the coupling matrices $H_{\mathrm{L},1}$ and $S_{\mathrm{L},1}$ are usually small compared to the onsite terms, 
we can keep the result only to lowest order in these matrices, so that we can write
\begin{eqnarray}
\Sigma_{\mathrm{L;0}}&\approx&
S_{\mathrm{L},1}^\dagger \left(S_{\mathrm{L},0}-\Sigma_\mathrm{L;S}\right)^{-1}   \left(H_{\mathrm{L},0}\right)\nonumber\\
&&\left(S_{\mathrm{L},0}-\Sigma_\mathrm{L;S}\right)^{-1} S_{\mathrm{L},1}\nonumber\\
&&-H_{\mathrm{L},1}^\dagger \left(S_{\mathrm{L},0}-\Sigma_\mathrm{L;S}\right)^{-1} S_{\mathrm{L},1}\nonumber\\
&&-S_{\mathrm{L},1}^\dagger \left(S_{\mathrm{L},0}-\Sigma_\mathrm{L;S}\right)^{-1} H_{\mathrm{L},1}.
\end{eqnarray}
However this remaining term is still of second order in the coupling matrices, so that it is usually small, 
and we can therefore neglect it altogether and approximate it as $\Sigma_{\mathrm{L;0}}\approx0$. We can perform an analogous expansion 
for the right self-energy, for which we can then write $\Sigma_\mathrm{R}\approx z\ \Sigma_{\mathrm{R};S}$, 
with $\Sigma_{\mathrm{R;S}}=S_{\mathrm{R},1} \left(S_{\mathrm{R},0}-\Sigma_\mathrm{R;S}\right)^{-1} S_{\mathrm{R},1}^\dagger$.

With this expansion for the leads' self-energies, we can now expand $\bar{g}_\mathrm{NI}$ in Eq. (\ref{eq:barGB}) to second order in $1/z$ as
\begin{eqnarray}
\bar{g}_\mathrm{NI}(z)&\approx&\left[z \bar{S}_\mathrm{B}-\bar{H}_\mathrm{B}-z \Sigma_{\mathrm{L};S} -z \Sigma_{\mathrm{R};S}\right]^{-1}\nonumber\\
&\approx&\frac{1}{z} M_1 + \frac{1}{z^2} M_2+ O\left(\frac{1}{z^3}\right),
\label{eq:limitsG}
\end{eqnarray}
with
\begin{eqnarray}
M_1&=&\left(\bar{S}_\mathrm{B}- \Sigma_{\mathrm{L};S} -\Sigma_{\mathrm{R};S}\right)^{-1}\label{eq:m1},\\
M_2&=&M_1
\bar{H}_\mathrm{B}
M_1.
\label{eq:m2}
\end{eqnarray}
The asymptotic limit of the Hybridization function for large $z$ can then be evaluated by inserting Eq. (\ref{eq:limitsG}) into Eq. (\ref{eq:bardelta2}).

If the Hamiltonian and overlap matrices are real, then also $M_1$ and $M_2$ are real. 
In this case, if the energy $z$ is purely imaginary, $z=i\omega_n$ (with $\omega_n$ a Matsubara frequency or any other real number), 
then the leading term of the imaginary part of $\bar{\Delta}_\mathrm{AI}(z)$ 
is proportional to $M_1$ and decays as $-1/\omega_n$, while the real part is proportional to $M_2$ and decays as $-1/\omega_n^2$. 
On the other hand, in the case of a very large real energy with a small imaginary part $\delta$ added, $z=E + i \delta$, 
the real part of the leading term of $\bar{\Delta}_\mathrm{AI}(z)$ is proportional to $M_1$ and decays as $1/E$, 
while the leading imaginary part is also proportional to $M_1$ and decays as $-\delta/E^2$. 

\subsection{Electronic current}
\label{sec:appendixcurrent}

The current flowing from the left electrode into the EM, $I_\mathrm{L}$, is given by\cite{Meir}
\begin{eqnarray}
I_\mathrm{L}&=&\frac{ie}{h}\int dE \;\mathrm{Tr}\left[f_\mathrm{L}(E)\bar{\Gamma}_\mathrm{L}(E) \left(\bar{G}^\mathrm{MB}(E)-\bar{G}^{\mathrm{MB}\supi{\dagger}}(E)\right)+\right.\nonumber\\
&&\left.\bar{\Gamma}_\mathrm{L}(E)\bar{G}^{\mathrm{MB},<}(E)\right].
\end{eqnarray}
The integral runs over real energies, $E$, so that $\bar{G}^\mathrm{MB}(E)$ and $\bar{G}^{\mathrm{MB}\supi{\dagger}}(E)$ are the retarded and advanced Green's functions, respectively, and $\bar{G}^{\mathrm{MB},<}(E)$ is the lesser Green's function\cite{Meir}. $\bar{\Gamma}_\mathrm{L}(E)$ is defined in Eq. (\ref{eq:gammal}).

In an analogous way the current flowing from the EM into the right electrode, $I_\mathrm{R}$, is
\begin{eqnarray}
I_\mathrm{R}&=&-\frac{ie}{h}\int dE \;\mathrm{Tr}\left[f_\mathrm{R}(E)\bar{\Gamma}_\mathrm{R}(E) \left(\bar{G}^\mathrm{MB}(E)-\bar{G}^{\mathrm{MB}\supi{\dagger}}(E)\right)+\right.\nonumber\\
&&\left.\bar{\Gamma}_\mathrm{R}(E)\bar{G}^{\mathrm{MB},<}(E)\right],
\label{eq:ir1}
\end{eqnarray}
where $f_R$ is the Fermi Dirac distribution at the chemical potential of the right electrode, and $\bar{\Gamma}_\mathrm{R}(E)$ is defined in Eq. (\ref{eq:gammar}).

With $\bar\Gamma^\mathrm{MB}(E)$ defined in Eq. (\ref{eq:gammamb}), the following relation is fulfilled
\begin{eqnarray}
&\bar{G}&^\mathrm{MB}(E)-\bar{G}^{\mathrm{MB}\supi{\dagger}}(E)=\label{eq:grelations}\\
&=&-i\bar{G}^\mathrm{MB}(E)\left[\bar{\Gamma}_\mathrm{L}(E)+\bar{\Gamma}_\mathrm{R}(E)+\bar{\Gamma}^\mathrm{MB}(E)\right]\bar{G}^{\mathrm{MB}\supi{\dagger}}(E)\nonumber\\
&=&-i\bar{G}^{\mathrm{MB}\supi{\dagger}}(E)\left[\bar{\Gamma}_\mathrm{L}(E)+\bar{\Gamma}_\mathrm{R}(E)+\bar{\Gamma}^\mathrm{MB}(E)\right]\bar{G}^\mathrm{MB}(E),\nonumber
\end{eqnarray}
and the lesser GF is given by
\begin{eqnarray}
\bar{G}^{\mathrm{MB},<}&(&E)=\nonumber\\
&&{\bar{G}^\mathrm{MB}}\left[i f_\mathrm{L}(E)\bar{\Gamma}_\mathrm{L}(E)+i f_\mathrm{R}(E)\bar{\Gamma}_\mathrm{R}(E)+\right.\nonumber\\
&&\left.i \bar{F}^\mathrm{MB}(E)\bar{\Gamma}^\mathrm{MB}(E)\right]\bar{G}^{\mathrm{MB}\supi{\dagger}}(E),
\label{eq:glesser}
\end{eqnarray}
where we have introduced the occupation matrix of the AI, $\bar{F}_\mathrm{MB}(E)$\cite{Ness}.
Using relations (\ref{eq:grelations}) and (\ref{eq:glesser}) in Eq. (\ref{eq:ir1}), one obtains the expression for the current in Eqs. (\ref{eq:ir},\ref{eq:ielastic},\ref{eq:irai}).

In an analogous way one obtains the current $I_\mathrm{L}$ as sum of a component transmitted from the left electrode into the right electrode, $I_\mathrm{L,R}$, and a component flowing from left electrode into the AI, $I_\mathrm{L,AI}$,
\begin{equation}
I_\mathrm{L}=I_\mathrm{L,R}+I_\mathrm{L,AI}.
\label{eq:il}
\end{equation}
The value of $I_\mathrm{L,R}$ can be evaluated as
\begin{eqnarray}
I_\mathrm{R,L}&=&\frac{e}{h}\int dE \;\left(f_\mathrm{L}(E)-f_\mathrm{R}(E)\right)T_\mathrm{L,R}(E),
\label{eq:ielasticL}
\end{eqnarray}
with
\begin{eqnarray}
T_\mathrm{L,R}(E)&=&\mathrm{Tr}\left[\bar{\Gamma}_\mathrm{L}(E)  \bar{G}^\mathrm{MB}(E)\bar{\Gamma}_\mathrm{R}(E){\bar{G}^{\mathrm{MB}\supi{\dagger}}}(E) \right].\label{eq:treL}
\end{eqnarray}

The incoherent component of the current flowing from the left electrode to the AI is given by
\begin{eqnarray}
I_\mathrm{L,AI}&=&\frac{e}{h}\int dE \;\mathrm{Tr}\left[
\left(f_\mathrm{L}(E)-\bar{F}^\mathrm{MB}(E)\right)
\right.\nonumber\\
&&\left.\bar{\Gamma}^\mathrm{MB}(E) \bar{G}^{\mathrm{MB}\supi{\dagger}}(E)\bar{\Gamma}_\mathrm{L}(E) \bar{G}^\mathrm{MB}(E)\right].
\label{eq:ilai}
\end{eqnarray}

\subsection{Summary of projection algorithm}
\label{sec:appendixsum}
Summarizing, the projection method proceeds as follows:

\begin{enumerate}

\item
Define a set of interacting orbitals, $\left\{\psi_i\right\}$.

\item
Determine the extensions of EM, ER, IR, and AI; with this information subdivide the total $S$ and $H$ matrices according to Eq. (\ref{eq:ssmall}),
and set the basic projection matrices $U_\mathrm{ER}$, $U_\mathrm{NI}$, and $U_\perp$.

\item
Calculate $\tilde{S}_\mathrm{AI}={U_\mathrm{ER}}^\dagger S_\mathrm{ER} U_\mathrm{ER}$  and  $\tilde{H}_\mathrm{AI}={U_\mathrm{ER}}^\dagger H_\mathrm{ER} U_\mathrm{ER}$.

\item
Calculate $\tilde{S}_\mathrm{AI}^{-\frac{1}{2}}$, and with it $U_{\mathrm{AI}\psi}$ by solving the eigenvalue system  
$\left(\tilde{S}_\mathrm{AI}^{-\frac{1}{2}}\tilde{H}_\mathrm{AI}\tilde{S}_\mathrm{AI}^{-\frac{1}{2}}\right)U_{\mathrm{AI}\psi}=U_{\mathrm{AI}\psi}\epsilon_{\mathrm{AI},D}$. 
Then evaluate $W_{2,\mathrm{AI}}=\tilde{S}_\mathrm{AI}^{-\frac{1}{2}}U_{\mathrm{AI}\psi}$.

\item
With the stored quantities calculate $W_\mathrm{AI}$ and $W_\mathrm{NI}$, by using Eqs. (\ref{eq:wai}) and (\ref{eq:wni}); 
with these quantities the total projection matrix, $W$, is known through Eq. (\ref{eq:wshort}).

\item
The inverse projection matrix, $W^{-1}$, is obtained from Eq. (\ref{eq:winvsimple}).

\item
Required blocks of the transformed matrices, $\bar{S}$ and $\bar{H}$, can then be calculated with the relations (\ref{eq:smatexp}) and (\ref{eq:hmatexp}).

\item
The hybridization function of the AI, $\bar{\Delta}_\mathrm{AI}(z)$, is obtained by using Eq. (\ref{eq:bardelta2}), and its Green's function is obtained from Eq. (\ref{eq:gbar2}) if needed.

\item
Optionally, the asymptotic limits of $\bar{\Delta}_\mathrm{AI}(z)$ for large energies are obtained from Eqs. (\ref{eq:limitsG}--\ref{eq:m2}) 
and Eq. (\ref{eq:bardelta2}).

\item
Once the many-body self-energy of the AI, $\bar{\Sigma}^\mathrm{MB}_{AI}(z)$, is calculated, its inverse projection is obtained by using Eq. (\ref{eq:sigmambinv}); 
this then allows to evaluate the full interacting GF, $G^\mathrm{MB}(z)$, of the EM in the original basis by using Eq. (\ref{eq:gmb}). 
Note however that to evaluate the transmission this inverse projection is not required.

\end{enumerate}

\end{document}